\documentclass[10pt,journal,compsoc]{IEEEtran}

\ifCLASSOPTIONcompsoc
  \usepackage[nocompress]{cite}
\else
  \usepackage{cite}
\fi

\usepackage{cite}
\usepackage{amsmath,amssymb,amsfonts}

\usepackage{array}
\usepackage{lipsum}

\usepackage{algorithm}
\usepackage{algpseudocode}

\usepackage{xspace}
\usepackage{listings}

\usepackage{multirow}
\usepackage{subcaption}
\usepackage{graphicx}
\usepackage{booktabs}
\usepackage{textcomp}
\usepackage{xcolor}

\usepackage{marginnote}
\usepackage[backgroundcolor=white,bordercolor=blue,linecolor=blue]{todonotes}

\def\BibTeX{{\rm B\kern-.05em{\sc i\kern-.025em b}\kern-.08em
    T\kern-.1667em\lower.7ex\hbox{E}\kern-.125emX}}
\begin{document}

\title{Mutation Analysis for Cyber-Physical Systems: Scalable Solutions and Results in the Space Domain}

\author{Oscar~Cornejo, Fabrizio~Pastore,~\IEEEmembership{Member,~IEEE,} and~Lionel~C.~Briand,~\IEEEmembership{Fellow,~IEEE}%

\IEEEcompsocitemizethanks{\IEEEcompsocthanksitem O. Cornejo, F.
Pastore, and L. Briand are affiliated with SnT Centre for Security, Reliability and Trust, University of Luxembourg, Luxembourg.\protect\\
E-mail:\{oscar.cornejo,fabrizio.pastore,lionel.briand\}@uni.lu
\IEEEcompsocthanksitem L. Briand also holds a faculty appointment with school of EECS, University of Ottawa.}%
\thanks{Manuscript received December 20, 2020; revised June 16, 2021.}}

\newcommand{\SAIL}{\emph{ESAIL}\xspace}
\newcommand{\GCSP}{\emph{LIBN}\xspace}
\newcommand{\CSP}{\emph{LIBCSP}\xspace}
\newcommand{\PARAM}{\emph{LIBP}\xspace}
\newcommand{\UTIL}{\emph{LIBU}\xspace}
\newcommand{\CITSAIL}{~\cite{ESAIL}}
\newcommand{\MLFS}{\emph{MLFS}\xspace}

\newcommand{\MPTS}{\emph{MASS-reduced} test suite\xspace}
\newcommand{\MPTSs}{\emph{MASS-reduced} test suites\xspace}

\newcommand{\GomSpace}{GomSpace\xspace}
\newcommand{\LuxSpace}{LuxSpace\xspace}
\newcommand{\ONE}{GSL\xspace}
\newcommand{\TWO}{LXS\xspace}
\newcommand{\CITONE}{~\cite{GSL}}
\newcommand{\CITTWO}{~\cite{LXS}}
\newcommand{\ESA}{ESA\xspace}
\newcommand{\LAUNCH}{on September 2020~\cite{ESAILlaunch}}
\newcommand{\OPENCSP}{the open source CubeSat Space Protocol (\CSP) library~\cite{CSP}}

\newcommand{\YAGO}{Yago Isasi Parache}
\newcommand{\ExaE}{ExactEarth}

\newcommand{\EduardoSpace}{GomSpace Luxembourg\xspace}
\newcommand{\YagoSpace}{LuxSpace\xspace}
\markboth{ACCEPTED FOR PUBLICATION ON IEEE Transactions on Software Engineering}%
{Cornejo \MakeLowercase{\textit{et al.}}: Mutation Testing for Cyber-Physical Systems}

\newcommand{\APPR}{\emph{MASS}\xspace}

\newcommand{\JMR}[2]{\textcolor{black}{#2}}
\newcommand{\NEWFSCI}[1]{\textcolor{black}{#1}}

\newcommand{\JMRCHANGE}[1]{\textcolor{black}{#1}}
\newcommand{\UPDATE}[1]{\textcolor{black}{#1}}

\newcommand{\CHANGED}[1]{\textcolor{black}{#1}}

\newcommand{\CHANGEDOCT}[1]{\textcolor{black}{#1}}

\newcommand{\CHANGEDNOV}[1]{\textcolor{black}{#1}}

\newcommand{\TODO}[1]{\textcolor{black}{#1}}

\newcommand{\MR}[1]{\textcolor{black}{#1}}

\newcommand{\FIXME}[1]{\textcolor{black}{#1}}

\IEEEtitleabstractindextext{
\begin{abstract}
On-board embedded software developed for spaceflight systems (\emph{space software}) must adhere to stringent software quality assurance procedures. 
For example, verification and validation activities are typically performed and assessed by third party organizations. To further minimize the risk of human mistakes, space agencies, such as the European Space Agency (ESA), are looking for automated solutions for the assessment of software testing activities, which play a crucial role in this context. Though space software is our focus here, it should be noted that such software shares the above considerations, to a large extent, with embedded software in many other types of cyber-physical systems.

Over the years, mutation analysis has shown to be a promising solution for the automated assessment of test suites; 
\JMRCHANGE{it consists of measuring the quality of a test suite in terms of the percentage of injected faults leading to a test failure.}
A number of optimization techniques, addressing scalability and accuracy problems, have been proposed to facilitate the industrial adoption of mutation analysis.
However, to date, two major problems prevent space agencies from enforcing mutation analysis in space software development. First, there is uncertainty regarding the feasibility of applying mutation analysis optimization techniques in their context. Second, most of the existing techniques either can break the real-time requirements common in embedded software or cannot be applied when the software is tested in Software Validation Facilities, including CPU emulators and sensor simulators.

In this paper, we enhance mutation analysis optimization techniques to enable their applicability to embedded software and propose a pipeline that successfully integrates them to address scalability and accuracy issues in this context, as described above.
Further, we report on the largest study involving embedded software systems in the mutation analysis literature. 
Our research is part of a research project funded by ESA ESTEC involving private companies (\EduardoSpace and \YagoSpace) in the space sector. These industry partners provided the case studies reported in this paper;
they include an on-board software system managing a microsatellite currently on-orbit, a set of libraries used in deployed cubesats, and a mathematical library certified by ESA.

\end{abstract}

}

\maketitle

\IEEEdisplaynontitleabstractindextext
\IEEEpeerreviewmaketitle

\IEEEraisesectionheading{\section{Introduction}\label{sec:introduction}}

\IEEEPARstart{F}{rom} spacecrafts to ground stations, software has a prominent role in space systems; for this reason, the success of space missions depends on the quality of the system hardware as much on the dependability of its software. Mission failures due to insufficient software sanity checks~\cite{Schiaparelli} are unfortunate examples, pointing to the necessity for systematic and predictable quality assurance procedures in space software. 

Existing standards for the development of space software regulate software quality assurance and emphasize its importance. 
The most stringent regulations are the ones that concern flight software, i.e., embedded software installed on spacecrafts, our target in this paper.
In general, software testing plays a prominent role among  quality assurance activities for space software, and standards put a strong emphasis on the quality of test suites. For example, the European Cooperation for Space Standardization (ECSS) provides detailed guidelines for the definition and assessment of test suites~\cite{ecss80C,ecss40C}. 

Test suites assessment 
is typically based on code inspections performed by space authorities and
independent software validation and verification (ISVV) activities, which include the verification of test procedures and data \JMR{3.6}{(e.g., ensure that all the requirements have been tested and that representative input partitions have been covered~\cite{ISVV})}. Though performed by specialized teams, such assessment is manual and thus error prone and time-consuming. \textbf{Automated and effective methods to evaluate the quality of the test suites are thus necessary.}

Since one of the primary objectives of software testing is to identify the presence of software faults, an effective way to assess the quality of a test suite consists of artificially injecting faults in the software under test and verifying the extent to which the test suite can detect them. 
This approach is known as \emph{mutation analysis}~\cite{DeMillo78}. 
In mutation analysis, faults are automatically injected in the program through automated procedures referred to as mutation operators. Mutation operators enable the generation of faulty software versions that are referred to as \emph{mutants}.  
Mutation analysis helps evaluate the effectiveness of a test suite, \JMRCHANGE{for a specific software system,} based on its mutation score, which is the percentage of mutants leading to test failures.

Despite its potential, mutation analysis is not widely adopted by industry in general and space system development in particular. The main reasons include its limited scalability and the pertinence of the mutation score as an adequacy criterion~\cite{papadakis2016threats}. Indeed, for a large software system, the number of generated mutants might prevent the execution of the test suite against all the mutated versions. Also, the generated mutants might be either 
semantically equivalent to the original software~\cite{madeyski2013overcoming} or redundant with each other~\cite{Shin:TSE:DCriterion:2018}.
Equivalent and redundant mutants may bias the mutation score as an adequacy criterion. 

The mutation analysis literature has proposed several optimizations to address problems related to scalability and mutation score pertinence. 
For example, scalability problems are addressed by approaches that sample mutants~\cite{zhang2013operator}~\cite{gopinath2015hard}
or solutions that prioritize and select the test cases to be executed for each mutant~\cite{zhang2013faster}. Equivalent and redundant mutants can be detected by comparing the code coverage of the original program and its mutants~\cite{grun2009impact,schuler2010covering,schuler2013covering,schuler2009efficient}. 
However, these approaches 
have not been evaluated on industrial, embedded systems
and there are no feasibility studies concerning the integration of such optimizations and their resulting, combined benefits.
For example, we lack mutants sampling solutions that accurately estimate the mutation score in the presence of reduced test suites;
indeed, mutant sampling, which comes with a certain degree of inaccuracy, may lead to inaccurate results when applied to reduced test suites that do not have the same effectiveness of the original test suite.

Finally, there is no work on identifying and assessing approaches that are feasible and effective for embedded software in general and space software in particular. Such software is very different than other types of software systems (e.g., Java graphical libraries or Unix utility programs) in several ways. More precisely, space software -- like many other types of embedded software within cyber-physical systems (CPS) \cite{KapinskiCPS2016,ZhouCPS2018} -- \JMR{3.5}{presents a combination of characteristics that altogether substantially limit the applicability of existing mutation testing optimizations.}
First, this type of software normally contains many functions to deal with signals and data transformation, which 
\JMRCHANGE{may diminish the effectiveness of approaches---both compiler-based and coverage-based---to identify equivalent and redundant mutants. Indeed, the proportion of equivalent and redundant mutants detected by compiler optimization approaches may vary for different software systems~\cite{papadakis2015trivial,delgado2018evaluation}. In addition, coverage-based approaches~\cite{grun2009impact,schuler2010covering,schuler2013covering} may not be effective with systems performing a large number of mathematical operations.}  
Second, embedded software for CPS is thoroughly tested with test suites (e.g., to satisfy functional safety standards) that may take hours to execute\JMR{3.5}{\footnote{Although this might be also true for other types of software, in the context of CPS it is due to a large input space including inputs in the continuous domain (e.g., signals) combined with long test execution times (e.g., to observe a specific signal shape or to verify diverse combinations of inputs).}}, thus exacerbating scalability problems.
Finally, it requires dedicated hardware, \JMRCHANGE{software emulators\footnote{Software emulators are used to test software compiled for specific hardware architectures.}}, or simulators\JMR{3.1}{\footnote{In this paper, we use the term simulator to indicate software that model complex environments, including physical phenomena and hardware.}}, which affect the applicability of optimizations that make use of multi-threading or other OS functions~\cite{untch1993mutation}. 
\JMR{1.2}{The reliance on dedicated hardware, emulators, and simulators also prevents the use of static program analysis to detect equivalent mutants~\cite{holling2016nequivack,riener2011test,Chekam2021}}.
\JMR{1.2 3.1}{Such characteristics are common to embedded software in other types of CPS domains including avionics, automotive, and industry 4.0 (e.g., robotics systems).}

\JMR{3.1}{In this paper, \emph{we define and evaluate a mutation analysis pipeline} to assess the quality of test suites targeting embedded software developed for spaceflight systems or the many other CPS with similar characteristics.
We call our mutation analysis pipeline: 
\emph{Mutation Analysis for Space Software} (\APPR).
To account for the above-mentioned limitations, we propose and sometimes adapt a set of mutation analysis optimizations.
More precisely,} \APPR combines (i) trivial compiler optimizations, to remove equivalent and redundant mutants after their generation, 
(ii) mutation sampling~\cite{zhang2013operator}~\cite{gopinath2015hard}, to reduce the number of mutants to execute,
(iii) code coverage information regarding the original software~\cite{zhang2013faster}, to prioritize and select test cases,
and (iv) code coverage information regarding the mutated files~\cite{schuler2013covering}, to further detect equivalent and redundant mutants.
\APPR is based on analyses that are feasible with large, real-time embedded software, including an incremental compilation pipeline to scale up the generation of mutants and the collection of coverage data focused on mutated files only.
To provide statistical guarantees about the accuracy of mutants sampling, we propose to sample mutants by relying on sequential analysis based on the fixed size confidence interval approach (FSCI)~\cite{Frey:FixedWidthSequentialConfidenceIntervals:AmericanStatistician:2010}. Also, we extend FSCI to provide accurate results in the presence of reduced test suites.
Finally, to effectively use code coverage information for test suite prioritization and equivalent/redundant mutants detection, we \JMR{3.16} apply four different distance metrics to compare coverage results between test cases: Jaccard, Ochiai, Euclidean distance, and Cosine similarity.

To evaluate the effectiveness of the proposed approach, we rely on a space software benchmark provided by our industry partners, which are the European Space Agency~\cite{ESA}, 
\EduardoSpace{} (\JMR{3.16}{\ONE}), a manufacturer and supplier of nanosatellites\CITONE, and
\YagoSpace{} (\TWO), a developer of infrastructure products (e.g., microsatellites) and solutions for space\CITTWO.
Our benchmark consists of (1) the on-board embedded 
{software system (service layer, high-level drivers, application layer)} for \SAIL{}\CITSAIL{}, a maritime microsatellite launched into space \LAUNCH{}, 
(2) a set of libraries designed for Cubesats, including a network-layer delivery library, a utility library, and a satellite configuration library,
and (3) a mathematical library for flight software~\cite{MLSF}.
To the best of our knowledge, \textbf{this is the first study proposing and assessing a mutation analysis optimization pipeline in the context of testing embedded software in cyber-physical systems}. 
\JMR{1.2 3.1}{Though our benchmark does not include embedded software for other CPS domains, we believe that its characteristics, as discussed above, make it representative of embedded software across many CPS domains, which has never been considered in mutation analysis studies (see Section~\ref{sec:related}).}

In our empirical evaluation, we assess the validity, in the space software context, of the reported scientific findings concerning the state-of-the-art mutation analysis optimizations integrated into \APPR. 
Further, we evaluate the feasibility of their integration into the \APPR pipeline by reporting on the accuracy of the estimated mutation score and the execution time savings obtained with \APPR.
Our results show that (1) different compiler optimization options are complementary for the detection of trivially equivalent or duplicate mutants; we confirm related work findings about the possibility of discarding 30\% of the mutants that way.
(2) The proposed FSCI-based mutant sampling strategy outperforms state-of-the-art strategies; indeed, it is the only approach that minimizes the number of mutants selected while providing accuracy guarantees both when a full test suite or a prioritized subset of it are executed. (3) The proposed test suite selection and prioritization approach enables an impressive reduction of mutation analysis time (above \JMRCHANGE{70\%}), thus making mutation analysis feasible also for large and complex space software. (4) Small differences in code coverage enable the identification of nonequivalent mutants. 

\JMR{1.1 2.3}{To summarize, our contributions include the following:
\begin{itemize}
\item A \textbf{mutation analysis pipeline} %
targeting \textbf{embedded software for CPS} that innovatively combines (a) state-of-the-art techniques (i.e., equivalent mutants detection with trivial compiler optimizations), (b) improvements to techniques proposed in related work (i.e., coverage-based prioritization and selection of test cases~\cite{zhang2013faster},
along with coverage-based detection of equivalent and redundant mutants~\cite{schuler2013covering}), and (c) new techniques for mutant sampling. 
\item To \textbf{address} \textbf{scalability problems}, which are acute in the case of embedded software for CPS, we propose FSCI-based mutant sampling. In addition, we also apply a test suite prioritization and selection strategy~\cite{zhang2013faster} adapted to work in the CPS context (i.e., to deal with test cases with long execution times and in the presence of mutants sampling). .
\item To \textbf{address}  problems related to the \textbf{pertinence of mutation scores}---which is an open problem in the case of CPS since it is not possible to rely on static program analysis to detect equivalent mutants,
 we adapt a strategy based on code coverage~\cite{schuler2013covering} to work with embedded software while avoiding intrusive monitoring (e.g., collecting return values).
\item To evaluate the effectiveness of \APPR, which integrates strategies that may interfere with each other and had never been evaluated with embedded software for CPS, we rely on an \textbf{industrial benchmark} including space software currently on orbit.
\end{itemize}}

The paper proceeds as follows. Section~\ref{sec:background} presents state-of-the-art mutation analysis techniques and assesses their applicability to space software. Section~\ref{sec:approach} presents our proposed pipeline for the mutation analysis of embedded software. Section~\ref{sec:evaluation} presents our empirical evaluation in the space domain. Section~\ref{sec:related} presents related work. Section~\ref{sec:conclusion} concludes the paper.

\section{Background and Applicability of state-of-the-art mutation analysis techniques to space software}
\label{sec:background}

In this section, we discuss the applicability of state-of-the-art mutation analysis optimizations in the context of space software. 
Mutation analysis can drive the generation of test cases, which is referred to as \emph{mutation testing} in the literature.
A detailed overview of mutation testing and analysis solutions and optimizations can be found in recent surveys~\cite{jia2010analysis,papadakis2019mutation}.

\subsection{Mutation Adequacy and Mutation Score computation}
\label{background:adequacy}
A mutant is said to be killed if at least one test case in the test suite fails when exercising the mutant.
Mutants that do not lead to the failure of any test case are said to be live.
Three conditions should hold for a test case to kill a mutant: \emph{reachability} (i.e, the test case should execute the mutated statement), \emph{necessity} (i.e., the test case should reach an incorrect intermediate state after executing the mutated statement), and \emph{sufficiency} (i.e., the final state of the mutated program should differ from that of the original program)~\cite{offutt1997automatically}.

The mutation score, i.e., the percentage of killed mutants, is a quantitative measure of the quality of a test suite. Recent studies have shown that achieving a high mutation score improves significantly the fault detection capability of a test suite
~\cite{papadakis2018mutation},  
\JMR{1.6}{
a result which contrasts with that of structural coverage measures~\cite{Chekam:17}. However, a very high mutation score (e.g., above 75\%) is required to achieve a higher fault detection rate than the one obtained with other coverage criteria, such as statement and branch coverage~\cite{Chekam:17}. In other words, there exists a strong association between a high mutation score and a high fault revelation capability for test suites.}

The capability of a test case to kill a mutant also depends on the observability of the program state. To overcome the limitations due to observability, different strategies to identify killed mutants can be adopted; they are known as strong, weak, firm, and flexible mutation coverage~\cite{ammann2016introduction}. 
\JMR{1.11 3.9}{With strong mutation, to kill a mutant, there shall be an observable difference between the outputs of the original and mutated programs.  
With weak mutation, the state (i.e., the valuations of the program variables in scope) of the mutant shall differ from the state of the original program, after the execution of the mutated statement~\cite{Lee1994}. 
With firm mutation, the state of the mutant shall differ from the state of the original program at execution points between the first execution of the mutated statement and the termination of the program~\cite{Woodward88}. 
Flexible mutation coverage consists of checking if the mutated code leads to object corruption~\cite{mateo2012validating}}.
For space software, we suggest to rely on strong mutation because it is the only criterion that truly assesses the fault detection capability of the test suite; indeed, it relies on a mutation score that reflects the percentage of mutants leading to test failures. With the other mutation coverage criteria, a mutant is killed if the state of the mutant after  execution of the mutated statement differs from the one observed with the original code, without any guarantee that either the erroneous values in state variables will propagate or the test oracles will detect them.

\subsection{Mutation Operators}
\label{sec:related:operators}

Mutation analysis introduces small syntactical changes into the code (source code or machine code) of a program through a set of mutation operators that simulate programming mistakes.

The  \emph{sufficient set of operators} is widely used for conducting empirical evaluations ~\cite{offutt1996experimental,rothermel1996experimental,andrews2005mutation,kintis2017detecting}. 
The original sufficient set, defined by Offutt et al., is composed of the following operators: Absolute Value Insertion (ABS), Arithmetic Operator Replacement (AOR), Integer Constraint Replacement (ICR), Logical Connector Replacement (LCR), Relational Operator Replacement (ROR), and Unary Operator Insertion (UOI)~\cite{offutt1996experimental}.
Andrews et al.~\cite{andrews2005mutation} have also included the 
\emph{statement deletion operator} (SDL)~\cite{delamaro2014designing}, which ensures that every pointer-manipulation and field-assignment statement is tested.

The sufficient set of operators enables an accurate estimation of the mutation score of a test suite~\cite{siami2008sufficient}; furthermore, the mutation score computed with the sufficient set is a good estimate of the fault detection rate (i.e., the portion of real faults discovered) of a test suite~\cite{andrews2005mutation,Just:RealFaults:2014}. 

However, empirical work has shown that, to maximize the detection of real faults, a set of operators should be used in addition to the sufficient set: Conditional Operator Replacement (COR),
Literal Value Replacement (LVR), and Arithmetic Operator Deletion (AOD)~\cite{Kintis2018}. 

\CHANGED{The SDL operator has inspired the definition of mutation operators (e.g., \emph{OODL operators}) that delete portions of program statements, with the objective of replacing the sufficient set with a simpler set of mutation operators.
The OODL mutation operators include the delete Arithmetic (AOD), Bitwise (BOD), Logical (LOD), Relational (ROD), and Shift (SOD) operators.}
\JMR{3.8}{Empirical results show that deletion operators produce significantly fewer equivalent mutants\footnote{\JMRCHANGE{For example, statement deletion can lead to equivalent mutants only if statements are redundant, which is unlikely~\cite{Offut:2013}.}}}
\cite{delamaro2014designing,delamaro2014experimental} and, furthermore, 
test suites that kill mutants generated with both SDL and OODL operators kill a very high percentage of all mutants (i.e., 97\%)~\cite{delamaro2014experimental}. 

Another alternative to the sufficient set of operators is the generation of \emph{higher order mutants}, which result from the application of multiple mutation operators for each mutation~\cite{jia2009higher,kintis2010evaluating,offutt1992investigations,papadakis2010empirical}. However, higher order mutants are \CHANGED{easier to  kill
than the first order ones (i.e., less effective to assess test suites limitations)}~\cite{papadakis2010mutation,papadakis2019mutation}, and there is 
\CHANGED{limited empirical evidence regarding which mutation operators should be combined to resemble real faults and minimize the number of redundant mutants~\cite{papadakis2019mutation}.}

\subsection{Compile-time Scalability}
\label{sec:compile:time}

\JMR{1.12}{The potentially large size of the software under test, combined with the large number of available mutation operators, may make the compilation of all  mutants infeasible.}

To reduce the number of invocations to the compiler to one, \emph{mutant schemata} include all the mutations into a single executable~\cite{untch1993mutation}. 
With mutant schemata, the mutations to be tested are selected at run-time through configuration parameters. This may lead to a compilation speed-up of 300\% \cite{papadakis2010automatic}.

Another solution to address compile-time scalability issues consists of \emph{mutating machine code}  (e.g., binary code~\cite{becker2012xemu}, assembly language~\cite{crouzet2006sesame},
Java bytecode~\cite{ma2006mujava}, 
 and
.NET bytecode~\cite{derezinska2011object}), thus avoiding the execution of the compilation process after creating a mutant. 
A common solution consists of mutating the
 LLVM Intermediate Representation (IR) \cite{hariri2016evaluating}, 
which enables the development of mutants that work with multiple programming languages~\cite{hariri2019comparing} and facilitates the integration of optimizations based on dynamic program analysis~\cite{denisov2018mull}.

Unfortunately, the mutation of machine code 
may lead to mutants that are not representative of real faults \JMRCHANGE{(i.e., faults caused by human mistakes at development time)} because they are impossible to generate from the source code\JMR{3.10}{~\cite{denisov2018mull}.
For instance, a function invocation in the source code may lead to hundreds of machine code instructions (e.g., the function call \emph{std::vector::push\_back} leads to 200 LLVM IR instructions) and, consequently, some of the mutants derived from such instructions cannot be derived by mutating the source code.}
In the case of IR mutation, some of these impossible mutants can be automatically identified~\cite{denisov2018mull}; however,
the number of generated mutants tend to be higher at the IR level than at the source code level, which may reduce scalability~\cite{hariri2019comparing}.
 In addition, we have encountered three problems that prevented the application of 
 mutation analysis tools based on LLVM IR to our case study systems.
First, space software relies on compiler pipelines (e.g., RTEMS~\cite{RTEMS}) that include architecture-specific optimizations not supported by LLVM. 
Second, there is no guarantee that the executables generated by LLVM are equivalent to those produced by the original compiler.
 Third, efficient toolsets based on LLVM often perform mutations dynamically~\cite{denisov2018mull}, which is infeasible when the software under test needs to be executed within a dedicated simulator, a common situation with space software and many other types of embedded software in cyber-physical systems.

\subsection{Runtime Scalability}
\label{sec:scalability}

A straightforward mutation analysis process consists of executing the full test suite against every mutant; however, it may lead to scalability problems in the case of a large software under test (SUT) with expensive test executions.
\emph{Simple optimizations} that can be applied to space software consist of (S1) stopping the execution of the test suite when the mutant has been killed, (S2) executing only those test cases that cover the mutated statements~\cite{delamaro1996proteum}, and (S3) rely on timeouts to automatically detect infinite loops introduced by mutation~\cite{papadakis2019mutation}. 

\emph{Split-stream execution} consists of generating a modified version of the SUT that creates multiple processes (one for each mutant) only when the mutated code is reached \cite{king1991fortran,tokumoto2016muvm}, thus saving time and resources. Unfortunately, it cannot be applied in the case of space software that needs to run with simulators because, in general, the hosting simulator cannot be forked by the hosted SUT.

Another feasible solution consists of  \emph{randomly selecting a subset of the generated mutants}~\cite{zhang2010operator,gopinath2015hard,zhang2013operator}. 
Zhang et al. \cite{zhang2013operator} empirically demonstrated that a random selection of 5\% of the mutants is sufficient for 
estimating, with high confidence, the mutation score obtained with the complete mutants set.
Further,
they show that sampling mutants uniformly across different program elements (e.g., functions) %
leads to a more accurate mutation score prediction than sampling mutants globally in a random fashion. 
For large software systems that lead to thousands of mutants, random mutation analysis is the only viable solution. However, for very large systems such as the ones commonly found in industry, randomly selecting 5\% of the mutants may still be too expensive.

\CHANGEDOCT{Gopinath et al. estimate the number of mutants required for an accurate mutation score~\cite{gopinath2015hard}.
They rely on the intuition that, under the assumption of independence between mutants,
mutation analysis can be seen as a Bernoulli experiment in which the outcome of the test for a single mutant is a Bernoulli trial (i.e., mutant successfully killed or not) and, consequently, 
the mutation score should follow a binomial distribution.
They rely on Tchebysheff’s inequality~\cite{Tchebichef1867} to find a theoretical lower bound on the number of mutants required for an accurate mutation score. 
More precisely, they suggest that, with 1,000 mutants, 
the estimated mutation score differs from the real mutation score at most by 7 percentage points.
However, empirical results show that the binomial distribution provides a conservative estimate of the population variance and, consequently, 1,000 mutants enable in practice a more accurate estimate ($> 97\%$) of the mutation score than expected.}

\CHANGEDNOV{In the statistics literature, the correlated binomial model~\cite{Bahadur}, 
and related models~\cite{Kupper1978,NG:ModifiedBinomialDistributions:1989,VanDerGeest:2005} are used when Bernoulli trials are not independent~\cite{Zhang:CrrelatedFirearm:NIST:2019}. 
In our work, based on the results achieved by Gopinath et al., we assume that the degree of correlation between mutants is limited and the binomial distribution can be used to accurately estimate the mutation score,
which is supported by our empirical results (see Section~\ref{sec:evaluation}).
  In Appendix B,
  we verify the correctness of our assumptions 
by reporting the degree of association between trials and by comparing the  probability mass function for the binomial and the correlated binomial distributions, for all our subjects.}

The statistics literature also provides a number of approaches for the computation of a sample size (i.e., the number of mutants, in our context) that enables estimates with a given degree of accuracy~\cite{Krejcie,Cochran,Bartlett,Krishnamoorthy07}. 
For binomial distributions, the most recent work is that of Gonçalves et al.~\cite{Goncalves2012}, that
determines the sample size by
relying on heuristics for the computation of confidence intervals for binomial proportions. 
A confidence interval has a probability $p_c$ (the confidence level) of including the estimated parameter (e.g., the mutation score). 
Results show that the largest number of samples required to compute a 95\% confidence interval is 1,568. 

If used to drive the selection of mutants, both the approaches of Gopinath et al. and Gonçalves et al., which suggest sampling at least 1,000 mutants, may be impractical when mutants are tested with large system test suites.

\CHANGED{An alternative to computing the sample size before performing an experiment is provided by sequential analysis approaches, which determine the sample size while conducting a statistical test~\cite{waldSequential}. 
Such approaches do not perform worst-case estimates and may thus lead to smaller sample sizes. For example, the sequential probability ratio test, which can be used to test hypotheses, has been used in mutation analysis as a condition to determine when to stop test case generation (i.e., when the mutation score is above a given threshold)~\cite{Hsu:90}. In our context, we are interested in point estimation, not hypothesis testing; in this case, 
the sample size can be determined through a fixed-width sequential confidence interval (FSCI), i.e., by computing the confidence interval after every new sample and then stop sampling when the interval is within a desired bound~\cite{Frey:FixedWidthSequentialConfidenceIntervals:AmericanStatistician:2010,Chen2013,Yaacoub:OptimalStopping}. 
Concerning the method used to compute the confidence interval in FSCI, the statistics literature~\cite{Frey:FixedWidthSequentialConfidenceIntervals:AmericanStatistician:2010} reports that the Wald method~\cite{WaldMethodVollset} minimizes the sample size but requires an accurate variance estimate. We will therefore resort to a non-parametric alternative, which is Clopper-Pearson~\cite{ClopperPearson}.
Note that FSCI 
has never been applied to determine the number of mutants to consider in mutation analysis.}

Other solutions to address \emph{runtime scalability problems} in mutation analysis  aim to \emph{prioritize test cases} to maximize the likelihood of executing first those that kill the mutants~\cite{just2012using,papadakis2011automatically,zhang2013faster}. The main goal is to save time by preventing the execution of a large subset of the test suite, for each mutant.
Previous work aimed at prioritizing faster test cases~\cite{just2012using} but this 
 may not be adequate with system-level test suites whose test cases have similar, long execution times.
Approaches that rely on data-flow analysis to identify and prioritize the test cases that likely satisfy the killing conditions~\cite{papadakis2011automatically} are prohibitively expensive and are unlikely to scale to large systems.
Other work~\cite{zhang2013faster} combines three coverage criteria:
(1) the number of times the mutated statement is exercised by the test case,
 (2) the proximity of the mutated statement to the end of the test case 
 (closer ones have higher chances of satisfying the sufficiency condition)
 , and (3) 
 the percentage of mutants belonging to the same class file of the mutated statement that were already killed by the test case. 
Criterion (3) is also used to reduce the test suite size, by only selecting the test cases above a given percentage threshold. 
Unfortunately, only criterion (1) seems applicable in our context; indeed, criterion (2) is ineffective with system test cases whose results are checked after long executions, while criterion (3) may be inaccurate when only a random, small subset of mutants is executed, as discussed above.

\subsection{Detection of Equivalent Mutants}
\label{sec:background:equivalent}

\JMR{1.13}{A mutant is equivalent to the original program when they both generate the same outputs for the same inputs.}
Although identifying equivalent mutants is an undecidable problem~\cite{madeyski2013overcoming,Bugg:Correctness:82}, several heuristics have been developed to address it. 

The simplest solution consists of relying on \emph{trivial compiler optimisations}~\cite{papadakis2015trivial, kintis2017detecting,papadakis2019mutation}, i.e., compile both the mutants and the original program with compiler optimisations enabled and then determine whether their executables match. In C programs, compiler optimisations can reduce the total number of mutants by 28\%~\cite{kintis2017detecting}.

Solutions that identify equivalent mutants based on \emph{static program analysis} (e.g., concolic execution~\cite{holling2016nequivack,Chekam2021} and bounded model checking~\cite{riener2011test}) 
show promising results \JMR{2.3}{(e.g., to automatically identify non-equivalent mutants for batch programs~\cite{Chekam2021})} but
they rely on static analysis solutions that cannot work with system-level test cases that execute with hardware and environment simulators in the loop.
Indeed, (1) simulation results cannot be predicted by pure static analysis,  (2) concolic execution tools, which rely on LLVM, cannot be run if the SUT executable should be generated with a specific compiler (see Section~\ref{sec:compile:time}),
\JMR{2.3}{ (3) there are no solutions supporting the concolic execution of large software systems within simulation environments
(state-of-the-art techniques work with small embedded software~\cite{Herdt2019}), and (4) communication among components not based on direct method invocations (e.g., through network or databases) is not supported by existing toolsets.}

Alternative solutions rely on \emph{dynamic analysis} and compare data collected when testing the original software and the mutants~\cite{grun2009impact,schuler2010covering,schuler2013covering,schuler2009efficient}.
The most extensive empirical study on the topic shows that nonequivalent mutants can be detected by counting the number of methods (excluding the mutated method) that, for at least one test case, either (1) have statements that are executed at a different frequency with the mutant, (2) generate at least one different return value, or (3) are invoked at a different frequency~\cite{schuler2013covering}. To determine if a mutant is non-equivalent, it is possible to define a threshold indicating the smallest number of methods with such characteristics. A threshold of one identifies non-equivalent mutants with an average precision above 70\% and an average recall above 60\%. This solution outperforms more sophisticated methods relying on dynamic invariants~\cite{schuler2009efficient}. Also, coverage frequency alone leads to results close to the ones achieved by including all three criteria above~\cite{schuler2013covering}.
However, such approaches require some tailoring 
because collecting all required data 
(i.e., coverage frequency for every program statement, return values of every method, frequency of invocation of every method) has a computational and memory cost that may break real-time constraints.

\subsection{Detection of Redundant Mutants}
\label{sec:background:redundant}
Redundant mutants are either \emph{duplicates}, i.e., mutants that are equivalent with each other but not equivalent to the original program, or \emph{subsumed}, i.e., mutants that are not equivalent with each other but are killed by the same test cases. 

Duplicate mutants can be detected by relying on the same approaches adopted for equivalent mutants. 

According to Shin et al., subsumed mutants should not be discarded but analyzed to augment the test suite with additional test cases that fail with one mutant only~\cite{Shin:TSE:DCriterion:2018}. 
The augmented test suite has a higher
 fault detection rate than a test suite that simply satisfies mutation coverage; however, with large software systems the approach becomes infeasible because of the lack of scalable test input generation approaches.

\subsection{Summary}
\label{sec:back:summary}

We aim to rely on the sufficient set of operators since it has been successfully used to generate a mutation score that accurately estimates the fault detection rate for software written in C and C++, languages commonly used in embedded software.
\CHANGED{Further, since recent results have reported on the usefulness of both LVR and OODL operators to support the generation of test suites with high fault revealing power~\cite{Kintis2018}, the sufficient set may be extended to include these two operators as well.}

To speed up mutation analysis by reducing the number of mutants, we should consider the SDL operator alone \CHANGED{or in combination with the OODL operators}. However, such heuristic should be carefully evaluated to determine the level of confidence we can expect.

Among compile time optimizations, only mutant schemata appear to be feasible with space software.
Concerning scalability, simple optimizations (i.e., S1, S2, and S3 in Section~\ref{sec:scalability}) are feasible. Alternative solutions are the ones relying on mutant sampling and coverage metrics. 
However, to be applied in a safety or mission critical context, mutant sampling approaches should provide guarantees about the 
level of confidence one may expect. Currently, this can only be achieved with approaches requiring a large number of sampled mutants (e.g., $1,000$).  Therefore, sequential analysis based on FSCI, which minimizes the number of samples and provides accuracy guarantees, appears to be the most appropriate solution in our context.
Further, test suite selection and prioritization strategies based on code coverage require some tailoring to cope with real time constraints.

Equivalent mutants can be identified through trivial compiler optimizations and the analysis of coverage differences; however, it is necessary to define and evaluate appropriate coverage metrics. The same approach can be adopted to identify duplicate mutants. The generation of test cases that distinguish subsumed mutants is out of the scope of this work.

\JMR{1.1 3.3}{A high-level description of a possible mutation testing pipeline was proposed in a recent survey\footnote{The main objective of such pipeline was to walk the reader through the survey, not to propose a precise and feasible solution.}~\cite{papadakis2019mutation}. It consists of the following sequence of activities: select (sample) mutants, compile mutants, remove equivalent and redundant mutants, generate test inputs that kill mutants, execute mutants, compute mutation score, reduce test suites and prioritize test cases. 
Unfortunately, such pipeline does not enable the integration of many optimizations proposed above, which further motivates our work. For example, it cannot support FSCI-based sampling, which requires mutants sampling to be coupled with mutants execution. Also, it does not envision the detection of equivalent and redundant mutants based on code coverage. Moreover, it only partially addresses scalability issues since test suite reduction and prioritization are performed after mutation analysis. 
Further, it includes a test input generation step that is not feasible in the context of CPS.
Finally, it has never been implemented and therefore its feasibility has not been evaluated.}

\section{Space Software Mutation Analysis Pipeline}
\label{sec:approach}

\begin{figure}[tb]
\begin{center}
\includegraphics[width=\columnwidth]{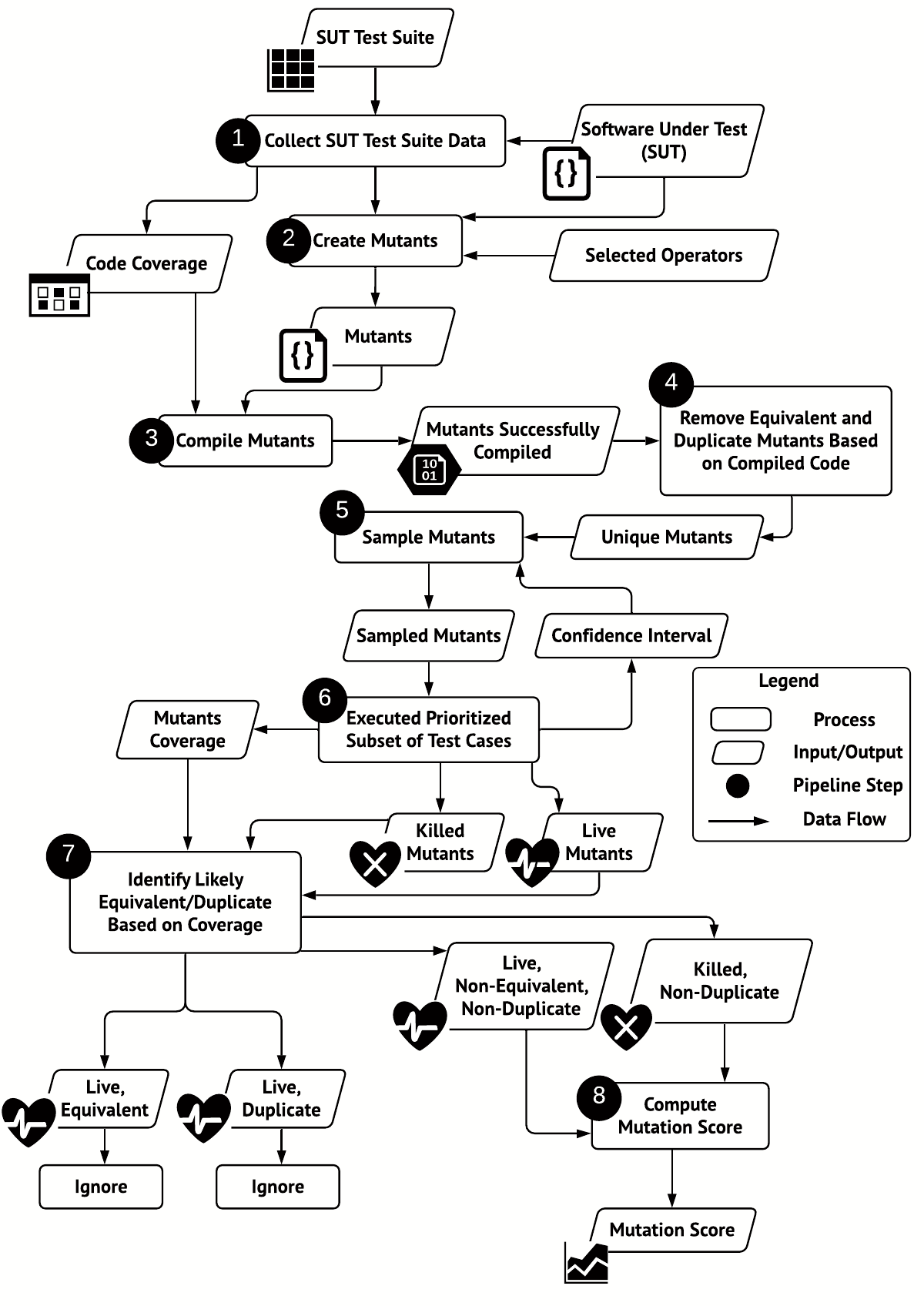}
\caption{Overview of the proposed Mutation Analysis Pipeline}
\label{fig:approach}
\end{center}
\end{figure}

Figure~\ref{fig:approach} provides an overview of the mutation analysis process that we propose, \APPR, based on the discussions and decisions in the previous section. Its goal is to propose a comprehensive solution for making mutation analysis applicable to embedded software in industrial cyber-physical systems. \JMR{3.4}{The ultimate goal of \APPR is to assess the effectiveness of test suites with respect to detecting violations of functional requirements.}
 
\JMR{3.3}{Different from the mutation analysis pipeline presented in related work~\cite{papadakis2019mutation}, \APPR enables the integration of all mutation analysis optimization techniques that are feasible in our context to address scalability and pertinence problems (see Section~\ref{sec:back:summary}). 
\APPR consists of eight steps: (Step 1) Collect SUT Test Suite Data, (Step 2) Create Mutants, (Step 3) Compile Mutants, (Step 4) Remove Equivalent and Duplicate Mutants Based on Compiled Code, (Step 5) Sample Mutants,  (Step 6) Execute Prioritized Subset of Test Cases, 
(Step 7) Identify Likely Equivalent / Duplicate mutants Based on Coverage, and
(Step 8) Compute the Mutation Score. Different from related work, \APPR enables FSCI-based sampling by iterating between mutants sampling (Step 5) and test cases execution (Step 6). 
Also, it integrates test suite prioritization and reduction (Step 6) before the computation of the mutation score.
Finally, it includes methods to identify likely equivalent and duplicate mutants based on code coverage (Step 7).}
We describe each step in the following paragraphs. 

\subsection{Step 1: Collect SUT Test Data}

In Step 1, the test suite is executed against the SUT 
and code coverage information is collected. 
More precisely, we rely on the combination of gcov~\cite{GCOV}
and GDB~\cite{GDB}, enabling the collection of coverage information for embedded systems without a file system~\cite{THANASSIS}.

\subsection{Step 2: Create Mutants}

In Step 2, we automatically generate mutants for the SUT by relying on a set of selected mutation operators.
In \APPR, based on the considerations provided in Section~\ref{sec:related:operators}, we rely on an extended sufficient set of mutation operators, which are listed in Table~\ref{table:operators}.
In addition, in our experiments, we also evaluate the feasibility of relying only on the SDL operator, combined or not with OODL operators, instead of the entire sufficient set of operators.

\newcommand{\op}{\mathit{op}}
\newcommand{\ArithmeticSet}{ \texttt{+}, \texttt{-}, \texttt{*}, \texttt{/}, \texttt{\%} }
\newcommand{\LogicalSet}{ \texttt{&&}, \texttt{||} }
\newcommand{\RelationalSet}{ \texttt{>}, \texttt{>=}, \texttt{<}, \texttt{<=}, \texttt{==}, \texttt{!=} }
\newcommand{\BitWiseSet}{ \texttt{\&}, \texttt{|}, \land }
\newcommand{\ShiftSet}{ \texttt{>>}, \texttt{<<} }

\begin{table}[tb]
\caption{Implemented set of mutation operators.}
\label{table:operators} 
\centering
\scriptsize
\begin{tabular}{|@{}p{4mm}@{}|@{}p{1cm}@{\hspace{1pt}}|@{}p{7.1cm}@{}|}
\hline
&\textbf{Operator} & \textbf{Description$^{*}$} \\
\hline
\multirow{7}{*}{\rotatebox{90}{\emph{Sufficient Set}}}&ABS               & $\{(v, -v)\}$	\\
\cline{2-3}
&AOR               & $\{(\op_1, op_2) \,|\, \op_1, \op_2 \in \{ \ArithmeticSet \} \land \op_1 \neq \op_2 \} $       \\
&    			  & $\{(\op_1, \op_2) \,|\, \op_1, \op_2 \in \{\texttt{+=}, \texttt{-=}, \texttt{*=}, \texttt{/=}, \texttt{\%} \texttt{=}\} \land \op_1 \neq \op_2 \} $       \\
\cline{2-3}
&ICR               & $\{i, x) \,|\, x \in \{1, -1, 0, i + 1, i - 1, -i\}\}$           \\
\cline{2-3}
&LCR               & $\{(\op_1, \op_2) \,|\, \op_1, \op_2 \in \{ \texttt{\&\&}, || \} \land \op_1 \neq \op_2 \}$            \\
&				  & $\{(\op_1, \op_2) \,|\, \op_1, \op_2 \in \{ \texttt{\&=}, \texttt{|=}, \texttt{\&=}\} \land \op_1 \neq \op_2 \}$            \\
&				  & $\{(\op_1, \op_2) \,|\, \op_1, \op_2 \in \{ \texttt{\&}, \texttt{|}, \texttt{\&\&}\} \land \op_1 \neq \op_2 \}$            \\
\cline{2-3}
&ROR               & $\{(\op_1, \op_2) \,|\, \op_1, \op_2 \in \{ \RelationalSet \}\}$            \\
&				  & $\{ (e, !(e)) \,|\, e \in \{\texttt{if(e)}, \texttt{while(e)}\} \}$ \\
\cline{2-3}
&SDL               & $\{(s, \texttt{remove}(s))\}$            \\
\cline{2-3}
&UOI               & $\{ (v, \texttt{--}v), (v, v\texttt{--}), (v, \texttt{++}v), (v, v\texttt{++}) \}$            \\   
\hline
\hline
\multirow{5}{*}{\rotatebox{90}{\emph{OODL}}}&AOD               & $\{((t_1\,op\,t_2), t_1), ((t_1\,op\,t_2), t_2) \,|\, op \in \{ \ArithmeticSet \} \}$       \\ 
\cline{2-3}
&LOD               & $\{((t_1\,op\,t_2), t_1), ((t_1\,op\,t_2), t_2) \,|\, op \in \{ \texttt{\&\&}, || \} \}$       \\ 
\cline{2-3}
&ROD               & $\{((t_1\,op\,t_2), t_1), ((t_1\,op\,t_2), t_2) \,|\, op \in \{ \RelationalSet \} \}$       \\ 
\cline{2-3}
&BOD               & $\{((t_1\,op\,t_2), t_1), ((t_1\,op\,t_2), t_2) \,|\, op \in \{ \BitWiseSet \} \}$       \\ 
\cline{2-3}
&SOD               & $\{((t_1\,op\,t_2), t_1), ((t_1\,op\,t_2), t_2) \,|\, op \in \{ \ShiftSet \} \}$       \\ 
\hline
\hline
\multirow{2}{*}{\rotatebox{90}{\emph{Other}}}&LVR			& $\{(l_1, l_2) \,|\, (l_1, l_2) \in \{(0,-1), (l_1,-l_1), (l_1, 0), $\\
&&\hspace{5mm}$(\mathit{true}, \mathit{false}), (\mathit{false}, \mathit{true})\}\}$           \\
\hline
\end{tabular}

$^{*}$Each pair in parenthesis shows how a program element is modified by the mutation operator on the left; we follow standard syntax~\cite{Kintis2018}. Program elements are literals ($l$), integer literals ($i$), boolean expressions ($e$), operators ($\op$), statements ($s$), variables ($v$), and terms ( $t_i$, which might be either variables or literals).
\end{table}

To automatically generate mutants, we have extended SRCIRor~\cite{hariri2018srciror} to include all the 
operators in Table~\ref{table:operators}. 
After mutating the original source file, our extension saves the mutated source file and keeps track of the mutation applied. Our toolset is available under the ESA Software Community Licence Permissive~\cite{ESAlicence} at the following URL \textbf{https://faqas.uni.lu/}.

\subsection{Step 3: Compile Mutants}
\label{sec:appr:compile}

In Step 3, we 
compile mutants by relying on an optimized compilation procedure that leverages the build system of the SUT. To this end, we have developed a toolset that, for each mutated source file: (1) backs-up the original source file, (2) renames the mutated source file as the original source file, (3) runs the build system (e.g., executes the command \texttt{make}), (4) copies the generated executable mutant in a dedicated folder, (5) restores the original source file. 

Build systems \JMR{1.15}{(e.g., GNU make~\cite{MAKE} driving the GCC~\cite{GCC} compiler)} create one object file for each source file to be compiled and then link these object files together into the final executable. 
\CHANGED{After the first build, in subsequent builds, 
build systems
recompile only the modified files and link them to the rest.}
For this reason, our optimized compilation procedure, which modifies at most two source files for each mutant (i.e., the mutated file and the file restored to eliminate the previous mutation), can reuse almost all the compiled object files in subsequent compilation runs, thus speeding up the compilation of multiple mutants. The experiments conducted with our subjects (Section~\ref{sec:evaluation}) have shown that 
\CHANGED{our optimization is sufficient to make the compilation of mutants feasible for large projects. Other state-of-the-art solutions introduce additional complexity (e.g., change the structure of the software under test~\cite{untch1993mutation}) that does not appear to be justified by scalability needs.}

\subsection{Step 4: Remove Equivalents and Duplicates}

In Step 4, we rely on trivial compiler optimizations to identify and remove equivalent and duplicate mutants. 
More precisely, for every available compiler optimization level (e.g., O0, O1, O2, O3, O4, Os, and Ofast for GCC), or a subset selected by engineers, \APPR re-executes Step 3 and stores the SHA-512 hash summaries of all the mutant and original executables~\cite{SHA512}.
To detect equivalent mutants, \APPR compares the hash summaries of the mutants with that of the original executable. To detect duplicate mutants but avoid combinatorial explosion, \APPR focuses its comparison of hash summaries on pairs of mutants belonging to the same source file (restricting the scope of the comparison is common practice~\cite{kintis2017detecting}).
Hash comparison allows us to (1) determine the presence of equivalent mutants (i.e., mutants having the same hash as the original executable), and (2) identify duplicate mutants (i.e., mutants with the same hash). %
Equivalent and duplicate mutants are then discarded. We compare hash summaries rather than executable files because it is much faster, an important consideration when dealing with a large number of mutants.
The outcome of Step 4 is a set of \emph{unique mutants}, i.e., mutants with compiled code that differs from the original software and any other mutant.

\subsection{Step 5: Sample Mutants}

In Step 5, \APPR samples the mutants to be executed to compute the mutation score. 
\JMR{1.8 3.3}{\APPR does not selectively generate mutants but samples them from the whole set of successfully compiled, nonequivalent, and nonduplicated mutants (result of Steps 2 to 4). This choice aims to avoid sampling bias which may result from the presence of such mutants; indeed, there is no guarantee that these mutants, if they were discarded after being sampled, would be uniformly distributed across program statements. Our choice does not affect the feasibility of \APPR since Steps 2 to 4 have negligible cost (see Section~\ref{sec:evaluation}).}

Our pipeline supports different sampling strategies: \emph{proportional uniform sampling}, \emph{proportional method-based sampling},  \emph{uniform fixed-size sampling}, and \emph{uniform FSCI sampling}. 

The strategies \emph{proportional uniform sampling} and \emph{proportional method-based sampling} were selected based on the results of Zhang et al.~\cite{zhang2013operator}, who compared eight strategies for sampling mutants. 
The former was the best performing strategy and consists of sampling mutants evenly across all functions of the SUT, i.e., sampling $r\%$ mutants from each set of mutants generated inside the same function.
The latter consists of randomly selecting $r\%$ mutants from the complete mutants set. This is included in our study because it is simpler to implement and showed to be equivalent to stratified sampling strategies, based on recent work~\cite{gopinath2015hard}.

The \emph{uniform fixed-size sampling} strategy stems from the work of Gopinath et al.~\cite{gopinath2015hard} and consists of selecting a fixed number $N_M$ of mutants for the computation of the mutation score. Based their work, with 1,000 mutants, one can guarantee an accurate estimation of the mutation score.

In this paper, we introduce the \emph{uniform FSCI sampling} strategy that determines the sample size dynamically, while exercising mutants, based on a fixed-width sequential confidence interval approach.
With \emph{uniform FSCI sampling}, we introduce a cycle between Step 6 and Step 5, such that a new mutant is sampled only if deemed necessary.
 More precisely, \APPR iteratively selects a random mutant from the set of unique mutants and exercises it using the SUT test suite. 
Based on related work, we assume that the mutation score computed with a sample of mutants follows a binomial distribution (see Section~\ref{sec:scalability}).
For this reason, to compute the confidence interval for the FSCI analysis, we rely on the Clopper-Pearson method since it is reported to provide the best results (see Section~\ref{sec:scalability}).
Mutation analysis (i.e., sampling and testing a mutant) stops when the confidence interval is below a given threshold $T_{\mathit{CI}}$ (we use $T_{\mathit{CI}}=0.10$ in our experiments). More formally, given a confidence interval 
$[\mathit{L}_{S};\mathit{U}_{S}]$, with $\mathit{L}_{S}$ and $\mathit{U}_{S}$ indicating the lower and upper bound of the interval, mutation analysis stops when the following condition holds:
\begin{equation}
\label{eq:CI:T}
(\mathit{U}_{S}-\mathit{L}_{S})<T_{\mathit{CI}}.
\end{equation}

Unfortunately, the assumption about the estimated mutation score following a binomial distribution may not hold when a subset of the test suite is executed for every mutant (which could happen in Step 6). Without going into the details behind the implementation of Step 6, which is described in Section~\ref{sec:step:prioritize}, 
we can expect that a reduced test suite may not be able to kill all the mutants killed by the entire test suite, i.e., the estimated mutation score may be affected by negative bias. Consequently, over multiple runs, the mean of the estimated mutation score may not be close to the \emph{actual mutation score} (i.e., the mutation score computed with the entire test suite exercising all the mutants for the SUT)
 but may converge to a lower value. 
To compute a correct confidence interval that includes the actual mutation score of the SUT, we thus need to take into account this negative bias.

To study the effect of negative bias on the confidence interval, we address first the relation between the actual mutation score and the mutation score computed with the reduced test suite when the entire set of mutants for the SUT is executed. 
A mutant killed by the entire test suite has a probability $P_{\mathit{KErr}}$ of not being killed by the reduced test suite.
The probability $P_{\mathit{KErr}}$  can be estimated as the proportion of mutants (erroneously) not killed by the reduced test suite 
\begin{equation}
P_{\mathit{KErr}} = \frac{|E_R|}{|M|}
\end{equation}
with 
$E_R$ being the subset of mutants that are killed by the entire test suite but not by the reduced test suite, and $M$ being the full set of mutants for the SUT.

The mutation score for the reduced test suite ($\mathit{MS}_R$) can be computed as

\begin{equation}
\small
\mathit{MS}_R=\frac{|K|-|E_R|}{|M|}=\frac{|K|}{|M|}-\frac{|E_R|}{|M|}=\mathit{MS}-\frac{|E_R|}{|M|}=\mathit{MS}-P_{\mathit{KErr}}
\end{equation}

where $K$ is the set of mutants killed by the whole test suite, $M$ is the set of all the mutants of the SUT,  and $\mathit{MS}$ is the actual mutation score. Consequently, the actual mutation score can be computed as 
\begin{equation}
\label{eq:MS}
\mathit{MS}=\mathit{MS}_R+P_{\mathit{Err}_R}
\end{equation}

We now discuss the effect of a reduced test suite on the confidence interval for a mutation score estimated with mutants sampling. When mutants are sampled and tested with the entire test suite, the actual mutation score is expected to lie in the confidence interval $[\mathit{L}_{S};\mathit{U}_{S}]$. 
\CHANGED{In the presence of a reduced test suite, we can still rely on the Clopper-Pearson method to compute the confidence interval $\mathit{CI}_R=[\mathit{L}_{R};\mathit{U}_{R}]$.
However, }
we have to take into account the probability of an error in the computation of the mutation score $\mathit{MS}_R$;  $\mathit{MS}_R$ can be lower than $\mathit{MS}$ and, based on Equation~\ref{eq:MS}, we expect the actual mutation score to lie in 
\JMR{NEW}{an interval that is shifted with respect to the interval for $\mathit{MS}_R$:}

\begin{equation}
\label{eq:CI}
\mathit{CI}=[\mathit{L}_{R}+P_\mathit{KErr};\mathit{U}_{R}+P_{\mathit{KErr}}]
\end{equation}

We can only estimate  $P_{\mathit{KErr}}$ since computing it would require the execution of all the mutants with the complete test suite, thus undermining our objective of reducing test executions. 
To do so, we can randomly select a subset $M_R$ of mutants, on which to execute the entire test suite and identify the mutants killed by the reduced test suite. %
The size of the set $M_R$ should be lower than the number of mutants we expect FSCI sampling to return, 
otherwise sampling would not provide any cost reduction benefit.
Since, for every mutant in $M_R$, we can determine if it is erroneously reported as not killed by the reduced test suite R,
we can 
estimate the probability $P_{\mathit{KErr}}$ as the percentage of such mutants.
As for the case of the mutation score, 
we assume that the binomial distribution provides a conservative estimate of the variance for $P_{\mathit{KErr}}$. 

We can estimate the confidence interval for $P_{\mathit{KErr}}$ using one of the methods for binomial distributions.
We rely on the Wilson score method because it is known to perform well with small samples~\cite{Newcombe:Wilson:1998}. 
The value of $P_{\mathit{KErr}}$ will thus lie within $\mathit{CI}_E=[\mathit{L}_{E};\mathit{U}_{E}]$,  with $\mathit{L}_{E}$ and $\mathit{U}_{E}$ indicating the lower and upper bounds of the interval.

\NEWFSCI{Based on Equation~\ref{eq:CI}, 
the confidence interval to be used with FSCI sampling in the presence of a reduced test suite should thus be }
\begin{equation}
\label{eq:CI:FSCI}
\mathit{CI}=[\mathit{L}_{R}+\mathit{L}_{E};\mathit{U}_{R}+\mathit{U}_{E}]
\end{equation}

\JMRCHANGE{The estimated mutation score is the value lying in the middle of the interval.}

\UPDATE{Since the width of the confidence interval CI (hereafter, $|CI|$) results from the sum of $|\mathit{CI}_R|$ and $|\mathit{CI}_E|$,
mutation sampling with a reduced test suite may lead to the execution of a larger set of mutants.}

\UPDATE{Based on Equations~\ref{eq:CI:T} and~\ref{eq:CI:FSCI}, $|\mathit{CI}_R| \le T_{\mathit{CI}} - |\mathit{CI}_E|$.
Consequently, when $|\mathit{CI}_E|>T_{\mathit{CI}}$, the reduced test suite cannot lead to sufficiently accurate results. 
Also, a large $|\mathit{CI}_E|$ may prevent the identification of accurate results with a feasible number of mutants. For example, Clopper-pearson may require up to 1568 samples for a confidence interval below 0.05~\cite{Goncalves2012}}. 
\UPDATE{
We shall thus identify a threshold ($T_{\mathit{CE}}$) for the confidence interval $|\mathit{CI}_E|$ that enables 
accurate estimates 
with a small sample size (e.g., in the worst case, with less than 1000 samples, the sample size for related work).
For this reason, starting from a minimal number of samples to estimate $P_{\mathit{KErr}}$ (150 in our experiments), \APPR keeps estimating $P_{\mathit{KErr}}$ until it yields $|\mathit{CI}_E| \le T_{\mathit{CE}}$. 
In our experiments we set $T_{\mathit{CE}} = 0.035$.
To select $T_{\mathit{CE}}$, we have identified a reasonable minimal mutation score to be expected in space software (i.e., 65\%) and identified, based on confidence interval estimation methods with finite population correction factor~\cite{BasicBusinessStatistics}, 
the minimal value for $|\mathit{CI}_E|$ that requires a number of samples below 850 (i.e., $1000-150$).
}

When it is not possible to estimate $|\mathit{CI}_E| \le T_{\mathit{CE}}$ \UPDATE{or when the number of samples required to estimate $|\mathit{CI}_E| \le T_{\mathit{CE}}$ is sufficient to accurately estimate the mutation score,} the test suite can be prioritized but not reduced and the confidence interval is computed using the traditional Clopper-Pearson method, i.e., $[\mathit{L}_{S};\mathit{U}_{S}]$.

\subsection{Step 6: Test Prioritization}
\label{sec:step:prioritize}

In Step 6, we execute a prioritized subset of test cases. 
We select only the test cases that satisfy 
the reachability condition (i.e., cover the mutated statement) and  execute them in sequence.
Similarly to the approach of Zhang et al. \cite{zhang2013faster}, we define the order of execution of test cases based on their estimated likelihood of killing a mutant.
\CHANGED{However, in our work, this likelihood is estimated differently since, as discussed above, the measurements they rely on are not applicable in the context of system-level testing and complex cyber-physical systems 
(see Section~\ref{sec:scalability}).} \CHANGEDNOV{In contrast, to minimize the impact of measurements on real-time constraints, we only collect code coverage information for a small part of the system.}

To reduce the number of test cases to be executed with a mutant, 
we should first execute the ones that more likely satisfy the necessity condition. 
This might be achieved by executing a test case that exercises the mutated statement with variable values not observed before. 
Unfortunately, in our context, the size of the SUT and its real-time constraints prevent us from recording all the variable values processed during testing. 

Therefore, we rely on code coverage to determine if two test case executions exercise the mutated statement with diverse variable values. Such coverage is collected by efficient procedures provided by compilers, thus having lower impact on execution performance than other types of dynamic analysis solutions (e.g., tracing variable values).
Since, because of control- and data-flow dependencies, a different set of input values may lead to differences in code coverage, 
the latter helps determine if two or more test cases likely exercise a mutated statement with different variable values. %
To increase the likelihood that the observed differences in code coverage are due to the use of different variable values to exercise the mutated statement, we restrict the scope of code coverage analysis 
to the functions belonging to the component (i.e., the source file) that contains the mutated statement.
Indeed, such functions typically present several control- and data-flow dependencies, thus 
augmenting the likelihood that a coverage difference is due to the execution of the mutated statement with a diverse set of values. Also, collecting code coverage for a small part of the system further reduces the impact of our analysis on system performance.

Based on related work, we have identified two possible strategies to characterize test case executions based on code coverage:
\begin{itemize}
\item[S1] Compare the sets of source code statements that have been covered by test cases~\cite{grun2009impact}.
\item[S2] Compare the number of times each statement has been covered by test cases~\cite{schuler2013covering}.
\end{itemize}

To determine how dissimilar two test cases are and, consequently, how likely they are to exercise the mutated statement with different values, we rely on widely adopted distance metrics. 
In the case of S1, we rely on the Jaccard and Ochiai indices, which are two similarity indices for binary data which have been successfully used to compare program executions based on code coverage~\cite{Zou:Ochiai:2019,Keller:Jaccard:2017,Briand:2019}. Given two test cases $T_A$ and $T_B$, the Jaccard  ($D_J$) and Ochiai ($D_O$) distances are computed as follows:

$D_J(T_a,T_b)=1-\frac{|C_a \cap C_b|}{|C_a \cup C_b|}$ \hspace{2mm} $D_O(T_a,T_b)=1-\frac{|C_a \cap C_b|}{\sqrt{|C_a| * |C_b|}}$, 
where $C_a$ and $C_b$ are the set of covered statements exercised by $T_a$ and $T_b$, respectively.

In the case of S2, we compute the distance between two test cases by relying on the euclidean distance ($D_E$) and the cosine similarity distance ($D_C$), two popular distance metrics used in machine learning. Given two vectors $V_A$ and $V_B$, whose elements capture the number of times a statement has been covered by test cases $T_A$ and $T_B$, the distances $D_E$ and $D_C$ can be computed as follows:

$D_E=\sqrt{\sum_{i=1}^{n}(A_i-B_i)^2}$

$D_C= 1-\frac{\sum_{i=1}^{n}A_i*B_i}{\sqrt{\sum_{i=1}^{n}{A_i}^2}*\sqrt{\sum_{i=1}^{n}{B_i}^2}}$,
where $A_i$ and $B_i$ refer to the number of times the i-th statement had been covered by $T_A$ and $T_B$, respectively.

Figure~\ref{alg:prioritize} shows the pseudocode of our algorithm for selecting and prioritizing test cases. It generates as output
a prioritized test suite (\emph{PTS}).
Based on the findings of Zhang et al. \cite{zhang2013faster}, we first select the test case that exercises the mutated statement the highest number of times (Line~\ref{alg:prioritize:first}) \CHANGED{ and add it to the prioritized test suite (Line~\ref{alg:prioritize:add}).}
Then, in the next iterations, the test case selected is the one with the largest distance from the closest test case already selected (Lines~\ref{alg:prioritize:selectStart} to~\ref{alg:prioritize:selectEnd}). 
When two or more test cases have the same distance, we select randomly among the test cases that exercise the mutated statement the most.

The algorithm iterates as long as it identifies a test case
showing a difference in code coverage from the 
already selected test cases (Line~\ref{alg:prioritize:until}).

Test cases are then executed in the selected order. During  execution, we collect code coverage information and identify killed and live mutants.

\newcommand{\INDA}{10}
\newcommand{\INDB}{15}
\newcommand{\INDC}{5}

\begin{figure}[tb]

\begin{algorithmic}[1]

\scriptsize

\Require \emph{TS}, the test suite of the software under test
\Require \emph{Cov}, coverage information, for each test case
\Require \emph{ms}, the mutated statement
\Ensure \emph{PTS}, a list of test cases to be executed, sorted by priority

\State $\mathit{TS}_m \gets$ subset of $\mathit{TS}$ that cover the mutated statement $\mathit{ms}$, based on \emph{Cov} \label{alg:prioritize:select}
\State $\mathit{PTS} \gets \mathit{new} \mathit{list}$ \textcolor{darkgray}{//this list is initially empty}
\State $\mathit{PTS} \gets$ based on \emph{Cov} select from $\mathit{TS_m}$ the test case $t$ that exercises $\mathit{ms}$ more times \label{alg:prioritize:first}
\State $\mathit{PTS} \gets \mathit{PTS} \cup t$ \textcolor{darkgray}{//include first the test case selected above}  \label{alg:prioritize:add}

\State \textbf{repeat} \label{alg:prioritize:repeat}
\State \hspace{\INDC mm} \textbf{for each} $n$ in ($\mathit{TS}_m$ - $\mathit{PTS}$) \textcolor{darkgray}{, i.e., is the set of test cases not already added to $\mathit{PTS}$} \label{alg:prioritize:notSel}
\State \hspace{\INDA mm} \textbf{for each} $t$ in $\mathit{PTS}$
\State \hspace{\INDB mm} compute the distance between $t$ and $n$
\State \hspace{\INDA mm} identify $t_n$ i.e., the test case $t$ with the minimal $d$ \label{alg:prioritize:minD}
\State \hspace{\INDC mm} among all the $t_n$ identified, select the one with the highest distance $d$ \label{alg:prioritize:selectStart}
\State \hspace{\INDC mm} \textbf{if} $d > 0$ \textcolor{darkgray}{//there is at least a test case with a different coverage}
\State \hspace{\INDA mm} \textcolor{darkgray}{//note: $n$ is the test case in the set ($\mathit{TS}_m$ - $\mathit{PTS}$) closer to $t_n$}
\State \hspace{\INDA mm} $\mathit{PTS} \gets \mathit{PTS} \cup n$ \label{alg:prioritize:selectEnd}
\State \textbf{until} $d > 0$ \label{alg:prioritize:until}

\end{algorithmic}
\vspace{-3mm}
\caption{PrioritizeAndReduce: Algorithm for prioritizing test cases}
\label{alg:prioritize}
\end{figure}

\subsection{Step 7: Discard Mutants}
\label{sec:algostepSeven}

In this step, we identify likely nonequivalent and likely nonduplicate mutants by relying on code coverage information \CHANGED{collected in the previous step}.

Similarly to related work~\cite{schuler2013covering}, 
we identify nonequivalent and nonduplicate mutants based on a threshold, which we will empirically investigate in Section~\ref{sec:empirical:thrshold}.

In our case, consistently with previous steps of \APPR,
we compute normalized distances based on the distance metrics $D_J$, $D_O$, $D_E$, and $D_C$. A mutant is considered nonequivalent when the distance from the original program is above the threshold $T_E$, for at least one test case.
Similarly, a mutant is considered nonduplicate when the distance from every other mutant is above the threshold $T_D$, for at least one test case. For the identification of nonequivalent mutants, we consider live mutants only. To identify nonduplicate mutants, we consider both live and killed mutants; however, to avoid combinatorial explosion, we compare only mutants belonging to the same source file (indeed, mutants belonging to different files are unlikely to be redundant). 
Killed mutants that lead to the failure of different test cases are not duplicate, regardless of their distance.

Thresholds $T_E$ and $T_D$ should enable the identification of mutants that are guaranteed to be nonequivalent and nonduplicate. In particular, we are interested in the set of \emph{live, nonequivalent, nonduplicate mutants} (hereafter, $\mathit{LNEND}$) and the set of \emph{killed, nonduplicate mutants} (hereafter, $\mathit{KND}$). With such guarantees, the mutation score can be adopted as an adequacy criterion in safety certification processes. For example, certification agencies may require safety-critical software to reach a mutation score of 100\%, which is feasible in the presence of nonequivalent mutants. 

\subsection{Step 8: Compute Mutation Score}
\label{sec:appr:score}

The mutation score (MS) is computed as the percentage of killed nonduplicate mutants 
over the number of nonequivalent, nonduplicate mutants identified in Step 7):

\begin{equation}
\label{equation:ms}
\mathit{MS} = \frac{|\mathit{KND}|}{|\mathit{LNEND}|+|\mathit{KND}|}
\end{equation}

\section{Empirical Evaluation}
\label{sec:evaluation}

\JMR{1.3 2.1}{
Our empirical evaluation aims to assess the effectiveness of the techniques integrated into \APPR to address scalability and pertinence problems (i.e., Steps 2, 4, 5, 6, 7, and 8, in Figure~\ref{fig:approach}). Our objectives include (RQ1) confirming, in our context, trivial compiler optimization results observed in related work (Step 4), (RQ2) identifying the most effective solution for mutants sampling (Step 5), (RQ3) comparing mutants generation strategies implemented by \APPR (Step 2), evaluating the (RQ4) accuracy and (RQ5) effectiveness of the strategies proposed  for test suite prioritization (Step 6), and (RQ6) evaluating the accuracy of the strategy for the identification of likely equivalent/duplicate mutants (Step 7). Finally, we aim to (RQ7) compare the mutation score computed by \APPR  (Step 8) with the mutation score computed without \APPR optimizations.
In the following, we provide our research questions and describe the \APPR steps they aim to evaluate in more detail.}

\begin{itemize}

    \item[RQ1] \JMRCHANGE{(Step 4)} What are the cost savings provided by compiler optimization techniques detecting equivalent and duplicate mutants?
    We wish to determine what is the percentage of mutants reported as being equivalent and duplicate by compiler optimization techniques. After accounting for the additional compilation time entailed by such techniques, we want to identify the optimal subset of compilation options to be used in Step 4 of \APPR.

    \item[RQ2] \JMRCHANGE{(Step 5)} Can a randomly selected subset of mutants be used to accurately estimate the mutation score obtained from the entire set of mutants? 
    \CHANGED{We attempt to evaluate four mutants sampling strategies: \emph{proportional uniform sampling}, \emph{proportional method-based sampling},  \emph{uniform fixed-size sampling}, and \emph{uniform FSCI sampling}. More precisely, we aim to determine the best configuration for each sampling strategy (i.e., sampling ratio, sample size, and confidence interval). Furthermore, we need to identify which strategy offers the best trade-off between the number of mutants to be tested and accuracy.}

    \item[RQ3] \JMRCHANGE{(Step 2)} Do mutants generated with deletion operators (i.e., SDL and OODL) lead to a mutation score that accurately estimates the mutation score of the entire set of mutants?  
    We want to determine if we can minimize the number of selected mutants by only relying on deletion operators. To do so, we compare the mutation score generated with SDL and OODL operators with the mutation score based on all available mutation operators.

    \item[RQ4] \JMRCHANGE{(Step 6)} Can a prioritized subset of test cases that maximizes test suite diversity be used to accurately estimate the mutation score of the entire test suite?
    We investigate how the various distance metrics used in the PrioritizeAndReduce algorithm implemented by \APPR  (Step 6) compare in terms of accuracy.

    \item[RQ5] \JMRCHANGE{(Step 6)} To what extent different test suite prioritization strategies can speed up the mutation analysis process? We investigate the execution time reduction achieved by different distance metrics used in the PrioritizeAndReduce algorithm.

    \item[RQ6] \JMRCHANGE{(Step 7)} Is it possible to identify thresholds, based on code coverage information, that enable the detection of nonequivalent and nonduplicate mutants? We investigate 
    the accuracy of our strategy 
    based on threshold values for the best distance metric  (\APPR  Step 7).

    \item[RQ7] \JMRCHANGE{(Step 8)} How does the mutation score computed by \APPR relate to the mutation score of the original test suite based on the complete set of mutants? In other words, is there any tradeoff between the gains in scalability due to \APPR and  mutation score accuracy? We therefore analyze the difference between the \APPR mutation score, which is obtained with a subset of the test suite and excludes likely equivalent and duplicate mutants, and the mutation score obtained with the entire set of mutants tested with the full test suite.

\end{itemize}

\subsection{Subjects of the study}
\label{sec:empirical:subjects}

To perform our experiments, we considered five software artifacts (hereafter, \emph{subjects}), each one developed by one of the aforementioned industry partners for different satellites: \SAIL{}\emph{-CSW} (central software), \UTIL{}, \GCSP{}, \PARAM{}, and \MLFS{}.
\CHANGED{They are representative of common types of flight software--- that are also typically present in other cyber-physical systems---including on-board controllers (\SAIL{}\emph{-CSW}), libraries providing features related to the application layer (\PARAM{}), as well as networking (\GCSP{}), utility (\UTIL{}), and mathematical functions (\MLFS{}{}).}

\emph{\SAIL{}} is a microsatellite developed by \TWO{}  in a Public-Private-Partnership with ESA and \ExaE{}. The Payload is an AIS Receiver for ship and vessel detection from space. 
For our empirical evaluation, we considered the onboard central control software of \SAIL{} (hereafter, simply \SAIL{}\emph{-CSW}), which consists of 924 source files with a total size of 187,116 LOC. 
\SAIL{}\emph{-CSW} is verified by unit test suites and system test suites that run in different facilities (e.g., Software Validation Facility~\cite{Isasi2019}, FlatSat~\cite{Eickhoff:Simulate}, Protoflight Model~\cite{ecssHB10A}). 
Except for the test suite running in the Software Validation Facility (SVF), which is a simulator for the onboard hardware~\cite{Isasi2019}, the other test suites require dedicated hardware.
The SVF simulates both the target hardware and the satellite units (e.g., a magnetometer connected to the Attitude Determination And Control Subsystem unit).
For this study, we considered the SVF test suite, which
consists of a total of 384 carefully selected test cases targeting mainly functional and interface requirements of the system. 
Other requirements (e.g., timing, robustness, and performance requirements) are verified by the other system test suites.
Unit test suites are used for preliminary development stages and later to ensure 
higher level of code coverage for critical modules on the target hardware.
For this study, we could not consider all the available test suites because of hardware availability; also, our evaluation required repeated executions of the provided test suites, which would not have been practically feasible with dedicated hardware devices in the loop (see Section~\ref{experimnt:setup}).
The SVF test suite already takes 10 hours to execute.

\GCSP{}, \PARAM{}, and \UTIL{}  are utility libraries developed by \ONE\footnote{We use anonymized acronyms according to \ONE policy.}.
\emph{\GCSP{}} is a network protocol library including low-level drivers (e.g., CAN, I2C).
{\PARAM{}} is a light-weight parameter system designed for \ONE satellite subsystems. 
{\UTIL{}} is a utility library providing cross-platform APIs for use in both embedded systems and Linux development environments.

The Mathematical Library for Flight Software (\MLFS{}{})
implements mathematical functions qualified for flight software (it complies with ECSS criticality category B).
 
The first four columns of Table~\ref{table:caseStudies} provide additional details.  
These software components differ in size and complexity; they range from 3,179 (\PARAM{}) to 74,155 (\SAIL{}\emph{-CSW}) LOC (see column \emph{LOC} in Table~\ref{table:caseStudies}). We also provide information concerning a subset of \SAIL{}\emph{-CSW} (i.e., \SAIL{}$_S$) that is introduced in the following paragraphs.

All the test suites considered in our study are characterized by high statement coverage  as required by space software standards (e.g., category C software requires statement adequacy according to ECSS standards~\cite{ecss80C}). 
However, in our study, we do not consider dedicated test suites that require the target hardware to be executed \CHANGED{because of scalability issues, costs, and hardware safety. Indeed, our experiments imply the execution of a large number of test cases (see Section~\ref{experimnt:setup}) that cannot be parallelized when real hardware is required, as only one or few hardware components are available because of their high cost. Also, hardware often needs to be manually set-up, which would make our experiments prohibitively expensive. Finally, the automatically generated mutants may damage the hardware}.
\CHANGED{In the case of \GCSP{}, \PARAM{}, and \UTIL{}, we considered unit and integration test suites that exercise the SUT in the development environment (a Linux-based system).
For \MLFS{}{}, we considered a unit test suite achieving modified condition/decision coverage (MC/DC) coverage~\cite{chilenski1994applicability}.
Since we exclude test cases that must be executed with hardware in the loop, the test suites considered in our study do not achieve 100\% statement coverage, except for \MLFS{}{}. Working with test suites that do not achieve statement adequacy should not affect the validity of our findings because we apply mutation only to statements that are covered by the test suite.
}

\begin{figure}[tb]
\begin{center}
\includegraphics[width=\columnwidth]{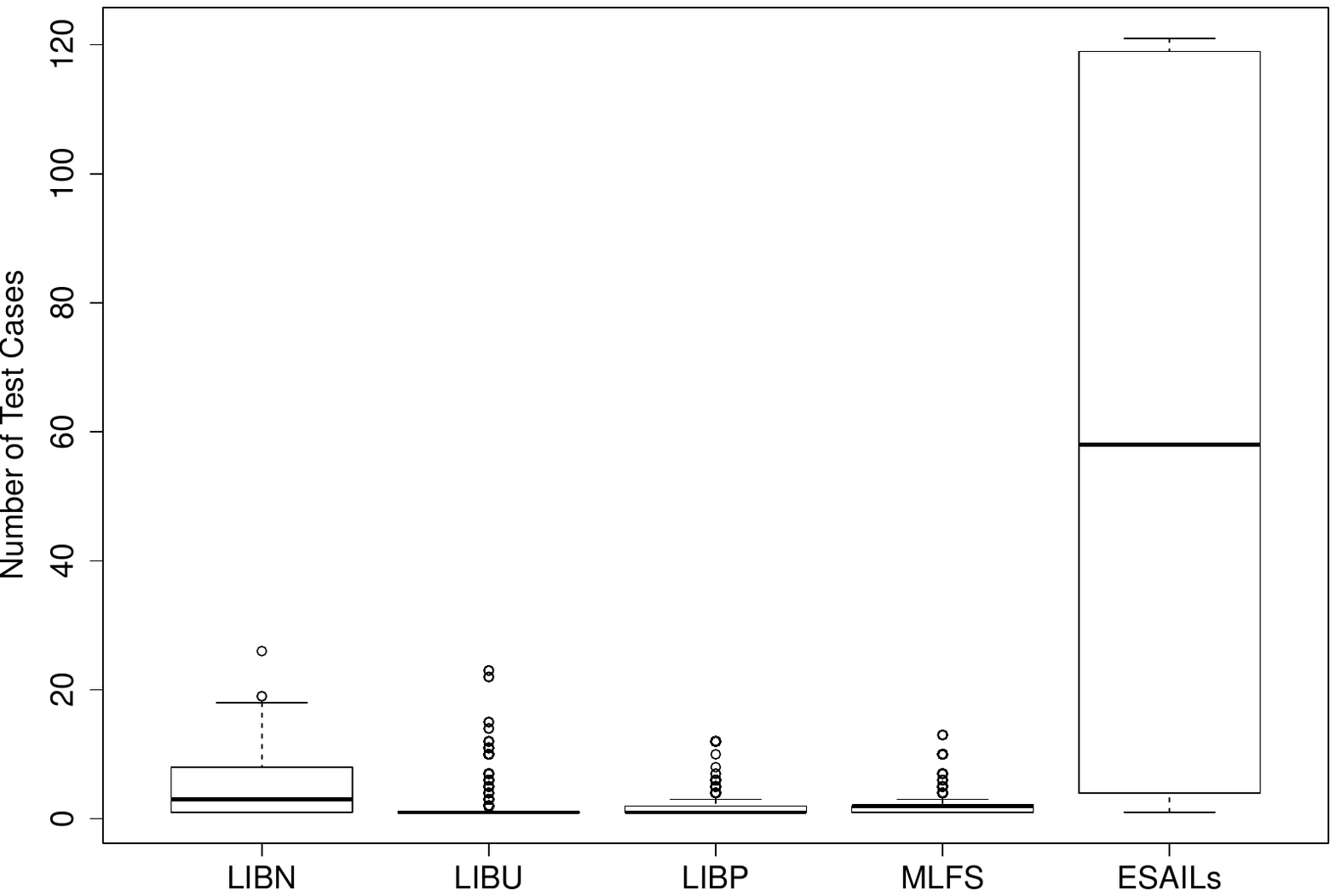}
\caption{Distribution of test cases exercising each statement.}
\label{fig:tastCaseDist}
\end{center}
\end{figure}

\NEWFSCI{The test suites considered in our experiments differ 
regarding  
the distribution of test cases exercising each statement (see Figure~\ref{fig:tastCaseDist}). 
For unit and integration test suites, test cases focus on a subset of functionalities and input ranges, as a result, the number of test cases exercising a same statement is expectedly low, between one and 18.
For the \SAIL{}$_S$ system test suite, whose test cases exercise multiple functionalities (e.g., periodic tasks), the number of test cases exercising a same statement is much higher, between one and 121, with a median equal to 58.
These numbers further highlight the diversity across our subjects. 
}

To address some of our research questions, all the mutants must be executed against the test suite, which is not feasible for the case of \SAIL{}\emph{-CSW} due to its large size and test suite. For this reason, we have identified a subsystem of \SAIL{}\emph{-CSW} (hereafter, \emph{\SAIL{}}$_{S}$) that consists of a set of files, selected by \TWO engineers, that are representative of the different functionalities in \SAIL{}\emph{-CSW}: service/protocol layer functions, critical functions of the satellite implemented in high-level drivers, application layer functions.
Details about $\mathit{\SAIL{}}_{S}$ are reported in Table~\ref{table:caseStudies}.

Except for \SAIL{}\emph{-CSW}, all subjects
are compiled to generate executables for the development environment OS (Linux); we rely on the Gnu Compiler Collection (GCC)  for Linux X86~\cite{GCC} versions 5.3 and 6.3 for \MLFS{}{} and ONE, respectively. \SAIL{}\emph{-CSW} is compiled with the
LEON/ERC32 RTEMS Cross Compilation System, which includes the GCC C/C++ compiler version 4.4.6 for  RTEMS-4.8 (Sparc architecture)~\cite{RTEMS}.

It is important to note that the technical and test suite characteristics described above are very common in embedded software across many industry domains and cyber-physical systems, thus suggesting our results can be generalizable beyond space software.

\begin{table}[tb]
\caption{Descriptions of subject artifacts.}
\label{table:caseStudies} 
\scriptsize
\centering
\begin{tabular}{|
@{\hspace{1pt}}p{13mm}
@{\hspace{2pt}}|
@{\hspace{1pt}}>{\raggedleft\arraybackslash}p{8mm}@{\hspace{1pt}}|
@{\hspace{1pt}}>{\raggedleft\arraybackslash}p{18mm}@{\hspace{1pt}}|
@{\hspace{1pt}}>{\raggedleft\arraybackslash}p{20mm}@{\hspace{1pt}}|
@{\hspace{1pt}}>{\raggedleft\arraybackslash}p{24mm}@{\hspace{1pt}}|
p{20mm}|}
\hline
\textbf{Subject}&\textbf{LOC}&\textbf{Test suite type}&\textbf{\# Test cases}&\textbf{Statements} \textbf{coverage}\\
\hline
\mbox{\SAIL{}\emph{-CSW}}& 74,155 & System& 384 & 90.38\% \\
\SAIL{}$_S$& 2,235 & System& 384 & 95.36\%\\
\GCSP{}& 9,836 & Integration& 89 & 63.10\%\\
\PARAM{}& 3,179 & Integration& 170 & 77.60\%\\
\UTIL{}& 10,576 & Unit& 201 & 83.20\%\\
\MLFS{}{}& 5,402 & Unit& 4042 & 100.00\%\\
\hline
\end{tabular}

\end{table}

\begin{table}[tb]
\caption{Generated and compiled mutants per subject.}
\label{table:mutants} 
\scriptsize
\begin{tabular}{|
@{\hspace{1pt}}p{13mm}
@{\hspace{2pt}}|
>{\raggedleft\arraybackslash}p{10mm}@{\hspace{1pt}}|
>{\raggedleft\arraybackslash}p{10mm}@{\hspace{1pt}}|
>{\raggedleft\arraybackslash}p{12mm}@{\hspace{1pt}}|
>{\raggedleft\arraybackslash}p{12mm}@{\hspace{1pt}}|
>{\raggedleft\arraybackslash}p{10mm}|}
\hline
\textbf{}&\multicolumn{1}{c|}{\textbf{Mutants}}&\multicolumn{1}{c|}{\textbf{MGT}}&\multicolumn{1}{c|}{\textbf{Mutants}}&\multicolumn{1}{c|}{\textbf{\% of}}&\multicolumn{1}{c|}{\textbf{MCT}}\\
\textbf{Subject}&\multicolumn{1}{c|}{\textbf{generated}}&\multicolumn{1}{c|}{\textbf{(sec)}}&\multicolumn{1}{c|}{\textbf{compiled}}&\multicolumn{1}{c|}{\textbf{compiled}}&\multicolumn{1}{c|}{\textbf{(sec)}}\\ 
&                    &    & &\multicolumn{1}{c|}{\textbf{mutants}}&\multicolumn{1}{c|}{ }\\ 
\hline
\mbox{\SAIL{}\emph{-CSW}}& 142,763 & 182 &121,848& 85.35\% & 151,234\\
\SAIL{}$_S$& 7,212 & 9& 5,347 & 74.14\% & 7,640\\
\GCSP{}& 8,666 & 12 &7,878&90.91\% & 11,425\\
\PARAM{}& 7,252 & 7 &6,440&88.80\% & 9,392\\
\UTIL{}& 22,295 & 28 &20,268&90.91\% & 30,624\\
\MLFS{}{}& 31,526 & 20 &28,069&89.03\% &3,157\\
\hline
\textbf{Total*}& 212,502 & 249 & 184,503 &86.82\% &205,832\\ 
\hline
\end{tabular}
\vspace{1mm}

*We ignore \SAIL{}$_S$ from the total counting because it is a subset of \SAIL{}\emph{-CSW}.
\textbf{Legend:} MGT = mutants generation time, MCT = mutants compilation time.
\end{table}

\subsection{Experimental Setup}
\label{experimnt:setup}

To perform the empirical evaluation, we have implemented the \APPR pipeline in a toolset that is available under the ESA Software Community Licence Permissive~\cite{ESAlicence} at the following URL \textbf{https://faqas.uni.lu/}
\footnote{MASS can be retrieved also using the following DOI: https://doi.org/10.5281/zenodo.5235941}.
For the implementation of mutation operators, we extended the SRCiror toolset~\cite{hariri2018srciror}.
In our analysis, we consider all the operators reported in Table~\ref{table:operators}.

Related studies~\cite{zhang2010operator,zhang2013operator} are performed by relying on mutation adequate test suites (i.e., test suites that kill all non-equivalent mutants). 
Such test suites are typically automatically generated using static analysis~\cite{papadakis2012mutation}.
\JMR{3.13}{Since we cannot leverage static analysis to generate mutation adequate test suites (see Section~\ref{sec:background:equivalent}), 
we rely on the original test suites provided with the subjects.}
As a result, to perform our study, we mutate only the statements that are covered by the considered test suites.

\CHANGED{For every subject, we generated mutants by executing the \APPR toolset on Linux OS running on a MacBook Pro with 2,3 GHz 8-Core Intel Core i9. 
Table~\ref{table:mutants} reports, for every subject, the total number of mutants that were generated, their generation time, the number of mutants successfully compiled, the proportion of compiled mutants with respect to the overall number of mutants generated, and the time required to compile mutants using one compiler optimization level only (i.e., the one originally selected by engineers for each subject).
The generation of mutants is fast, it takes at most {182} seconds on the largest subject (\SAIL{}\emph{-CSW}). On average, across subjects, it takes {11} milliseconds to generate a single mutant. 
The proportion of successfully compiled mutants is large (i.e., 86.82\% overall), though it varies from 85.35\% for \SAIL{}\emph{-CSW} to 90.91\% for \GCSP{} and \UTIL{}.
This proportion is in line with the ones reported in related work, though there is variation. For example, industrial systems 
have shown lower success rates (e.g., 81.13\% for safety-critical software components~\cite{Baker2013}) than open source, batch utilities (e.g., 96.30\% for Coreutils~\cite{hariri2016evaluating}).}
 
\CHANGED{As shown in Table~\ref{table:mutants}, the time required to compile all the mutants ranges from 3,157 to 151,234 seconds, for an average of 0.96 seconds required to compile a single mutant, across subjects. Because of our selective compilation strategy (see Section~\ref{sec:appr:compile}), the time required by our pipeline to compile a single mutant is significantly lower than the one required by state-of-the-art approaches, which is around 4.6 seconds per mutant, even when software components have a lower number of LOC %
than our subjects~\cite{kintis2017detecting}.
}

\CHANGED{To collect the data required to address research questions RQ2 to RQ8, for every subject and unique mutant generated by Step 3, we have executed all the test cases covering the mutated statement. Table~\ref{table:magnitude} provides, for every subject, the overall number of executed test cases and the total execution time required for our experiments. In total, the entire experiment took 1,912,662 minutes (31,878 hours or more than 1300 days). Test  execution time depends on the number of mutants, the number of executed test cases, and the test suite level (e.g., system test suites exercise more complex scenarios than unit or integration test suites).} 

\CHANGED{To be able to execute test cases for 31,878 hours, we performed our experiments using the HPC cluster of the University of Luxembourg~\cite{HPC}.
The HPC cluster consists of Intel Xeon E5-2680 v4 (2.4 GHz) nodes. To perform our experiments, we tested 100 mutants in parallel, each one on a dedicated node.}

\begin{table}[tb]
\caption{Scale of experiments.}
\label{table:magnitude} 
\scriptsize
\centering
\begin{tabular}{|
@{\hspace{1pt}}p{15mm}
@{\hspace{2pt}}|
@{\hspace{1pt}}>{\raggedleft\arraybackslash}p{30mm}@{\hspace{1pt}}|
@{\hspace{1pt}}>{\raggedleft\arraybackslash}p{35mm}@{\hspace{1pt}}|
p{20mm}|}
\hline
\textbf{Subject}&\textbf{Total test cases executed}&\textbf{Total execution time (minutes)}\\
\hline
\SAIL{}$_S$&  302,158 & 1,624,336\\
\GCSP{}&  771,274 & 33,756\\
\PARAM{}&  1,094,800 & 7,205\\
\UTIL{}&  4,481,295 & 57,732\\
\MLFS{}{}&  170,303,452 & 189,633\\
\hline
\textbf{Total}&  176,952,979 & 1,912,662\\
\hline
\end{tabular}

\end{table}

In the following sections, we 
\JMR{3.11}{discuss statistical significance}
using a non-parametric Mann Whitney U-test (with $\alpha$ = 0.05)~\cite{Arcuri:practicalGuide:ICSE:2015}.  We discuss effect size based on Vargha and Delaney’s $A_{12}$ statistics, a non-parametric effect size measure~\cite{VDA,Arcuri:practicalGuide:ICSE:2015}. Based on $A_{12}$, effect size is considered small when $0.56 \le A_{12} < 0.64$, medium when $0.64 \le A_{12} < 0.71$, large when $A_{12} \ge 0.71$. Otherwise the compared samples are considered equivalent, that is to be drawn from the same population~\cite{VDA}.

\subsection{RQ1}

\begin{table*}[htb]
\caption{RQ1. Proportion (\%) of Trivially-Equivalent and Trivially-Redundant Mutants Detected by Compiler Optimizations.}
\label{table:results:compilerOptimizations} 
\scriptsize
\centering

\begin{tabular}{|
p{12mm}|
@{\hspace{1pt}} >{\raggedleft\arraybackslash}p{8mm}@{\hspace{1pt}}|
@{\hspace{1pt}}>{\raggedleft\arraybackslash}p{11mm}@{\hspace{1pt}}|
@{\hspace{1pt}}>{\raggedleft\arraybackslash}p{8mm}@{\hspace{1pt}}|
>{\raggedleft\arraybackslash}p{6mm}@{\hspace{1pt}}|
>{\raggedleft\arraybackslash}p{6mm}@{\hspace{1pt}}|
 >{\raggedleft\arraybackslash}p{6mm}@{\hspace{1pt}}|
 >{\raggedleft\arraybackslash}p{6mm}@{\hspace{1pt}}|
 >{\raggedleft\arraybackslash}p{6mm}@{\hspace{1pt}}|
 >{\raggedleft\arraybackslash}p{6mm}@{\hspace{1pt}}|
>{\raggedleft\arraybackslash}p{8mm}@{\hspace{1pt}}|
@{\hspace{1pt}} >{\raggedleft\arraybackslash}p{11mm}@{\hspace{1pt}}|
@{\hspace{1pt}} >{\raggedleft\arraybackslash}p{7mm}@{\hspace{1pt}}|
 >{\raggedleft\arraybackslash}p{6mm}@{\hspace{1pt}}|
 >{\raggedleft\arraybackslash}p{6mm}@{\hspace{1pt}}|
 >{\raggedleft\arraybackslash}p{6mm}@{\hspace{1pt}}|
 >{\raggedleft\arraybackslash}p{6mm}@{\hspace{1pt}}|
 >{\raggedleft\arraybackslash}p{6mm}@{\hspace{1pt}}|
 >{\raggedleft\arraybackslash}p{6mm}@{\hspace{1pt}}|
}

\hline
\textbf{Subject} & \multicolumn{9}{c|}{\textbf{Equivalent}} & \multicolumn{9}{c|}{\textbf{Duplicate}} \\
\textbf{}
&\textbf{All}&\textbf{Overall}\textbf{\%}&\textbf{-O0-3}&\textbf{-O0}&\textbf{-O1} & \textbf{-O2} & \textbf{-O3} & \textbf{-Os} & \textbf{-Of} 
&\textbf{All}&\textbf{Overall}\textbf{\%}&\textbf{-O0-3}&\textbf{-O0}&\textbf{-O1} & \textbf{-O2} & \textbf{-O3} & \textbf{-Os} & \textbf{-Of}
\\
\hline
\mbox{\SAIL{}\emph{-CSW}} &8,861&7.27 &7.13&34.18 &95.25 &96.15 &95.61 &97.44 &-&35,133&28.83&27.74&39.94 &59.80 &61.96 &61.52 &62.46 &-\\

\GCSP{} &701&8.90 &8.90&25.11 &97.43 &72.90 &74.04 &44.94 &47.93 &2,655&33.70&27.63&37.55 &66.14 &50.17 &57.55 &63.65 &62.98 \\
\PARAM{} &450& 6.99 &6.89&33.56 &95.11 &94.00 &94.44 &96.89 &94.44 &2,076&32.24&31.44&43.79 &63.10 &64.84 &64.84 &64.93 &64.84 \\
\UTIL{} &1,366&6.74 &6.65&28.84 &93.48 &90.26 &91.87 &91.29 &91.87 &4,392&21.67&21.26&47.65 &70.40 &75.59 &76.71 &77.03 &76.71 \\
\MLFS{}{} &361 &1.29 &1.04&31.86 &63.71 &77.84 &78.12 &85.04 &81.16 &6,356&22.64&20.13&37.54 &51.04 &56.95 &57.08 &58.72 &61.28 \\
\hline
\textbf{Total}  &11,739&6.36 &6.22&32.14 &89.95 &93.39 &89.93 &95.50 &17.86 &50,612&27.43&25.99&40.46 &60.09 &62.37 &62.83 &62.96 &20.66 \\
\hline

\end{tabular}

\end{table*}

\begin{table*}[htb]
\caption{RQ1.  Proportion (\%) of Univocal-Trivially-Equivalent and Univocal-Trivially-Redundant Mutants Detected.}
\label{table:results:compilerOptimizationsUnivocal} 
\scriptsize
\centering

\begin{tabular}{|
p{12mm}|
@{\hspace{1pt}} >{\raggedleft\arraybackslash}p{7mm}@{\hspace{1pt}}|
@{\hspace{1pt}}>{\raggedleft\arraybackslash}p{19mm}@{\hspace{1pt}}|
>{\raggedleft\arraybackslash}p{6mm}@{\hspace{1pt}}|
>{\raggedleft\arraybackslash}p{6mm}@{\hspace{1pt}}|
 >{\raggedleft\arraybackslash}p{6mm}@{\hspace{1pt}}|
 >{\raggedleft\arraybackslash}p{6mm}@{\hspace{1pt}}|
 >{\raggedleft\arraybackslash}p{6mm}@{\hspace{1pt}}|
 >{\raggedleft\arraybackslash}p{6mm}@{\hspace{1pt}}|
>{\raggedleft\arraybackslash}p{7mm}@{\hspace{1pt}}|
@{\hspace{1pt}} >{\raggedleft\arraybackslash}p{19mm}@{\hspace{1pt}}|
 >{\raggedleft\arraybackslash}p{6mm}@{\hspace{1pt}}|
 >{\raggedleft\arraybackslash}p{6mm}@{\hspace{1pt}}|
 >{\raggedleft\arraybackslash}p{6mm}@{\hspace{1pt}}|
 >{\raggedleft\arraybackslash}p{6mm}@{\hspace{1pt}}|
 >{\raggedleft\arraybackslash}p{6mm}@{\hspace{1pt}}|
 >{\raggedleft\arraybackslash}p{6mm}@{\hspace{1pt}}|
}

\hline
\textbf{Subject} & \multicolumn{8}{c|}{\textbf{Univocal-Equivalent}} & \multicolumn{8}{c|}{\textbf{Univocal-Duplicate}} \\
\textbf{}
&\textbf{All}&\textbf{\%} \textbf{of} \textbf{Equivalent}&\textbf{-O0}&\textbf{-O1} & \textbf{-O2} & \textbf{-O3} & \textbf{-Os} & \textbf{-Of} 
&\textbf{All}&\textbf{\%} \textbf{of} \textbf{Duplicate}&\textbf{-O0}&\textbf{-O1} & \textbf{-O2} & \textbf{-O3} & \textbf{-Os} & \textbf{-Of}
\\
\hline
\mbox{\SAIL{}\emph{-CSW}} &237& 2.67&0.00 &4.64 &0.84 &19.83 &\textbf{74.68} &0.00 &827&2.35&7.26	&11.61	&3.26	&28.78	&\textbf{49.09}	&0.00  \\

\GCSP{} &112 & 15.98 & 0.00 & \textbf{97.32} & 0.00 & 2.68 & 0.00 & 0.00 &202& 7.61&0.99&3.47&0.00&0.00&\textbf{48.02}&47.52  \\
\PARAM{} &11 & 2.44&0.00 &45.45 &0.00 &0.00 &\textbf{54.55} &0.00 &44& 2.12&0.00&38.64&6.82&0.00&\textbf{54.55}&0.00 \\
\UTIL{} &96& 7.03&1.04 &77.08 &2.08 &0.00 &\textbf{19.79} &0.00 &148& 3.37&4.05&34.46&4.05&0.00&\textbf{54.73}&2.70 \\
\MLFS{}{} &65& 18.01&0.00 &9.23 &0.00 &0.00 &60.00 &\textbf{30.77} &471& 7.41&0.85&2.55&0.85&0.00&37.37&\textbf{58.39} \\
\hline
\textbf{Total}  &521& 4.44&0.19 &18.43 &2.11 &9.02 &\textbf{50.67} &3.84 &1,692& 3.34&4.26&10.82&2.36&14.07&\textbf{46.34}&22.16 \\
\hline

\end{tabular}

\end{table*}

\begin{table}[h]
\caption{RQ1. Compilation Time Required by Compiler Optimization Techniques.}
\label{table:results:compilerOptimizationsTime} 
\scriptsize
\begin{tabular}{|
@{\hspace{1pt}}p{12mm}|
@{\hspace{1pt}}>{\raggedleft\arraybackslash}p{8mm}@{\hspace{1pt}}|
>{\raggedleft\arraybackslash}p{8mm}@{\hspace{1pt}}|
>{\raggedleft\arraybackslash}p{8mm}@{\hspace{1pt}}|
>{\raggedleft\arraybackslash}p{8mm}@{\hspace{1pt}}|
>{\raggedleft\arraybackslash}p{8mm}@{\hspace{1pt}}|
>{\raggedleft\arraybackslash}p{8mm}@{\hspace{1pt}}|
 >{\raggedleft\arraybackslash}p{9mm}@{\hspace{1pt}}|
}
\hline
\textbf{Subject} & 
\multicolumn{7}{c|}{\textbf{Time required to compile all the mutants(sec)}}\\
\textbf{}&
\textbf{-O0}&\textbf{-O1} & \textbf{-O2} & \textbf{-O3} & \textbf{-Os} & \textbf{-Of}
&\textbf{All O* (hours)} 
\\
\hline
\mbox{\SAIL{}\emph{-CSW}} & 135,455 & 132,528 & 149,088 & 145,620 & 151,234 & N/A&198\\
\GCSP{} & 11,053 & 11,256 & 11,079 & 11,425 & 10,299 & 11,424 &18\\
\PARAM{} & 9,036 & 9,246 & 9,352 & 9,392 & 8,794 & 9,442 &16\\
\UTIL{} & 25,228 & 26,914 & 28,564 & 30,624 & 26,845 & 29,583 &47\\
\MLFS{}{} & 2,176 & 2,509 & 3157 & 3,167 & 3,052 & 3,164 &5\\
\hline
\textbf{Total}&182,948	&182,453	&201,240	&200,228	&200,224	&53,613 & 284\\
\hline
\end{tabular}

\end{table}

\subsubsection*{Design and measurements}

RQ1 aims to determine the cost savings provided by 
trivial compiler optimization techniques~\cite{papadakis2015trivial,kintis2017detecting}. To do so, we assess the number of equivalent and duplicate mutants discarded by these techniques and the additional costs introduced by the augmented compilation process. Since the mutants detected by these techniques 
are a subset of 
the overall set of equivalent and duplicate mutants, we refer to them as \emph{trivially equivalent} and \emph{trivially duplicate} mutants.

For every subject, we compile every mutant six times, each time with a different optimization level enabled. We consider all the available optimization levels for the GCC compiler, which are \emph{-O0}, \emph{-O1}, \emph{-O2}, \emph{-O3}, \emph{-Os},
and \emph{-Ofast}~\cite{GCCopt}. Level \emph{-O0} indicates that no optimization is applied. Levels \emph{-O1}, \emph{-O2}, \emph{-O3}, and \emph{-Ofast}, in this order, enable an increasing number of optimization options (e.g., level \emph{-Ofast} includes all the optimizations of level \emph{-O3} plus two additional ones, which are \emph{-ffast-math} and \emph{-fallow-store-data-races}). Level \emph{-Os} enables all \emph{-O2} optimizations except those that increase code size. This is the first reported experiment including options \emph{-Os} and \emph{-Ofast} in such an analysis.

To identify the most effective compiler optimization level, we consider the percentage of trivially equivalent and trivially duplicate mutants they detect. Also, to discuss how complementary different optimization levels  are and, therefore, whether they should be combined, we report the number of mutants identified as equivalent and duplicate by each compiler optimization only (hereafter called \emph{univocal-trivially-equivalent mutants} and \emph{univocal-trivially-duplicate mutants}). 

\CHANGED{To further assess the different optimization levels, we analyze the distribution of trivially equivalent and duplicate mutants across the different mutation operators considered in our study.}
Last, we compare our results with the ones reported in related work~\cite{papadakis2015trivial}.

\CHANGED{By default, subjects are compiled with  different compiler optimization options, \emph{-Os} for \SAIL{}\emph{-CSW}, \emph{-O2} for \MLFS{}{}, \emph{-O3} for \GCSP{}, \PARAM{}, and \UTIL{}.}
To estimate the costs entailed by the augmented compilation process, we collect, for every subject, the time required for compiling all their mutants with each of the five optimization levels enabled. To compile mutants, we rely on the optimized compilation process implemented in Step 3 of \APPR (see Section~\ref{sec:appr:compile}).

\subsubsection*{Results}

Table~\ref{table:results:compilerOptimizations} provides the results concerning the detection of trivially equivalent and trivially duplicate mutants. We report the total number of such mutants detected for each subject (column \emph{All}), their percentage with respect to the set of mutants successfully compiled (column \emph{Overall \%}), and the percentages obtained with the options included in related work (i.e., by discarding the equivalent and duplicate mutants detected by \emph{-O0}, \emph{-O1}, \emph{-O2}, and \emph{-O3}, reported in column \emph{-O0-3}). 
 The proportion of trivially equivalent mutants detected with optimization levels O0-O3 (6.22\%) is in line with related work~\cite{papadakis2015trivial} (7\%) and the range observed for the different subjects (i.e., 1.04\% to 8.90\%) largely overlaps with the range observed in related work (2\%-10\%). 
 For trivially duplicate mutants, instead, we observe a slightly larger set of duplicate mutants when compared to related work. Optimization levels O0-O3 determine that 25.99\% of the mutants are trivially duplicate, while in related work the average is around 21\%. Finally,  
 optimization levels \emph{-Os} and \emph{-Of} enable the detection of additional trivially equivalent and duplicate mutants, thus leading to an average of \JMR{3.12}{6.36\% and 27.43\% of the mutants being discarded, respectively (see column \emph{Overall \%})}. In particular, \textbf{we observe that the optimization level \emph{-Os}, not evaluated by related work, is the most effective}. Our results confirm the effectiveness of compiler optimizations for removing a significant percentage of equivalent and duplicate mutants.
 
Table~\ref{table:results:compilerOptimizationsUnivocal} provides additional details about the trivially equivalent and duplicate mutants univocally detected by the different optimization options. In total, 4.44\% and 3.34\% of these mutants are univocally detected  by one optimization level (see columns \emph{\% of Equivalent} and \emph{\% of Duplicate}); moreover, since all optimization options contribute to the univocal detection of equivalent and redundant mutants, \textbf{it is preferable to rely on all the available compiler optimization options}. 
\UPDATE{Overall, the most effective optimization option is \emph{-Os}, which detects 50.67\% and 46.34\% of univocal-equivalent and univocal-duplicate mutants, respectively. It is followed by \emph{-O1}, detecting 18.43\% and 10.82\% of such mutants, respectively}. These results suggest that, when the number of compilation runs must be limited, then \emph{-Os} and \emph{-O1} should be prioritized over the other options. This is interesting since \textbf{stronger optimization levels such as \emph{-Ofast} and \emph{-O3} do not contribute more than \emph{-Os} to the detection trivially equivalent and duplicate mutants}. Surprisingly, the optimization level \emph{-O0}, which does not enable any compiler optimization option, can detect trivially equivalent and trivially redundant mutants not detected by other optimization levels. However, this seems to highly depend on the code surrounding the mutated statement. 
\JMR{3.14}{For example, in one \UTIL mutant the optimization level \emph{-O0} detected that \texttt{if ( ptr )} is equivalent to \texttt{if( ptr > NULL )}, with \texttt{ptr} being a pointer, which was not detected with other optimization levels; however, the same instructions are detected as equivalent by other optimization levels when they belong to other functions (i.e., when the code surrounding them is different than the one appearing in the \UTIL mutant).}

Details about the distribution of  trivially equivalent and  duplicate mutants per mutation operator are reported in Appendix A.

Table~\ref{table:results:compilerOptimizationsTime} provides the time required to compile the artifacts with the different optimization levels. Different from related work, which reports that optimization levels, in the worst case, lead an increase in compilation time by a factor of 5, \textbf{we do not observe a large difference in compilation time among the different optimization levels}. Indeed, in the worst case (i.e., option \emph{-Os}) this factor is $1.1$, an increase of 10\%. This directly results from the \APPR compilation pipeline, which minimizes the number of source files that need to be compiled. If developers can accept a compilation time increased by a factor of 5, as suggested in related work, all the compilation optimization levels can be applied, thus maximizing the number of equivalent and duplicate mutants being detected. In three out of five subjects, it takes less than a day to compile all the mutants with all the available optimization levels, which is acceptable, given the cost saved in subsequent steps. For the cases in which it may take multiple days, our practical solution consists in executing the compilation of the various mutants in parallel (e.g., on Cloud systems); for example, our toolset includes scripts to parallelize mutants compilation on HPC and cloud platforms. In the case of \SAIL{}\emph{-CSW}, the parallel compilation of 142,763 mutants, with the four available compilation options, can be performed in ~90 minutes using 100 nodes.

\subsection{RQ2 - Accuracy of Mutant Sampling Methods}
\label{sec:RQ2}
\subsubsection*{Design and measurements}

RQ2 aims to investigate  
to what extent the mutation score computed from a sample of mutants (hereafter, \emph{estimated mutation score}) accurately estimates the mutation score of the complete set of mutants (hereafter, \emph{actual mutation score}).

In our study, we consider the sampling strategies which are part of \APPR, as justified earlier: \emph{proportional uniform sampling}, \emph{proportional method-based sampling},  \emph{uniform fixed-size sampling}, and \emph{uniform FSCI sampling}.

Because of the complexity and size of space software, combined with its high test execution cost, we are interested in selecting a very small subset of mutants. 
For this reason, 
to evaluate  \emph{proportional uniform sampling} and \emph{proportional method-based sampling},
we consider sampling ratios ranging from $1\%$ to $10\%$, in steps of $1\%$. Further,
 we also cover the range 10\% to 100\%, in steps of $10\%$. To evaluate  \emph{uniform fixed-size sampling}, consistent with our earlier discussion, we consider a number of mutants in the range 100 to 1000, in steps of $100$.
Finally, to evaluate \emph{proportional method-based sampling}, we consider a threshold for the confidence interval (i.e., $T_{\mathit{CI}}$) that ranges from $0.05$ to $0.10$, in steps of $0.01$, with a confidence level of $95\%$, which is a common choice. The experiments conducted to address RQ2 entail the execution of the entire test suite for every sampled mutant. Executions with a prioritized and reduced test suite are addressed in Section~\ref{exp:accuracy:prioritize}. The evaluation of different values for $T_{\mathit{CI}}$ enable us to determine the costs associated with a more accurate estimation of the mutation score, in order to better understand the trade-offs.

We compute the actual mutation score of each system by executing the test suite against all the mutants that were successfully compiled, excluding mutants detected as being equivalent or duplicate by simple compiler optimization techniques (see RQ1). 
For each sampling ratio, to account for randomness,
we repeat the analysis 100 times, i.e., we compute the mutation score 100 times, based on 100 randomly selected subsets of mutants.
Since it is not feasible to test all the mutants generated for \SAIL{}\emph{-CSW}, as discussed above, we focus on $\mathit{\SAIL{}}_{S}$.

Our goal is to determine if the estimated mutation score is an accurate estimate of the actual mutation score.
This happens when the estimated mutation score differs from the actual mutation score for less than a small delta (hereafter, accuracy delta, $\delta_{acc}$) for a large percentage of the runs (e.g., 95\%).
We thus study the distribution of the difference between the estimated and actual mutation scores across all runs. More precisely, we estimate the 2.5\% and 97.5\% quantiles\footnote{We rely on linear interpolation using the type 8 
algorithm suggested in Hyndman and Fan~\cite{Hyndman1996}. It does not make assumptions about the underlying distribution.}.
Since these two quantiles delimit 95\% of the population,
we consider the mutation scores to be accurately estimated when they are within a pre-defined small range of the actual score [$-\delta_{acc}$;$+\delta_{acc}$].
In other words, we consider the estimated mutation score to be accurate when 
the absolute value of the largest difference between quantiles and the actual score is below $\delta_{acc}$.
Since the range of acceptable mutation score values is small (75\%-100\%, see Section~\ref{background:adequacy}), we decided to use a threshold of 5\%, which is more conservative than that reported in related work ~\cite{gopinath2015hard}. 

Below, we analyze $\delta_{acc}$ for varying sampling rates. To improve readability, we discuss the results concerning the different sampling strategies separately.

\subsubsection*{Results - proportional uniform sampling}

\begin{table}[tb]
\caption{Actual mutation scores across subjects.}
\label{table:results:accuracy:full} 
\scriptsize
\centering
\begin{tabular}{|
@{\hspace{1pt}}p{13mm}|
@{\hspace{1pt}}>{\raggedleft\arraybackslash}p{10mm}@{\hspace{1pt}}|
>{\raggedleft\arraybackslash}p{9mm}@{\hspace{1pt}}|
>{\raggedleft\arraybackslash}p{9mm}@{\hspace{1pt}}|
 >{\raggedleft\arraybackslash}p{25mm}@{\hspace{1pt}}|
}
\hline
\textbf{Subject}&\textbf{Mutants}&\textbf{Killed}&\textbf{Live}&\textbf{Mutation Score (\%)}\\ 
\hline
\SAIL{}$_S$ & 3,536 & 2,311  & 1,225  & 65.36 \\
\GCSP{}&4,982&3,270&1,712&65.64 \\
\PARAM{}&3,931&2,717&1,214&69.12 \\
\UTIL{} &14,574 & 10,376 & 4,198 & 71.20 \\
\MLFS{}{}&21,375&17,484&3,981&81.80 \\
\hline
\end{tabular}

\end{table}

\begin{table}[tb]
\caption{RQ2. Accuracy of proportional uniform sampling.}
\label{table:results:accuracy:regSampling} 
\scriptsize
\centering
\begin{tabular}{|
@{\hspace{1pt}}p{3mm}|
@{\hspace{1pt}}>{\raggedleft\arraybackslash}p{6mm}@{\hspace{1pt}}|
>{\raggedleft\arraybackslash}p{5mm}@{\hspace{1pt}}|
>{\raggedleft\arraybackslash}p{6mm}@{\hspace{1pt}}|
 >{\raggedleft\arraybackslash}p{5mm}@{\hspace{1pt}}|
  >{\raggedleft\arraybackslash}p{6.5mm}@{\hspace{1pt}}|
@{\hspace{1pt}}>{\raggedleft\arraybackslash}p{5mm}@{\hspace{1pt}}|
@{\hspace{1pt}}>{\raggedleft\arraybackslash}p{7mm}@{\hspace{1pt}}|
>{\raggedleft\arraybackslash}p{5mm}@{\hspace{1pt}}|
 >{\raggedleft\arraybackslash}p{6mm}@{\hspace{1pt}}|
  >{\raggedleft\arraybackslash}p{6mm}@{\hspace{1pt}}|
}
\hline
     & \multicolumn{2}{c|}{\textbf{\GCSP{}}} & \multicolumn{2}{c|}{\textbf{\PARAM{}}} & \multicolumn{2}{c|}{\textbf{\UTIL{}}} & \multicolumn{2}{c|}{\textbf{\MLFS{}}} & \multicolumn{2}{c|}{\textbf{\SAIL{}}$_S$} \\
\hline
\textbf{r=} & \textbf{\#M}&\textbf{$\delta_{acc}$}& \textbf{\#M}&\textbf{$\delta_{acc}$}& \textbf{\#M}&\textbf{$\delta_{acc}$}& \textbf{\#M}&\textbf{$\delta_{acc}$}& \textbf{\#M}&\textbf{$\delta_{acc}$}               \\
\hline
0.01 & 50 & 13.64    			 & 40 & 12.19    			& 146 & 7.54    		& 214 & \textbf{4.90} &   36    & 14.04\\
0.02 & 100 & 10.64    			 & 79 & 10.03    			& 292 & 6.15    		& 428 & \textbf{3.19} &   71    & 11.84\\
0.03 & 150 & 10.36   			 & 118 & 7.70     			& 438 & \textbf{4.20}    & 642 & \textbf{3.15} &   107    & 8.84\\
0.04 & 200 & 6.40     			 & 158 & 6.46     			& 583 & \textbf{3.28}    & 855 & \textbf{2.53} &   142    & 7.92 \\
0.05 & 250 & 7.07     			 & 197 & 6.98     			& 729 & \textbf{3.11}    & 1,069 & \textbf{2.58} &  177 & 6.39 \\
0.06 & 299 & 5.95    		 	 & 236 & 5.78     			& 875 & \textbf{2.92}    & 1,283 & \textbf{2.24} & 213      &  6.25\\
0.07 & 349 & 5.01     			 & 276 & 5.73     		    & 1,021 & \textbf{2.84}    & 1,497 & \textbf{2.24} &  248     & 5.20 \\
0.08 & 399 & 5.12    		     & 315 & 5.01 		    	& 1,166 & \textbf{3.24}    & 1,710 & \textbf{1.71} &  283     & 5.68\\
0.09 & 449 & \textbf{4.53}     	 & 354 & \textbf{3.48}      & 1,312 & \textbf{2.15}    & 1,924 & \textbf{1.73} &  319     & \textbf{4.55}\\
0.10  & 499 & \textbf{4.61}      & 394 & \textbf{4.36}      & 1,458 & \textbf{2.10}    & 2,138 & \textbf{1.55} &  354     & \textbf{5.26}\\
0.20  & 997 & \textbf{2.81}      & 787 & \textbf{3.24}      & 2,915 & \textbf{1.57}    & 4,275 & \textbf{1.09} &    708   & \textbf{3.52}\\
0.30  & 1,495 & \textbf{2.32}     & 1,180 & \textbf{2.24}     & 4,373 & \textbf{1.00}    & 6,413 & \textbf{0.80} &  1,061 & \textbf{2.56}\\
0.40  & 1,993 & \textbf{1.60}     & 1,573 & \textbf{1.64}     & 5,830 & \textbf{0.91}    & 8,550 & \textbf{0.74} &  1,415 & \textbf{2.04}\\
0.50  & 2,491 & \textbf{1.58}     & 1,965 & \textbf{1.49}     & 7,287 & \textbf{0.67}    & 10,688 & \textbf{0.48} & 1,768  & \textbf{1.50}\\
0.60  & 2,990 & \textbf{1.18}     & 2,358 & \textbf{1.25}     & 8,745 & \textbf{0.66}    & 12,825 & \textbf{0.45} & 2,122  & \textbf{1.28}\\
0.70  & 3,488 & \textbf{1.05}     & 2,750 & \textbf{1.07}     & 10,202 & \textbf{0.43}    & 14,963 & \textbf{0.34} & 2,476   & \textbf{1.16}\\
0.80  & 3,986 & \textbf{0.61}     & 3,143 & \textbf{0.88}     & 11,660 & \textbf{0.38}    & 17,100 & \textbf{0.25} & 2,829   & \textbf{0.92}\\
0.90  & 4,484 & \textbf{0.51}     & 3,534 & \textbf{0.43}     & 13,117 & \textbf{0.26}   & 19,238 & \textbf{0.17} & 3,183& \textbf{0.55}\\
\hline 
\end{tabular}
Note: \#M, number of mutants. Accurate results (i.e., $\delta_{acc} \le 5\%$) are in bold.
\end{table}

\begin{table}[tb]
\caption{RQ2. Accuracy of proportional method-based sampling.}
\label{table:results:accuracy:methodBased} 
\scriptsize
\centering
\begin{tabular}{|
@{\hspace{1pt}}p{3mm}|
@{\hspace{1pt}}>{\raggedleft\arraybackslash}p{6mm}@{\hspace{1pt}}|
>{\raggedleft\arraybackslash}p{5mm}@{\hspace{1pt}}|
>{\raggedleft\arraybackslash}p{6mm}@{\hspace{1pt}}|
 >{\raggedleft\arraybackslash}p{5mm}@{\hspace{1pt}}|
  >{\raggedleft\arraybackslash}p{6.5mm}@{\hspace{1pt}}|
@{\hspace{1pt}}>{\raggedleft\arraybackslash}p{5mm}@{\hspace{1pt}}|
@{\hspace{1pt}}>{\raggedleft\arraybackslash}p{7mm}@{\hspace{1pt}}|
>{\raggedleft\arraybackslash}p{5mm}@{\hspace{1pt}}|
 >{\raggedleft\arraybackslash}p{6mm}@{\hspace{1pt}}|
  >{\raggedleft\arraybackslash}p{6mm}@{\hspace{1pt}}|
}
\hline
     & \multicolumn{2}{c|}{\textbf{\GCSP{}}} & \multicolumn{2}{c|}{\textbf{\PARAM{}}} & \multicolumn{2}{c|}{\textbf{\UTIL{}}} & \multicolumn{2}{c|}{\textbf{\MLFS{}}} & \multicolumn{2}{c|}{\textbf{\SAIL{}}$_S$} \\
\hline
\textbf{r=} & \textbf{\#M}&\textbf{$\delta_{acc}$}& \textbf{\#M}&\textbf{$\delta_{acc}$}& \textbf{\#M}&\textbf{$\delta_{acc}$}& \textbf{\#M}&\textbf{$\delta_{acc}$}& \textbf{\#M}&\textbf{$\delta_{acc}$}               \\
\hline
0.01 & 19 & 23.53    			 & 15 & 22.45    			& 111 & 9.51    		 & 232 & 5.50 &  33& 13.84\\
0.02 & 75 & 11.67    			 & 77 & 11.36    			& 250 & \textbf{4.82}    & 447 & \textbf{4.29} &64& 16.18\\
0.03 & 131 & 6.88    			 & 120 & 8.82    			 & 422 & \textbf{4.76}   & 661 & \textbf{2.53} &104   &8.63\\
0.04 & 194 & 6.52    			 & 165 & 6.73    			 & 564 & \textbf{3.98}   & 881 & \textbf{2.91} &137 &9.93\\
0.05 & 258 & \textbf{4.90}     & 208 & 6.36     			& 731 & \textbf{3.97}    & 1,094 & \textbf{1.96} &178   &6.84\\
0.06 & 312 & \textbf{3.78}     & 254 & 5.54     			& 905 & \textbf{2.62}    & 1,306 & \textbf{1.86} &223& 6.18\\
0.07 & 368 & \textbf{4.21}     & 290 & 5.72     			& 1,045 & \textbf{2.11}    & 1,517 & \textbf{1.72} &254& 5.72\\
0.08 & 417 & \textbf{4.51}     & 335 & 5.22     			& 1,197 & \textbf{2.36}    & 1,733 & \textbf{1.59} & 287&\textbf{4.67}\\
0.09 & 466 & \textbf{4.61}     & 378 & \textbf{4.72}    	 & 1,353 & \textbf{2.51}    & 1,942 & \textbf{1.37} & 331      &\textbf{4.59}\\
0.1  & 515 & \textbf{3.03}     & 413 & \textbf{4.49}     	& 1,512 & \textbf{2.20}    & 2,159 & \textbf{1.23} &364&\textbf{3.87}\\
0.2  & 1,030 & \textbf{2.78}     & 811 & \textbf{2.95}     	& 2,963 & \textbf{1.75}    & 4,295 & \textbf{1.26} & 721  & \textbf{2.95}\\
0.3  & 1,523 & \textbf{1.95}     & 1,210 & \textbf{2.15}     & 4,446 & \textbf{1.38}    & 6,430 & \textbf{0.75} & 1,079&\textbf{2.76}\\
0.4  & 2,045 & \textbf{1.35}     & 1,605 & \textbf{1.52}     & 5,942 & \textbf{0.76}    & 8,576 & \textbf{0.72} & 1,432&\textbf{1.81}\\
0.5  & 2,518 & \textbf{1.10}     & 1,985 & \textbf{1.46}     & 7,361 & \textbf{0.79}    & 10,702 & \textbf{0.49} & 1,779&\textbf{1.42}\\
0.6  & 3,035 & \textbf{0.84}     & 2,395 & \textbf{1.19}     & 8,853 & \textbf{0.60}    & 12,863 & \textbf{0.46} & 2,139&\textbf{1.22}\\
0.7  & 3,550 & \textbf{0.82}     & 2,791 & \textbf{0.70}     & 10,334 & \textbf{0.46}    & 15,007 & \textbf{0.40} & 2,494&\textbf{1.03}\\
0.8  & 4,027 & \textbf{0.48}     & 3,178 & \textbf{0.70}     & 11,779 & \textbf{0.34}    & 17,134 & \textbf{0.23} & 2,849&\textbf{0.86}\\
0.9  & 4,527 & \textbf{0.39}     & 3,574 & \textbf{0.42}     & 13,228 & \textbf{0.24}    & 19,272 & \textbf{0.16} & 3,202&\textbf{0.62}\\
\hline
\end{tabular}
Note: \#M, number of mutants. Accurate results (i.e., $\delta_{acc} \le 5\%$) are in bold.
\end{table}

Table~\ref{table:results:accuracy:full} reports on the mutation scores obtained with the entire test suite for all subjects. 
As expected, the best mutation score is obtained for \MLFS{}{}, whose test suite achieves MC/DC coverage.

Table~\ref{table:results:accuracy:regSampling} provides accuracy results (column $\delta_{acc}$) for proportional uniform sampling for a range of sampling rates~($r$). 
To enable comparisons across sampling methods, Column \emph{\#M} reports the number of mutants sampled for each sampling rate.
As expected, a larger sampling rate leads to more accurate results (i.e., low $\delta_{acc}$). 
We notice that for test suites that ensure MC/DC coverage (i.e., \MLFS{}{}), even a very small sampling ratio (i.e., 0.01) guarantees a $\delta_{acc}$ below 5\%. However, to achieve an accurate mutation score estimate across all subjects, a minimum sampling rate of 0.09 is required.

In addition, we observe that, for $r=0.09$, the worst results (highest deltas) are observed for smaller projects, which indicates that \textbf{the estimation accuracy may not depend on the percentage of sampled mutants but on the size of the sample}; indeed, for most of the subjects, accurate results \CHANGED{(i.e., $\delta_{acc} < 5\%$)} are obtained with a number of mutants between 350 and 450. This aspect is further studied when considering  \emph{uniform fixed-size sampling} and \emph{uniform FSCI sampling}.

\subsubsection*{Results - proportional method-based sampling}

Table~\ref{table:results:accuracy:methodBased} shows the accuracy results for proportional method-based sampling. 
Interestingly, for two subjects (i.e., \GCSP{} and \UTIL{}), proportional method-based sampling leads to accurate estimates of the mutation score with a lower number of mutants than proportional uniform sampling (i.e., around 250).
However, to achieve accurate results with all subjects, we need a minimal sampling rate of $r=0.09$, as for proportional uniform sampling, which, in the case of method-based sampling, leads to a slightly higher number of mutants. For this reason, we do not see any benefit in using method-based sampling.

\subsubsection*{Results - uniform fixed-size sampling, uniform FSCI sampling}

\begin{table*}[htb]
\caption{RQ2. Accuracy with uniform fixed-size sampling and uniform FSCI sampling. }
\label{table:results:accuracy:FSCI:sampling} 
\scriptsize
\centering
\begin{tabular}{|
>{\raggedleft\arraybackslash}p{11mm}@{\hspace{1pt}}|
>{\raggedleft\arraybackslash}p{5mm}@{\hspace{1pt}}|
p{12mm}@{\hspace{1pt}}|
>{\raggedleft\arraybackslash}p{11mm}@{\hspace{1pt}}|
>{\raggedleft\arraybackslash}p{5mm}@{\hspace{1pt}}|
p{12mm}@{\hspace{1pt}}|
>{\raggedleft\arraybackslash}p{11mm}@{\hspace{1pt}}|
>{\raggedleft\arraybackslash}p{5mm}@{\hspace{1pt}}|
p{12mm}@{\hspace{1pt}}|
>{\raggedleft\arraybackslash}p{11mm}@{\hspace{1pt}}|
>{\raggedleft\arraybackslash}p{5mm}@{\hspace{1pt}}|
p{12mm}@{\hspace{1pt}}|
>{\raggedleft\arraybackslash}p{11mm}@{\hspace{1pt}}|
>{\raggedleft\arraybackslash}p{5mm}@{\hspace{1pt}}|
p{12mm}@{\hspace{1pt}}|
}
\hline
\multicolumn{3}{|c|}{\textbf{\GCSP{}}}      & \multicolumn{3}{c|}{\textbf{\PARAM{}}}      & \multicolumn{3}{c|}{\textbf{\UTIL{}}}       & \multicolumn{3}{c|}{\textbf{\MLFS{}}}          & \multicolumn{3}{c|}{\textbf{\SAIL{}}$_S$}       \\
\hline
\textbf{\#Mutants} & $\delta_{acc}$ & \textbf{Method}   & \textbf{\#Mutants} & $\delta_{acc}$ & \textbf{Method}   & \textbf{\#Mutants} & $\delta_{acc}$ & \textbf{Method}   & \textbf{\#Mutants} & $\delta_{acc}$ & \textbf{Method}   & \textbf{\#Mutants} & $\delta_{acc}$ & \textbf{Method} \\
\hline
100       & 9.88       & FIXED    & 100       & 10.17      & FIXED    & 100       & 7.32       & FIXED    & 100       & 7.80       & FIXED    &  100         &     8.89       &   FIXED     \\
200       & 5.86       & FIXED    & 200       & 6.12       & FIXED    & 200       & 5.73       & FIXED    & 200       & 5.20       & FIXED    &   200        &      6.14      &   FIXED     \\
300       & \textbf{4.48}       & FIXED    & 300       & 5.88       & FIXED    & 300       & 5.37       & FIXED    & 248       & \textbf{4.56}       & FSCI 0.1  &    300       &      5.53      &    FIXED    \\
364       & \textbf{4.42}       & FSCI 0.1  & 346       & \textbf{4.26}       & FSCI 0.1  & 333       & \textbf{4.73}       & FSCI 0.1  & 300       & \textbf{4.04}       & FIXED    &   366        &     \textbf{3.92}       &    FSCI 0.1    \\
400       & \textbf{5.49}       & FIXED    & 400       & \textbf{4.27}       & FIXED    & 400       & \textbf{4.45}       & FIXED    & 302       & \textbf{4.64}       & FSCI 0.09 &     400      &     \textbf{4.52}       &    FIXED    \\
447       & \textbf{3.70}       & FSCI 0.09 & 425       & \textbf{3.79}       & FSCI 0.09 & 409       & \textbf{4.08}       & FSCI 0.09 & 379       & \textbf{4.01}       & FSCI 0.08 &   449        &    \textbf{3.66}     &   FSCI 0.09    \\
500       & \textbf{3.85}       & FIXED    & 500       & \textbf{3.63}       & FIXED    & 500       & \textbf{3.80}       & FIXED    & 400       & \textbf{3.80}       & FIXED    &   500      &      \textbf{4.08}      &   FIXED     \\
564       & \textbf{3.53}       & FSCI 0.08 & 536       & \textbf{3.75}       & FSCI 0.08 & 514       & \textbf{3.94}       & FSCI 0.08 & 490       & \textbf{3.90}       & FSCI 0.07 &    567       &  \textbf{3.03}   &  FSCI 0.08      \\
600       & \textbf{3.65}       & FIXED    & 600       & \textbf{3.72}       & FIXED    & 600       & \textbf{3.29}       & FIXED    & 500       & \textbf{3.11}       & FIXED    &     600      &  \textbf{3.73}          &     FIXED   \\
700       & \textbf{3.00}       & FIXED    & 696       & \textbf{2.89}       & FSCI 0.07 & 668       & \textbf{3.99}       & FSCI 0.07 & 600       & \textbf{2.89}       & FIXED    &     700      &      \textbf{3.01}      &    FIXED    \\
734       & \textbf{3.62}       & FSCI 0.07 & 700       & \textbf{3.27}       & FIXED    & 700       & \textbf{3.30}       & FIXED    & 667       & \textbf{2.78}       & FSCI 0.06 &     738      &      \textbf{2.77}      &    FSCI 0.07    \\
800       & \textbf{2.90}       & FIXED    & 800       & \textbf{2.51}       & FIXED    & 800       & \textbf{3.26}       & FIXED    & 700       & \textbf{2.80}       & FIXED    &     800      &  \textbf{2.55}          &    FIXED    \\
900       & \textbf{3.09}       & FIXED    & 900       & \textbf{2.50}       & FIXED    & 900       & \textbf{3.04}       & FIXED    & 800       & \textbf{2.44}       & FIXED    &     900      &  \textbf{2.37}          &    FIXED    \\
994       & \textbf{3.00}       & FSCI 0.06 & 945       & \textbf{2.42}       & FSCI 0.06 & 906       & \textbf{3.28}       & FSCI 0.06 & 900       & \textbf{3.02}       & FIXED    &  998         &  \textbf{2.65}         &   FSCI 0.06     \\
1,000      & \textbf{2.41}       & FIXED    & 1,000      & \textbf{2.72}       & FIXED    & 1,000      & \textbf{2.31}       & FIXED    & 960       & \textbf{2.31}       & FSCI 0.05 & 1,000     &  \textbf{2.96}          &   FIXED     \\
1,422      & \textbf{2.44}       & FSCI 0.05 & 1,352      & \textbf{1.93}       & FSCI 0.05 & 1,298      & \textbf{2.72}       & FSCI 0.05 & 1,000      & \textbf{2.35}       & FIXED    &    1,429     &    \textbf{1.70}        &  FSCI 0.05 \\
\hline
\end{tabular}

Accurate results (i.e., $\delta_{acc} \le 5\%$) are in bold.
\end{table*}

Table~\ref{table:results:accuracy:FSCI:sampling} shows the accuracy results for uniform fixed-size sampling and uniform FSCI sampling. For each subject, we sort results according to the number of mutants. For FSCI sampling, we report the confidence interval threshold $T_\mathit{CI}$. 

The best results (i.e., lowest number of mutants with $\delta_{acc} \le 5\%$) are obtained using  FSCI sampling with $T_\mathit{CI}=0.10$. Predictably, FSCI sampling with $T_\mathit{CI}=0.10$ guarantees $\delta_{acc} \le 5\%$ (half of $T_\mathit{CI}$); indeed, by construction, if our assumptions 
on the limited correlation between mutants and the mutation score following a  binomial distribution hold (see Section~\ref{sec:scalability}),
FSCI sampling with $T_\mathit{CI}=0.10$ is expected to guarantee $\delta_{acc} \le 5\%$ (see Appendix
B
for further details on the distribution of the mutation score across subjects).

In addition, our results suggest that a limited number of mutants (between 300 and 400) is required to achieve the desired $\delta_{acc}$. This sample size is much lower than the (worst case) sample size proposed by Gopinath et al., which is 1,000~\cite{gopinath2015hard}. Also, our sample size is smaller 
than the one estimated, for a mutation score between 60\% and 80\%, by 
approaches
based on confidence-interval estimation, which is still around 1,000~\cite{Goncalves2012}. 
However, we confirm the finding of Gopinath et al., who demonstrated that the binomial distribution %
accurately estimates the mutation score~\cite{gopinath2015hard}.

To summarize, this is the first study demonstrating that \textbf{FSCI sampling is the 
best approach for obtaining the smallest sample size
while providing guarantees on the accuracy of mutation score estimates.} 
We therefore propose a better solution than that of Gopinath et al., who provide an upper bound for the number of mutants to be considered in uniform fixed-size sampling, since we have evidence suggesting that FSCI sampling helps to select a significantly smaller sample size for a desired confidence interval.

\subsection{RQ3 - SDL accuracy}

\subsubsection*{Design and measurements}

RQ3 assesses if mutants generated using only deletion operators can accurately estimate the mutation score of the complete mutants set.

To this end, we study the difference between the mutation score obtained by executing the entire test suite on the mutants generated with all the operators (i.e., the actual mutation score) and the mutation score obtained with either (1) the mutants generated with the SDL operator only, or (2) the mutants generated with both the SDL and OODL operators.
As for RQ2, to be accurate, the mutation score obtained with a subset of operators should differ by at most 5\%.

\subsubsection*{Results}

\begin{table}[htb]
\caption{Comparison of mutation scores obtained with mutants generated using all operators, the SDL operator only, and the SDL + OODL operators.}
\label{table:results:score:sdl:oodl} 
\scriptsize
\centering
\begin{tabular}{|
@{\hspace{1pt}}p{15mm}|
 >{\raggedleft\arraybackslash}p{8mm}@{\hspace{1pt}}|
  >{\raggedleft\arraybackslash}p{13mm}@{\hspace{1pt}}|
 >{\raggedleft\arraybackslash}p{6mm}@{\hspace{1pt}}|
  >{\raggedleft\arraybackslash}p{15mm}@{\hspace{1pt}}|
   >{\raggedleft\arraybackslash}p{15mm}@{\hspace{1pt}}|
}
\hline
\textbf{Subject}&\multicolumn{2}{c|}{\textbf{\# Mutants}}&\multicolumn{3}{c|}{\textbf{Mutation score}}\\ 
&SDL&SDL+OODL&ALL&SDL&SDL+OODL\\
\hline
$\mathit{\SAIL{}}_{S}$ &	701&	974& 65.36 & 61.91 (-3.45) & 63.45 (-1.91) \\
$\mathit{\GCSP{}}$ & 912	&1,546	&65.64 &70.72 (+5.08) &71.35 (+5.71)\\
$\mathit{\PARAM{}}$ & 731&1,324	&69.12 &64.84 (-4.28) &66.39 (+2.73)\\
$\mathit{\UTIL{}}$ 	 &2,341	&3,811	&71.20 & 73.26 (+2.06) &72.63 (+1.43)\\
$\mathit{\MLFS{}}$ &1,729	&	5,971	&81.80 &85.71 (3.91)& 88.03 (+6.23)\\
\hline
\end{tabular}

\end{table}

In Table~\ref{table:results:score:sdl:oodl}, column \emph{\# Mutants} shows, for each subject, the number of mutants generated with either the SDL operator or both the SDL and OODL operators. 
Column \emph{Mutation score} shows the mutation score obtained when using the entire test suite to exercise the mutants generated with either all the operators, the SDL operator only, or both the SDL and OODL operators. Between parentheses, we also report the difference between the mutation score obtained with all the operators and that obtained with a subset of operators. Results show that, for some of our subjects, the mutation score obtained with the SDL operator does not accurately estimate the mutation score obtained with a broader set of operators. Though these results do not invalidate related work~\cite{delamaro2014experimental}, whose focus is on the evaluation of the strength of SDL and OODL operators, it shows that \textbf{SDL and OODL operators should not be adopted to estimate the mutation score computed with a larger set of operators}. We leave the evaluation of the strength of SDL and OODL operators to future work.

\subsection{RQ4 - Mutation Score Accuracy with PrioritizeAndReduce}
\label{exp:accuracy:prioritize}

\subsubsection*{Design and measurements}

RQ4 assesses whether the mutation score obtained with the reduced and prioritized test suite generated by \APPR (hereafter, the \MPTS) accurately estimates the actual mutation  score.
To this end, we compare the accuracy obtained with the four distance metrics (i.e., $D_J$, $D_O$, $D_E$, and $D_C$) used by the proposed \emph{PrioritizeAndReduce} algorithm (Figure~\ref{alg:prioritize}). In addition, to determine to what extent our prioritization strategy based on code coverage contributes to the selection of test cases that kill mutants, we also compare the results obtained with a simple baseline that, for each mutant, randomly selects one test case among the ones that cover the mutant.

For all subjects, we consider (a) the complete set of mutants, (b) the reduced subset of mutants providing accurate results (i.e., the one obtained with FSCI sampling with $T_{\mathit{CI}}=0.10$). 
Based on RQ3 results, we exclude mutants generated with the SDL and SDL+OODL operators only. 
For FSCI sampling, since we evaluate the accuracy of a reduced test suite, we derive the confidence interval using Equation~\ref{eq:CI}.

For each subject and each distance metric, and for each of the two sets of mutants considered, we generated ten different \MPTSs. In the case of FSCI, since it randomly selects mutants, we considered ten different sets of mutants derived with distinct executions of the FSCI algorithm. For each \MPTS, we computed the mutation score obtained. 
Then, to determine if the mutation score of the \MPTS is accurate, we follow the same procedure adopted for RQ2, i.e., we rely on the 2.5/97.5 percentile distance from the actual mutation score.

\subsubsection*{Results}

\MR{
\begin{table}[]
\centering
\scriptsize
\caption{RQ4. Mutation score accuracy for the different strategies implemented by \emph{PrioritizeAndReduce}}
\label{table:results:PriritizeAndReduce} 
\begin{tabular}{|
p{14mm}@{\hspace{1pt}}|
p{15mm}@{\hspace{1pt}}|
>{\raggedleft\arraybackslash}p{10mm}@{\hspace{1pt}}|
>{\raggedleft\arraybackslash}p{8mm}@{\hspace{1pt}}|
>{\raggedleft\arraybackslash}p{8mm}@{\hspace{1pt}}|
>{\raggedleft\arraybackslash}p{8mm}@{\hspace{1pt}}|
>{\raggedleft\arraybackslash}p{8mm}@{\hspace{1pt}}|
}
\hline
           &          &\multicolumn{5}{c|}{$\delta_{acc}$ for different prioritization strategies}\\
\hline
\textbf{Subject} & \textbf{Mutants set} & Random & $D_J$ & $D_O$ & $D_E$ & $D_C$ \\
\hline
\multirow{2}{*}{\GCSP{}}    
    & ALL & 7.2055& 1.3455& 1.3100& 0.7300& 0.7300 \\
    &  FSCI 0.10  &  $>$5\% & 3.87& 3.80& 4.14& 4.14 \\
\hline
\multirow{2}{*}{\PARAM{}}    
    & ALL & 7.7927& 0  & 0& 0    & 0  \\
    & FSCI 0.10  & $>$5\% &3.22&3.22&3.22&3.22 \\
\hline
\multirow{2}{*}{\UTIL{}}    
    & ALL & 3.1400 &  0.0300 & 0.0300 & 0.0199 & 0.0199\\
    & FSCI 0.10  & $>$5\%  & 1.95 & 1.95   & 1.95 & 1.95 \\
\hline
\multirow{2}{*}{\MLFS{}}    
    & ALL &  6.721  &   0.3299  &   0.3299 &   0.0199 &    0.0300  \\
    & FSCI 0.10  & $>$5\%  & 2.97 & 2.97 & 2.85 & 2.85 \\
\hline
\multirow{2}{*}{\SAIL{}$_S$}    
    & ALL  & 38.8885  & 24.1688& 24.3650  &  4.0800   &  3.9833   \\
    & FSCI 0.10   &   $>$5\%   &   2.87  &  2.87  &  2.67 &   2.49\\
\hline
\end{tabular}
\end{table} 
}

Table~\ref{table:results:PriritizeAndReduce} provides the values of $\delta_{acc}$ obtained for the different subjects and distance metrics (i.e., the random baseline and the four distance metrics supported by \APPR). Rows named \emph{ALL} report the results obtained when executing the entire set of mutants, rows named \emph{FSCI}  report the results obtained  with the FSCI strategy.

\NEWFSCI{Unsurprisingly, for all the subjects except \SAIL{}$_S$, the mutation score computed with the entire set of mutants tested with the \MPTS (i.e., row ALL) is more accurate than the mutation score computed with a subset of the mutants tested with the same test suite (i.e., row FSCI). 
However, the mutation score estimated with FSCI is always accurate (i.e, $\delta_{acc}  < 5$).
In the case of \SAIL{}$_S$, where each statement is covered by a large number of test cases (see Section~\ref{sec:empirical:subjects}),
test suite reduction has a higher probability of retaining a test case that does not kill a mutant.
For this reason, executing a reduced test suite with a subset of mutants selected with FSCI, which estimates the error due to test suite reduction, 
leads to a more accurate mutation score than executing a reduced test suite with the entire set of mutants without estimating such error (i.e., row ALL).
}

The only distance metric that consistently leads to inaccurate estimates of the mutation score (i.e., $\delta_{acc}  > 5$) across subjects is the random baseline.
Based on a non-parametric Mann Whitney test, the difference between the random baseline (i.e., Random) and the four distance metrics implemented by \APPR (i.e., $D_*$) is always significant with a \textit{p}-value $< 0.05$. This indicates that \textbf{the \APPR distance metrics are necessary to accurately estimate the mutation score while reducing the number of test cases to  execute}.

Among the proposed distance metrics, 
$D_J$ and $D_O$ provide inaccurate results with \SAIL{}$_S$.
We conjecture the main reason is their inability to account for the number of times a statement is exercised by a test case. 
We believe this is an important factor as system test cases that repeatedly exercise mutated statements, with different variable values, are more likely to kill mutants than test cases exercising such statements only once (e.g., because of the uncertainty regarding the 
\JMR{3.16}{incorrect intermediate state
propagating to the state variables verified by test oracles, see Section~\ref{background:adequacy}}). On the other hand, unit and integration test cases, which exercise much simpler scenarios in other subjects, are more likely to kill a mutant when the mutated statement is executed only once (e.g., because the oracle is closer to the mutated statement). This is why $D_J$ and $D_O$ fare similarly to the other distance metrics for these subjects. 
In contrast, $D_C$ and $D_E$ are distance metrics ensuring an accurate estimate of the mutation score and providing the lowest $\delta_{acc}$. The differences between $D_E$ and $D_C$ are always statistically significant, \UPDATE{except for \MLFS{}{}}. However, there are no practically significant differences between them. 
Since $D_C$ provides a normalized score, which is required by Step 8, we select $D_C$ as the preferred metric to be used in \APPR.

\subsection{RQ5 - Time Savings with PrioritizeAndReduce}

\subsubsection*{Design and measurements}

RQ5 assesses to what extent  the \MPTS speeds up the mutation analysis process.

For each subject considered for RQ4, we measure the execution time taken by the \MPTS to execute on the mutants.
We compute time saving as the ratio of the difference in execution time from the original test suite over the time it requires to execute, for the set of mutants selected.
In particular, as in RQ4, we consider three scenarios (1) all the mutants are selected, (2) mutants are selected with FSCI sampling, (3) mutants are selected with FSCI sampling but we execute the entire test suite, as opposed to the \MPTS. By considering scenario (3), we can estimate both the time saved thanks to mutant sampling and the additional time saved when combining it with the \MPTS.
For the original test suite, to emulate a realistic mutation analysis process according to state-of-the-art solutions, we measure the time required to execute test cases until the mutant is killed (for live mutants it means that we execute the entire test suite). Also, we set a test case timeout equal to three times the duration of the original test case.

Since the execution time of test cases depends on multiple factors such as the underlying test harness, the development practices in place (e.g., verifying multiple scenarios in a single test case), and the type of testing conducted (e.g., unit, integration, system), we also compute 
the ratio of the number of test cases not executed by the \MPTS over the total number of test cases.

\subsubsection*{Results}

Table~\ref{table:time:original} reports, for every subject, the time required to test all the mutants with the entire test suite, in seconds. It also reports  the total number of test cases executed. 
We observe that mutation analysis requires a large amount of time. It takes between 13 and 73 hours for subjects tested with unit and integration test suites (which are faster to execute). When a system test suite needs to be executed (i.e., $\mathit{\SAIL{}_S}$), traditional mutation analysis becomes infeasible as shown by the 11,000 hours required to perform mutation analysis with $\mathit{\SAIL{}_S}$.

\begin{figure}[tb]
\begin{center}
\includegraphics[width=\columnwidth]{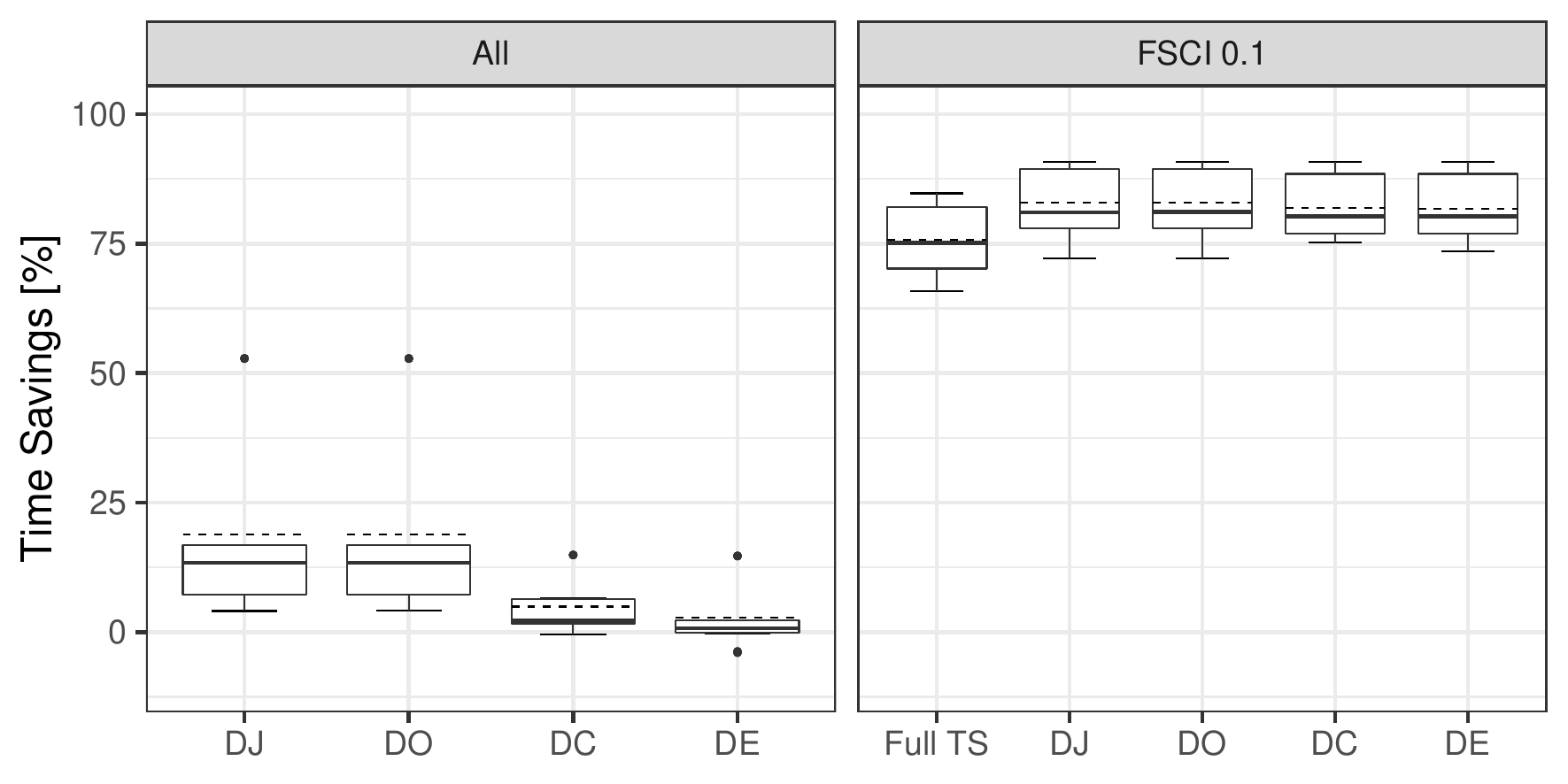}
\caption{Time savings for different sets of mutants, with the \MPTS being generated based on different distance measures.}
\label{fig:results:time:saving}
\end{center}
\end{figure}

\begin{figure}[tb]
\begin{center}
\includegraphics[width=\columnwidth]{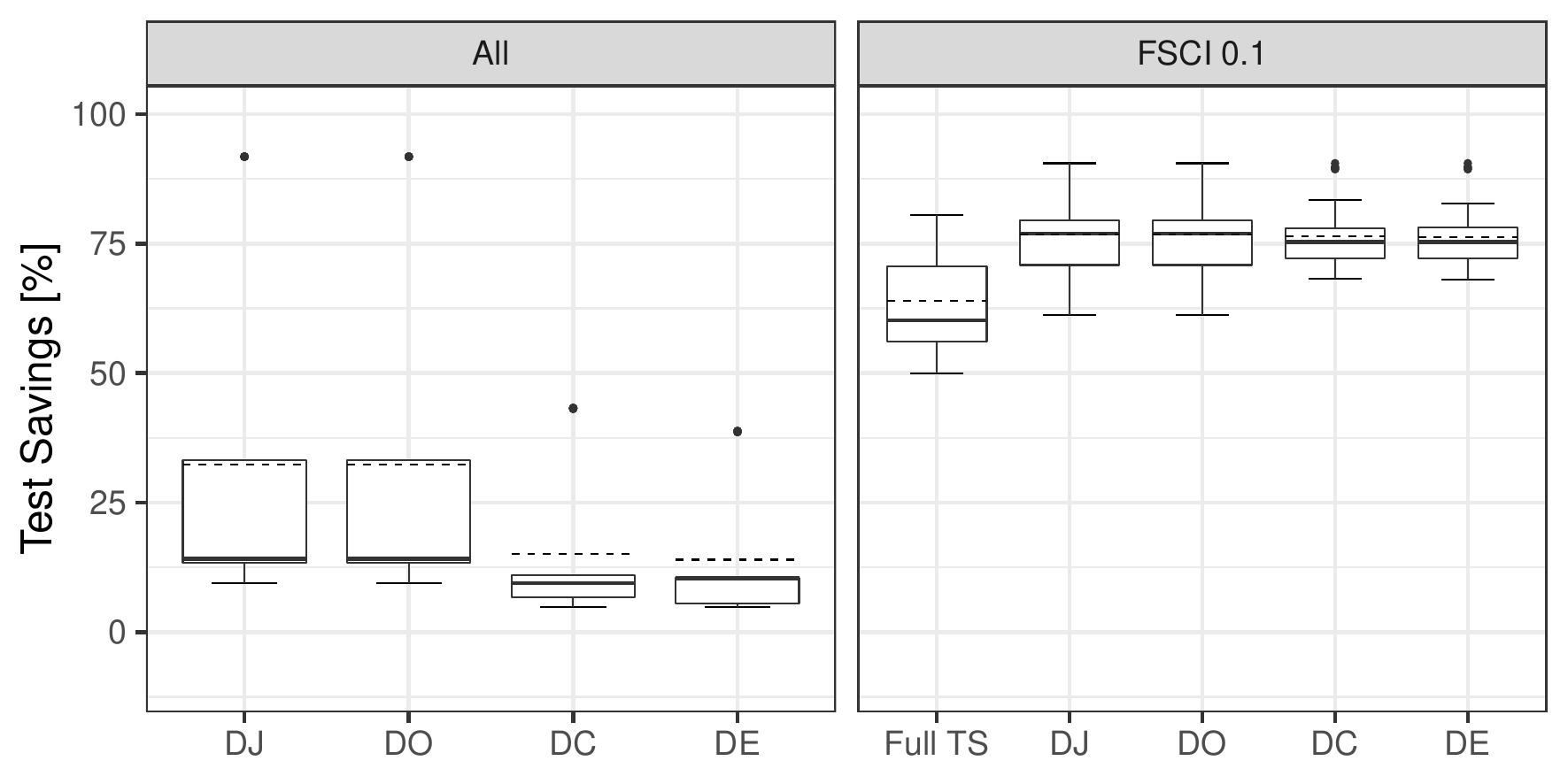}
\caption{Test case savings for different sets of mutants, with the \MPTS being generated based on different distance measures.}
\label{fig:results:test:saving}
\end{center}
\end{figure}

Figures~\ref{fig:results:time:saving} and~\ref{fig:results:test:saving} provide boxplots depicting 
the saving achieved when the \MPTS is executed with all the mutants and with the FSCI-selected mutants, in terms of execution time and test cases, respectively. Each observation corresponds to the saving obtained with one of the ten executions performed on each subject, for a specific configuration (i.e., distance $D_*$ and strategy adopted for selecting mutants). In Figures~\ref{fig:results:time:saving} and~\ref{fig:results:test:saving}, horizontal dashed lines show the average across all subjects, whiskers are used to identify outliers (i.e., they are placed below/above 1.5*Inter Quartile Range of the upper quartile/lower quartile). A detailed table including min, max, mean, and median values for each subject is provided in Appendix C.

Unsurprisingly, the smallest reduction in execution time and number of executed test cases is achieved when executing the \MPTS with all the mutants. 
Measuring such time reduction allows us to evaluate the benefits of test suite prioritization and selection when it is not combined with mutant selection.
Excluding outliers, execution time reduction goes from  \UPDATE{-0.52\%} %
to 16.82\% and appears to be correlated with the reduction in number of test cases to execute, which varies from 4.80\% to 33.17\%.
The largest reductions are obtained with $D_J$ and $D_O$ (the median is 13.36 and 13.40, respectively), which do not consider differences in coverage frequency, thus removing the largest number of test cases; however, $D_J$ and $D_O$ also lead to the worst accuracy results according to RQ4. Metrics $D_C$ and $D_E$, instead, lead to limited benefits in terms of time reduction.

A negative reduction indicates that the reduced and prioritized test suite increases the execution time of the mutation analysis process. This happens when (1) test cases are sorted in such a way that test cases that kill the mutants are executed later with respect to the original test suite, (2) test cases that kill the mutants but have long execution times (e.g., because they trigger a timeout) are executed before test cases with shorter execution times that kill the mutants. Negative reduction, however, affects only a few executions, thus showing that a reduced and prioritized test suite tends overall to be beneficial to the mutation analysis process.

Mutant sampling alone contributes to a high reduction in execution time; indeed, in Figure~\ref{fig:results:time:saving}, the boxplot \emph{FSCI 0.1-Full TS}, depicting the time saving for FSCI mutants tested with the entire test suite, shows minimum, median, and maximum values of 65.76\%, 75.10\%, and 84.72\%. Indeed, by reducing the number of mutants to execute, FSCI sampling significantly reduced the total number of test cases to execute within a [\JMRCHANGE{49.98\%} - 80.54\%] range (as shown by the boxplot in Figure~\ref{fig:results:test:saving}).

The highest reduction in execution time is achieved when combining the \MPTS with FSCI sampling. It ranges from 72.09\% to 90.83\%. $D_C$ and $D_E$, which are the approaches that guarantee accurate results, lead to an execution time reduction in the ranges \JMRCHANGE{[75.25\% - 90.83\%]} and \JMRCHANGE{[73.53\% - 90.83\%]}, respectively, an impressive achievement. 
Test case savings, 
\JMRCHANGE{as well, are above 65\%, [68.28\% - 83.45\%]} and \JMRCHANGE{[68.08\% - 82.70\%]} for $D_C$ and $D_E$, respectively. Based on savings results, there is no practical difference between $D_C$ and $D_E$.

To conclude, \textbf{we suggest to combine FSCI sampling with the \MPTS to minimize the time required by mutation analysis}. 
For $\mathit{\SAIL{}}_S$, 
\NEWFSCI{on average, when combining the \MPTS with FSCI sampling, mutation analysis time goes from 11,000 hours to 1,531 hours, which makes mutation analysis feasible in one week with 10 computing nodes.} Note that for safety or mission critical systems, such as satellites software, the cost of using computing nodes is minimal compared to the development cost of the entire system. Indeed, to test such systems, even paying for the computational power of 100 HPC nodes to make mutation analysis feasible in half a day, is economically justifiable. \JMR{2.2}{Otherwise, without \APPR, mutation analysis leads to 11,000 hours of test cases execution, thus being practically infeasible since it would require more than 100 days to be completed, even with 100 HPC nodes.}

Our results also show that when it is not feasible to collect coverage data for the mutants under test (a requirement to generate the \MPTS), \textbf{FSCI sampling alone, without a reduced test suite, may still provide a high reduction in execution time}. In the case of $\mathit{\SAIL{}}_S$, this leads to 2,920 mutation analysis hours, which require less than two days with 100 HPC nodes.

\begin{table}[tb]
\centering
\caption{Execution time and number of test cases executed when mutation analysis is based on the original test suite.}
\label{table:time:original} 
\scriptsize
\begin{tabular}{|
p{14mm}@{\hspace{2pt}}|
>{\raggedleft\arraybackslash}p{34mm}@{\hspace{1pt}}
>{\raggedleft\arraybackslash}p{10mm}@{\hspace{1pt}}|
>{\raggedleft\arraybackslash}p{12mm}@{\hspace{1pt}}|
}
\hline
\textbf{Subject}&\multicolumn{2}{c|}{\textbf{Execution time, seconds (hours)}}&\multicolumn{1}{c|}{\textbf{\# Test cases}}\\
\hline
\multirow{1}{*}{\SAIL{}$_S$}& 39,604,457  &(11,001) & 155,751 \\
\hline
\multirow{1}{*}{\GCSP{}}&  252,776 &(70) & 10,250\\
\hline
\multirow{1}{*}{\PARAM{}}&  47,949 &(13) & 6,629\\
\hline
\multirow{1}{*}{\UTIL{}}&  214,016 &(59) & 17,672\\
\hline
\multirow{1}{*}{\MLFS{}}&  171,790 &(47)& 28,159\\
\hline
\textbf{Total}& 40,290,988 &(11,191) & 218,461\\ 
\hline
\end{tabular}
\end{table}

\subsection{RQ6 - Precise Detection of Equivalent and Duplicate Mutants}
\label{sec:empirical:thrshold}
\subsubsection*{Design and measurements}

RQ6 investigates if it is possible to identify thresholds that enable the accurate identification of mutants that are nonequivalent ($T_E$) and nonduplicate ($T_D$), following the procedure described in Section~\ref{sec:algostepSeven}.

To determine $T_E$ and $T_D$, 
we rely on the optimal distance metric identified in RQ4 ($D_C$).
We analyze  precision and recall of the results obtained for  different values of $T_E$ and $T_D$.
To determine $T_E$, we measure
precision as the percentage of mutants with a distance above $T_E$ that are nonequivalent, recall as the percentage of nonequivalent mutants with a distance above $T_E$.
To determine $T_D$, we measure
precision as the percentage of mutant pairs with a distance above $T_D$ that are duplicate, recall as the percentage of duplicate mutant pairs that have a distance above $T_D$.

Since the quality of results might be affected by both test suite reduction (i.e., less coverage data may be available) and mutants sampling (e.g., less mutants might be sampled), consistent with the finding of previous RQs, we consider the following two configurations: 
\begin{itemize}
\item Execution of the original test suite with all the generated mutants (ALL)
\item Execution of the \MPTS with FSCI sampling (\APPR)
\end{itemize}

\begin{figure}[tb]
\begin{center}
\includegraphics[width=\columnwidth]{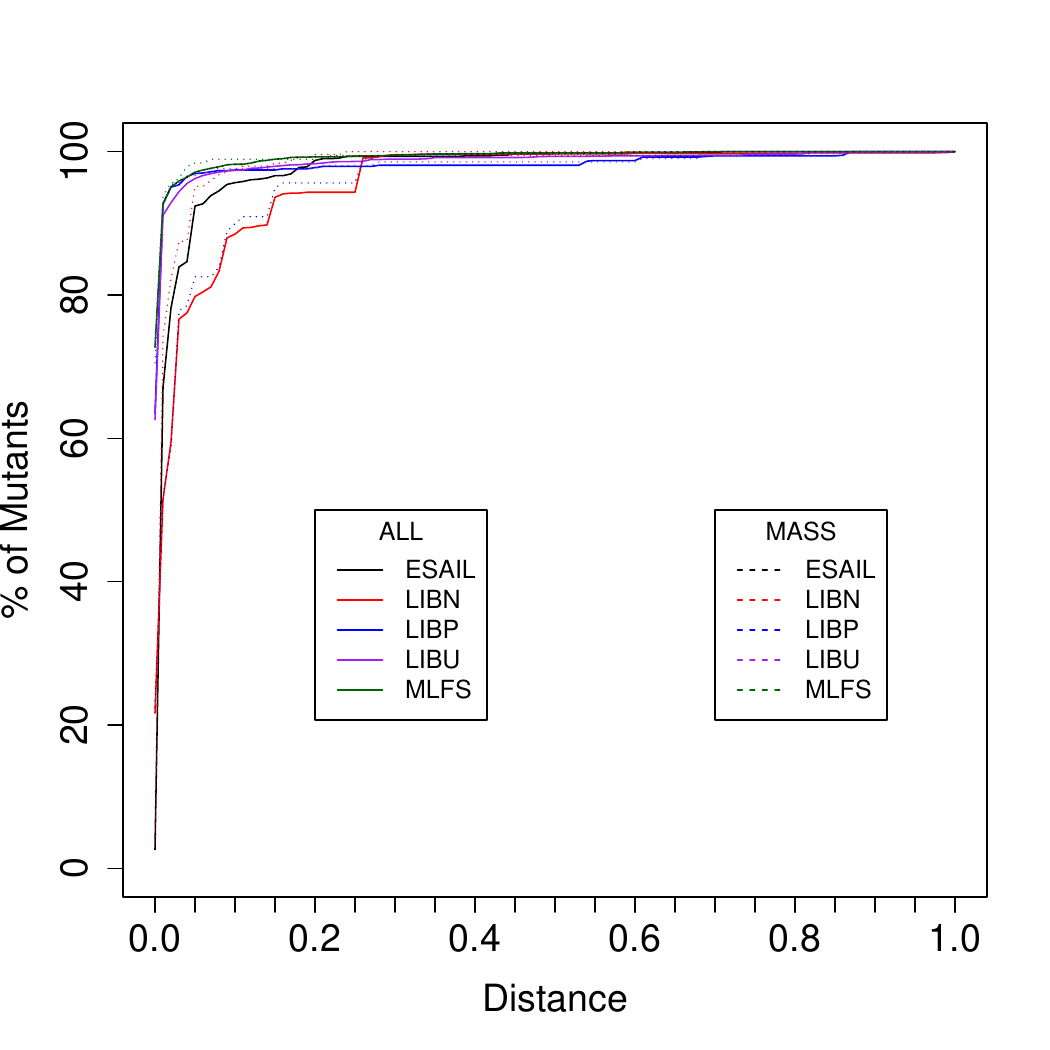}
\caption{Cumulative distribution of mutants over distance values computed to determine equivalent mutants.}
\label{fig:results:test:dde}
\end{center}
\end{figure}

\begin{figure}[tb]
\begin{center}
\includegraphics[width=\columnwidth]{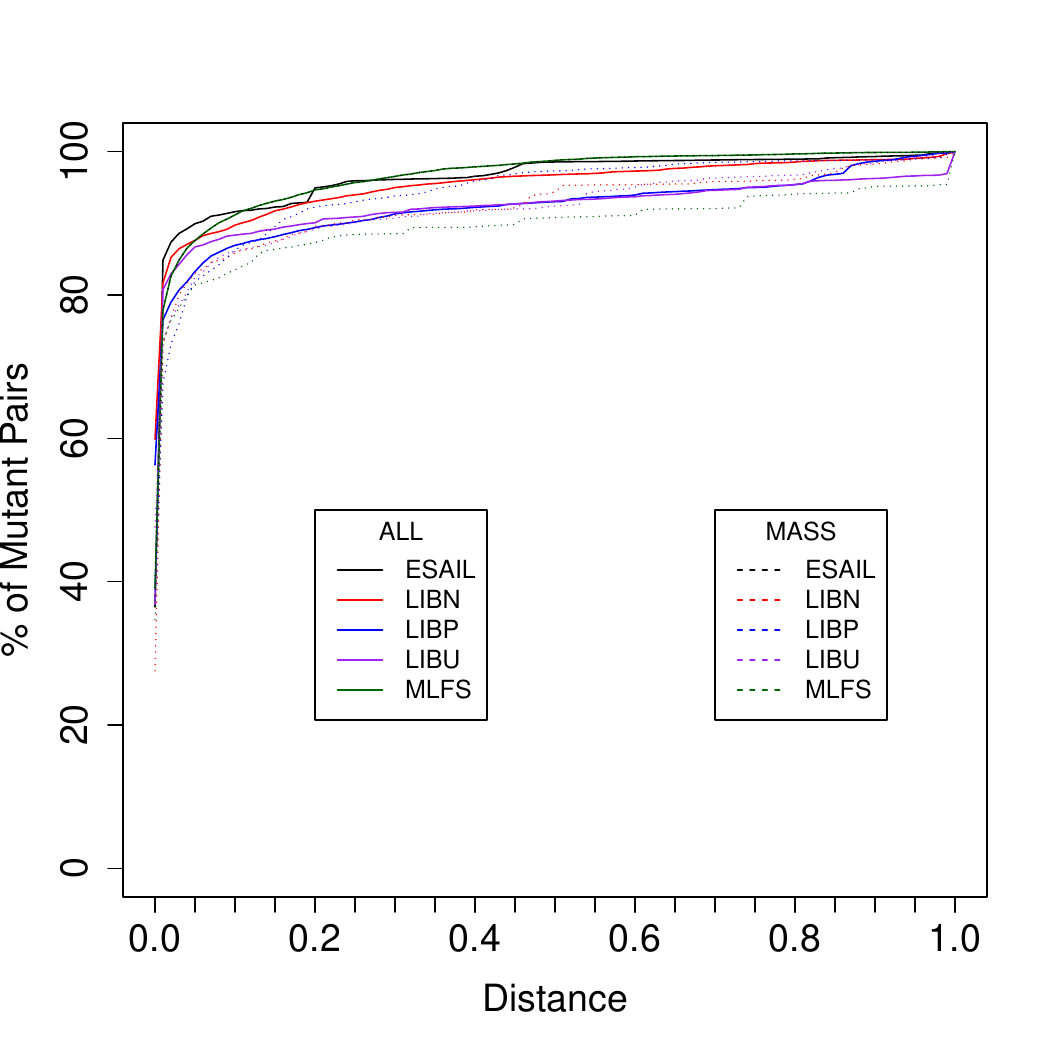}
\caption{Cumulative distribution of mutant pairs over distance values computed to determine duplicate mutants.}
\label{fig:results:test:ddd}
\end{center}
\end{figure}

We determine the values of $T_E$ and $T_D$ based on the analysis of the 
cumulative distribution of the distance values computed to determine equivalent and duplicate mutants, for the two configurations listed above.
Figures~\ref{fig:results:test:dde} and
~\ref{fig:results:test:ddd}
show the cumulative distribution --- the Y-axis shows the percentage of mutants and mutant pairs with a distance lower or equal to the value in the X-axis.
For both Figures~\ref{fig:results:test:dde} and
~\ref{fig:results:test:ddd} we can observe that the distribution of mutants is not uniform in the range 0-1 (otherwise we would have straight lines with 45 degree angle) but we observe a large proportion of mutants (Figures~\ref{fig:results:test:dde}) and mutant pairs (Figures~\ref{fig:results:test:ddd}) having small distances. For example,
Figure~\ref{fig:results:test:dde} shows that, across all subjects, more than 60\% of the mutants have a distance below 0.05 (i.e., with $x$ equal to $0.05$, the value of $y$ is above $60$). 

To evaluate precision and recall, we thus select values for $T_D$ and $T_E$ that 
either largely differ (i.e., $0.0$, $0.4$, and $0.8$) or delimit ranges including a large proportion of the mutants (i.e., $0$, $0.01$, and $0.05$). Table~\ref{table:results:proportion:mutants} reports the percentage of mutants and mutant pairs belonging to the ranges delimited by the selected values, for the two configurations considered in our study; the distribution of mutants in Table~\ref{table:results:proportion:mutants} is consistent with Figures~\ref{fig:results:test:dde} and
~\ref{fig:results:test:ddd}.

\begin{table}[h]
\caption{Distribution (\%) of mutants in distance ranges.}
\label{table:results:proportion:mutants} 
\scriptsize
\centering
\begin{tabular}{|
@{\hspace{1pt}}p{14mm}@{\hspace{1pt}}|
@{\hspace{1pt}}p{5mm}@{\hspace{1pt}}|
@{\hspace{1pt}}>{\raggedleft\arraybackslash}p{5mm}@{\hspace{1pt}}|
@{\hspace{1pt}}>{\raggedleft\arraybackslash}p{5mm}@{\hspace{1pt}}|
@{\hspace{1pt}}>{\raggedleft\arraybackslash}p{5mm}@{\hspace{1pt}}|
@{\hspace{1pt}}>{\raggedleft\arraybackslash}p{5mm}@{\hspace{1pt}}|
@{\hspace{1pt}}>{\raggedleft\arraybackslash}p{5mm}@{\hspace{1pt}}|
>{\raggedleft\arraybackslash}p{5mm}@{\hspace{1pt}}|
@{\hspace{1pt}}>{\raggedleft\arraybackslash}p{5mm}@{\hspace{1pt}}|
@{\hspace{1pt}}>{\raggedleft\arraybackslash}p{5mm}@{\hspace{1pt}}|
@{\hspace{1pt}}>{\raggedleft\arraybackslash}p{5mm}@{\hspace{1pt}}|
@{\hspace{1pt}}>{\raggedleft\arraybackslash}p{5mm}@{\hspace{1pt}}|
@{\hspace{1pt}}>{\raggedleft\arraybackslash}p{5mm}@{\hspace{1pt}}|
}
\hline
& \multicolumn{6}{c|}{\textbf{Distance for nonequivalent}}  & \multicolumn{6}{c|}{\textbf{Distance for nonduplicate}}  \\
\cline{2-13}
\textbf{}& (0.00 & (0.00 & (0.01& (0.05 & (0.4 & (0.8
& (0.00 & (0.00 & (0.01& (0.05 & (0.4 & (0.8\\
\textbf{Config.}& 0.00] & 0.01] & 0.05]& 0.40] & 0.8] & 1.0] 
& 0.00] & 0.01] & 0.05]& 0.40] & 0.8] & 1.0] \\
\hline
ALL   
& 53.9  & 29.8  & 10.1 & 5.5  & 0.5  & 0.2  
& 39.2  & 40.5  & 7.3 & 7.6  & 2.6  & 2.8  
\\
\APPR  
& 39.7  & 36.7  & 16.4  & 6.6 & 0.4  & 0.1 
& 33.7  & 37.9  & 10.5 & 10.5  & 3.8  & 3.3 
\\
\hline
\end{tabular}
\end{table}

To compute precision and recall for different values of $T_E$, since the distribution of mutants is not uniform, we rely on stratified sampling, as follows. 
We divide all the live mutants into six buckets, based on their distance from the original program, according to the ranges reported in Table~\ref{table:results:proportion:mutants}. 
We determine the ratio ($r_R$) of nonequivalent mutants in a specific range $R$ by randomly selecting 20 mutants (four for each subject) and inspecting them with the help of the engineers who developed the software. 
We rely on $r_R$ to estimate $e_{R}$, that is, the number of nonequivalent mutants in the entire set of mutants with a distance within the specific  range $R$
$$e_R = r_R * n_R$$
with $n_R$ being the number of mutants observed in the range $R$ for all the subjects\footnote{$n_R$ can be derived from Table~\ref{table:results:proportion:mutants}}.
Based on $e_R$, we estimate the number of nonequivalent mutants above a threshold and, consequently, compute precision and recall. We perform the analysis for both selected configurations,  ALL and \APPR.

Our aim is to identify a threshold value above which  we
maximize the number of nonequivalent mutants being selected (high recall) and maximize the number of equivalent mutants being discarded (high precision); since both precision and recall are equally important, we look for a threshold value that maximizes the harmonic mean of precision and recall (F-value).

To determine $T_D$, we repeated the same procedure as for $T_E$, except that we considered both killed and live mutants according to the procedure indicated in Section~\ref{sec:algostepSeven}.

In total, we manually inspected 410 mutants (186 mutants to detect equivalent mutants and 224 mutant pairs to detect duplicate ones), a larger number than that considered in related studies~\cite{schuler2013covering}. The number of inspected mutants is lower than the maximum of 480 (\{20 mutants + 20 mutant pairs\} x 6 buckets x 2 configurations) because the mutant distribution across ranges is not perfectly uniform (see Table~\ref{table:results:ratio:equivalent} for the number of observations per bucket).

\subsubsection*{Results}

The ratio ($r_R$) of nonequivalent mutants, for the distance ranges reported in Table~\ref{table:results:ratio:equivalent}, shows
similar results for both configurations (ALL and \APPR). 
The differences in distribution between mutants (Table~\ref{table:results:proportion:mutants}) and equivalent mutants  (Table~\ref{table:results:ratio:equivalent}), for both  configurations, is indeed not statistically significant and effect size is negligible.
Such similarity suggests that nonequivalent and nonduplicate mutants follow the same distribution for both configurations. This can be explained since 
FSCI sampling uniformly selects 
a subset of the mutants considered by the ALL configuration, which includes all mutants.

The 14 nonequivalent mutants leading to $d=0$ across the two configurations (seven for ALL, seven for \APPR) have the following characteristics.
Four mutants (29\%) invalidate data buffers' preconditions (e.g., an array size is indicated as larger than it should be). Since such faults are typically detected through profiling (e.g., by using Valgrind~\cite{Valgrind}), not detecting such mutants cannot be considered a major weakness of the approach. Seven mutants (50\%) affect variables that are not used in the mutated source file (i.e., the one for which we collect code coverage). 
Static analysis should, in principle, enable the identification of these mutants as nonequivalent. 
Three mutants (21\%) concern the deletion of clauses that are not tested by our test suites; these cases might be detected by our approach after combining statement coverage with additional coverage measures (e.g., clause coverage) to compute distances, but this is left to future work. Based on the above, the percentage of nonequivalent mutants that
may potentially indicate limitations of the test suite, cannot easily be detected by other means, 
and are ignored with $T_E$, when set to zero, is very low 
 (i.e., three out of 160, or 1.88\%). For this reason, \textbf{we consider the proposed $T_E$ threshold precise enough to be used for test suite evaluation in a safety context}.
 
Table~\ref{table:results:precision:equivalent} provides precision and recall obtained for different $T_E$ values; more precisely, we report the results obtained when all mutants are considered nonequivalent (i.e., $d\ge0$), along with the results obtained for $T_E$ being set to $0$, $0.01$, $0.05$, $0.4$, and $0.8$.

\begin{table}[tb]
\caption{RQ6. Ratio ($r_R$) of Nonequivalent/Nonduplicate Mutants per Distance Range.}
\label{table:results:ratio:equivalent} 
\scriptsize
\centering
\begin{tabular}{|
@{\hspace{1pt}}p{10mm}@{\hspace{1pt}}|
@{\hspace{1pt}}p{5mm}@{\hspace{1pt}}|
@{\hspace{1pt}}>{\raggedleft\arraybackslash}p{5mm}@{\hspace{1pt}}|
@{\hspace{1pt}}>{\raggedleft\arraybackslash}p{5mm}@{\hspace{1pt}}|
@{\hspace{1pt}}>{\raggedleft\arraybackslash}p{5mm}@{\hspace{1pt}}|
@{\hspace{1pt}}>{\raggedleft\arraybackslash}p{5mm}@{\hspace{1pt}}|
@{\hspace{1pt}}>{\raggedleft\arraybackslash}p{5mm}@{\hspace{1pt}}|
>{\raggedleft\arraybackslash}p{5mm}@{\hspace{1pt}}|
@{\hspace{1pt}}>{\raggedleft\arraybackslash}p{5mm}@{\hspace{1pt}}|
@{\hspace{1pt}}>{\raggedleft\arraybackslash}p{5mm}@{\hspace{1pt}}|
@{\hspace{1pt}}>{\raggedleft\arraybackslash}p{5mm}@{\hspace{1pt}}|
@{\hspace{1pt}}>{\raggedleft\arraybackslash}p{5mm}@{\hspace{1pt}}|
@{\hspace{1pt}}>{\raggedleft\arraybackslash}p{4mm}@{\hspace{1pt}}|
}
\hline
& \multicolumn{6}{c|}{\textbf{Distance range (nonequivalent)}}  & \multicolumn{6}{c|}{\textbf{Distance range (nonduplicate)}}  \\
\textbf{}& (0.00 & (0.00 & (0.01& (0.05 & (0.4 & (0.8
& (0.00 & (0.00 & (0.01& (0.05 & (0.4 & (0.8\\
\textbf{Config.}& 0.00] & 0.01] & 0.05]& 0.40] & 0.8] & 1.0] 
& 0.00] & 0.01] & 0.05]& 0.40] & 0.8] & 1.0] \\
\hline
ALL   
& 0.35  & 0.85  & 1.00 & 1.00  & 1.00  & 1.00  
& 0.95  & 1.00  & 1.00 & 1.00  & 1.00  & 1.00  
\\
& (20)  & (20)  & (20) & (20)  & (18)  & (12)  
& (20)  & (20)  & (20) & (20)  & (20)  & (20)  
\\
\APPR  
& 0.35  & 0.95  & 1.00  & 1.00 & 1.00  & N/A
& 1.00  & 1.00 & 1.00 & 1.00  & 1.00  & 1.00
\\
& (20)  & (20)  & (20) & (10)  & (6)  & (0)  
& (20)  & (20)  & (20) & (20)  & (18)  & (16)  
\\
\hline
\end{tabular}
\textbf{Note}: We report the number of observations in parenthesis.
\end{table}

\begin{table}[tb]
\caption{RQ6. Precision (P), Recall (R), and their harmonic mean (F) obtained for the Coverage-Based Detection of Nonequivalent/Nonduplicate Mutants.}
\label{table:results:precision:equivalent} 
\scriptsize
\centering
\begin{tabular}{|
@{\hspace{1pt}}p{6mm}@{\hspace{1pt}}|
@{\hspace{1pt}}p{2mm}@{\hspace{1pt}}|
@{\hspace{1pt}}p{5mm}@{\hspace{1pt}}|
@{\hspace{1pt}}>{\raggedleft\arraybackslash}p{5mm}@{\hspace{1pt}}|
@{\hspace{1pt}}>{\raggedleft\arraybackslash}p{6mm}@{\hspace{1pt}}|
@{\hspace{1pt}}>{\raggedleft\arraybackslash}p{6mm}@{\hspace{1pt}}|
@{\hspace{1pt}}>{\raggedleft\arraybackslash}p{6mm}@{\hspace{1pt}}|
@{\hspace{1pt}}>{\raggedleft\arraybackslash}p{6mm}@{\hspace{1pt}}|
>{\raggedleft\arraybackslash}p{5mm}@{\hspace{1pt}}|
@{\hspace{1pt}}>{\raggedleft\arraybackslash}p{5mm}@{\hspace{1pt}}|
@{\hspace{1pt}}>{\raggedleft\arraybackslash}p{6mm}@{\hspace{1pt}}|
@{\hspace{1pt}}>{\raggedleft\arraybackslash}p{6mm}@{\hspace{1pt}}|
@{\hspace{1pt}}>{\raggedleft\arraybackslash}p{6mm}@{\hspace{1pt}}|
@{\hspace{1pt}}>{\raggedleft\arraybackslash}p{6mm}@{\hspace{1pt}}|
}
\hline
&& \multicolumn{6}{c|}{\textbf{Threshold (nonequivalent)}}  & 
\multicolumn{6}{c|}{\textbf{Threshold (nonduplicate)}}  \\
\cline{3-14}
\textbf{Conf.}&&\tiny{d $\ge$ 0}& \tiny{d\textgreater0} &\tiny{d\textgreater0.01}& \tiny{d\textgreater0.05}& \tiny{d\textgreater0.4} & \tiny{d\textgreater0.8}
&\tiny{d $\ge$ 0}& \tiny{d\textgreater0} &\tiny{d\textgreater0.01}& \tiny{d\textgreater0.05}& \tiny{d\textgreater0.4} & \tiny{d\textgreater0.8}
\\

\hline
\multirow{2}{*}{ALL}&P   
& 0.62  & 0.90  & 1.00 & 1.00  & 1.00  & 1.00  
& 0.98  & 1.00  & 1.00 & 1.00  & 1.00  & 1.00  \\
&R   
& 1.00  & 0.71  & 0.28 & 0.10  & 0.01  & 0.01
& 1.00  & 0.62  & 0.21 & 0.13  & 0.05  & 0.03  
\\
&F   
& 0.77  & \underline{0.80}  & 0.44 & 0.19  & 0.03  & 0.01
& \underline{0.99}  & 0.77  & 0.34 & 0.23  & 0.10  & 0.05  
\\
\hline
\multirow{1}{*}{\emph{MA}}&P 
& 0.81  & 0.97  & 1.00  & 1.00 & 1.00 & N/A 
& 1.00  & 1.00 & 1.00 & 1.00  & 1.00  & 1.00
\\
\multirow{1}{*}{\emph{SS}}&R
& 1.00  & 0.89  & 0.33  & 0.11 & 0.01  & N/A
& 1.00  & 0.63 & 0.22 & 0.16  & 0.06  & 0.02
\\
&F   
& 0.90  & \underline{0.93}  & 0.50 & 0.20  & 0.02  & N/A
& \underline{1.00}  & 0.77  & 0.37 & 0.27  & 0.11  & 0.05  
\\
\hline
\end{tabular}
\end{table}

 We can observe that \textbf{$T_E$ set to zero enables the accurate detection of nonequivalent mutants}. Indeed, 
for $d>0$, we achieve the highest F-value, and the highest precision and recall, given that a value of $1.00$ cannot be achieved simultaneously for precision and recall.
These results are in line with related work~\cite{zhang2013faster} reporting that a difference in the frequency of execution of a single line of code (i.e., $d>0$) is indicative of a mutant not being equivalent to the original software.
Moreover, these results also indicate that \textbf{FSCI mutants sampling and \MPTS selection enable the accurate identification of nonequivalent mutants based on $T_E$}.

As for duplicate mutants, based on Table~\ref{table:results:precision:equivalent}, 
mutants are highly likely to be nonduplicate and thus \textbf{it is not possible to determine a threshold to identify duplicate mutants}. Indeed, among all the considered threshold values, the highest F-value is obtained when all the mutants are considered nonduplicate (i.e., $d\ge0$). These results are in line with related work~\cite{shin2017theoretical} showing that test suites are unlikely to distinguish nonredundant mutants (i.e., many nonduplicate and nonsubsumed mutants yield the same test results). 
With test suites that do not distinguish nonredundant mutants, it is very likely that nonduplicate mutants show the same coverage in addition to showing the same results. This is the reason why in Table~\ref{table:results:ratio:equivalent}, we observe a large percentage of nonduplicate mutants having the same coverage (i.e., $d=0$). For this reason, when no methods are available to automatically generate test cases that distinguish subsumed mutants (see Section~\ref{sec:background:redundant}), we suggest that all mutants should be considered as nonduplicate when computing the mutation score:

\begin{equation}
\label{equation:ms:exp}
\mathit{MS} = \frac{\mathit{KND}}{\mathit{LNE}+\mathit{KND}}
\end{equation}

where $\mathit{LNE}$ is the number of live, nonequivalent mutants.

\subsection{RQ7 - \APPR Mutation Score}

\subsubsection*{Design and measurements}

\JMR{3.16}{RQ7 investigates the extent to which  the mutation score estimated by MASS with Equation~\ref{equation:ms:exp}},
can accurately predict
the actual mutation score of the system.

To this end, we apply \APPR to the five subjects  described in Section~\ref{sec:empirical:subjects} and compute the mutation score according to equation~\ref{equation:ms:exp}.
 We compare the resulting mutation scores with those obtained with a traditional, non-optimized mutation analysis process that tests all the mutants with the entire test suite and do not discard likely equivalent mutants.
Since we have already demonstrated that FSCI, applied to a reduced and prioritized test suite, accurately estimates the mutation score (see RQ4), 
we discuss the percentage of live mutants that are discarded by means of $T_E$ and the effect it has on the mutation score.

\subsubsection*{Results}
From Table~\ref{table:results:mutationScore}, one can see that, on average, the percentage of live mutants that are discarded because considered equivalent is 42.28\%, which is in line with related work (i.e., 45\%~\cite{zhang2013faster}). Across our subjects, such percentage varies from 2.61\% (\SAIL{}$_S$) to 69.37\% (\MLFS{}{}), because of nondeterminism.
Indeed, complex embedded software, even when generating consistent functional results across multiple runs, may show nondeterminism 
in their nonfunctional properties (e.g., number of tasks started) 
when it is not possible to control the resources provisioned by the test environment.
For example, in our environment, \SAIL{}$_S$, which is a system including real-time tasks, show different code coverage for multiple executions of a same test case. The same happens for \GCSP{}, a network library, which may execute 
a different set of instructions
based on the current network usage (e.g., ports available on the host OS). 
Unsurprisingly, in our experiments, the subject having the largest number of predicted equivalent mutants removed is the mathematical library \MLFS{}{}, which should not be affected by nondeterministic behaviour due to real-time constraints or networking. 

To maximize the number of equivalent mutants detected by \APPR, it is therefore advisable to minimize the sources of nondeterminism. It can be achieved, for example, by executing test cases in a dedicated testing environment, which is standard practice for space software. However, since our analysis concerned the execution of a large, entire set of mutants, not only a sampled subset, we relied on a shared HPC environment. This may have introduced unexpected delays in the execution of the simulator and altered the number of available ports, thus exacerbating nondeterminism.

As expected, the removal of equivalent mutants results in the \APPR mutation score being higher than that computed with a traditional approach. On average, the score increased by 10.52 percentage points (i.e., from $70.62\%$ to $81.14\%$). 

To provide some additional insights about the software features that, according to \APPR mutation analysis results, 
warrant to be verified with additional test cases, we report on the characteristics of
manually inspected mutants having $d > 0$ in Table~\ref{table:results:ratio:equivalent}.
According to our analysis, live mutants concern (1) logging functions (11\%),  (2) code developed by third parties (5\%), (3) time operations (e.g., timeouts, 4\%), (4) thread synchronisations (e.g., mutex locks, 5\%), (5) memory operations (e.g., malloc and free operations, 20\%), and (6) the application logic (55\%). Most of these categories either do not need to be tested (cases 1 and 2), or concern operations that are difficult to test (cases 3, 4, and 5) and often verified by other means, e.g., test suites including hardware in the loop or through manual inspection. However, most of the live mutants concerning the application logic have enabled engineers to identify weaknesses  in test suites
(e.g., corner cases not being tested, scenarios testable with simulators but verified only by test suites with hardware in the loop), 
which further stresses the importance of mutation analysis in this context.
Furthermore, the manual inspection of these live mutants led to the identification of 
one previously undetected bug
\JMR{3.15} {since the test suite was not covering a specific combination of boolean clauses in a function, a problem that may occur even when MC/DC adequacy is achieved by test suites~\cite{Gay2016}.}

\begin{table}[htb]
\caption{RQ7. \APPR Mutation Score.}
\label{table:results:mutationScore} 
\scriptsize
\centering
\begin{tabular}{|
>{\raggedleft\arraybackslash}p{14mm}@{\hspace{1pt}}|
>{\raggedleft\arraybackslash}p{26mm}@{\hspace{1pt}}|
>{\raggedleft\arraybackslash}p{15mm}@{\hspace{1pt}}|
>{\raggedleft\arraybackslash}p{15mm}@{\hspace{1pt}}|
}
\hline
\textbf{Subject}&\textbf{Predicted Equivalent}&\multicolumn{2}{c|}{\textbf{Mutation Score (\%)}}\\
&\textbf{Mutants Removed (\%)}&\textbf{Traditional}&\textbf{\APPR}\\ 
\hline
\SAIL{}$_{S}$&2.61&65.36&65.95\\

\GCSP{}&21.67&65.64&70.92\\
\PARAM{}&63.43&69.12&85.95\\
\UTIL{}&54.34&71.20&84.41\\
\MLFS{}{}&69.37&81.80&93.49\\
\hline
$\textbf{Average}$&42.28&70.62&81.14\\
\hline
\end{tabular}

\end{table}

\subsection{Discussion}

\JMR{2.3}{Our results show that the \APPR pipeline helps to overcome mutation analysis limitations caused by common  characteristics of embedded software, present in space systems and, more generally, in similar CPS (e.g., avionics, automotive, and industry 4.0).}

\JMRCHANGE{Although the need for dedicated hardware and simulators prevent the applicability of optimizations that make use of multi-threading or other OS functions to minimize mutants compilation time~\cite{untch1993mutation}, we have shown that our selective compilation strategy (see Section~\ref{sec:appr:compile}) is sufficient to achieve an efficient mutant compilation process (see Section~\ref{experimnt:setup}).}

\JMRCHANGE{Trivial compiler optimization approaches are useful to eliminate a large proportion of mutants that are equivalent or duplicate. Based on our results, the presence of functions to deal with signals and data transformation does not limit the effectiveness of trivial compiler optimization approaches, which, across our subjects, enable the removal of 33,38\% of the mutants (62,351 out of 184,503). Our results are in line with empirical studies in the literature~\cite{papadakis2015trivial}. However, we show that for pure mathematical software (i.e., \MLFS{}{}), their effectiveness is significantly lower (i.e., 21\%, 6,717 out of 31,526).}

\JMRCHANGE{The time required to perform a traditional mutation analysis process that does not rely on mutants sampling (Step 5 of \APPR) is particularly high. Indeed it takes between 13 and 70 hours for unit and integration test suites and 11,001 hours for the system level test suite of \SAIL{}$_S$. These numbers confirm that the thorough testing required by critical CPS software, combined with long test execution times, may exacerbate the scalability problems of mutation analysis. 
\APPR applies FSCI-based mutation sampling and executes a prioritized subset of the test suite to address scalability issues. Our results show that such an optimized solution helps address scalability problems to a significant extent by reducing mutation analysis time by more than 70\% across subjects. In practice, for large software systems like \SAIL{}\emph{-CSW}, such reduction can make mutation analysis practically feasible; indeed, with 100 HPC nodes available for computation, \APPR can perform the mutation analysis of \SAIL{}\emph{-CSW} in half a day. In contrast, a traditional mutation analysis approach would take more than 100 days, thus largely delaying the development and quality assurance processes. Last, we demonstrated that FSCI-based sampling, a contribution of this paper, outperforms state-of-the-art mutants sampling strategies~\cite{zhang2013operator,gopinath2015hard} both in terms of mutation score accuracy and size of the selected mutant set.}

\JMRCHANGE{Finally, we confirm that a coverage-based approach (\APPR Step 7) enables the accurate identification of equivalent mutants, thus confirming related work's results~\cite{zhang2013faster} in our context. In addition, we demonstrate that such an approach still provides accurate results  
in the presence of optimizations (i.e., test suite reduction) that may affect the code coverage achieved by mutants. 
Coverage-based approaches, instead, are not effective in detecting likely duplicate mutants.}

\JMR{3.1}{This paper focused on investigating and identifying practical solutions to address the scalability of mutation analysis and the pertinence of mutation scores as an adequacy criterion in the context of embedded software for CPS. Important work remains concerning specific subjects in embedded software. For example, our work does not aim to assess if test suites can detect faults concerning the communication between heterogeneous components or the interoperability of different technologies and tools, two typical CPS problems. To address such issues, our future work includes the definition of mutation operators that alter the data exchanged by software components instead of their implementation.}

\subsection{Threats to validity}

In our experiments, despite their extremely large scale (test execution time over 30 thousands hours in total), the main threats to validity remains generalizability. 
To address this threat we have selected, as experimental subjects, software developed by our industry partners that is representative of different types of space software and, more generally, software in many cyber-physical systems. Our subjects include utility libraries (\UTIL{}, \GCSP{}, \PARAM{}), mathematical libraries (\MLFS{}{}), and embedded control software (\SAIL{}\emph{-CSW}) including service/protocol layer functions, satellite drivers, and application layer functions. To help generalize our results to different quality assurance practices, we considered different types of test suites: unit (\MLFS{}{} and \UTIL{}), integration (\PARAM{} and \GCSP{}), and system test suites requiring hardware and environment simulators (\SAIL{}\emph{-CSW}). \JMR{3.1}{Also, our research is monitored by ESA, who evaluates the applicability of the proposed solutions to a range of systems that go beyond the ones considered for our empirical evaluation.}

\section{Related Work}
\label{sec:related}

The mutation testing and analysis literature includes a number of relevant empirical studies involving large open-source systems.
Table~\ref{table:benchmarks} provides the list of C/C++ benchmarks considered in articles appearing in top software engineering conferences and journals, between 2013 and 2020\footnote{
We considered the following venues: IEEE International Conference on Software Testing, Validation and Verification (and its Workshops), IEEE/ACM International Conference on Software Engineering, IEEE/ACM International Conference on Automated Software Engineering, 
International Symposium on Software Testing and Analysis,
IEEE Transactions on Software Engineering,
ACM Transactions on Software Engineering and Methodology, 
IEEE Transactions on Reliability,
Elsevier Information and Software Technology Journal,
Elsevier Science Computer Programming Journal,	
Wiley Software Testing, Verification and Reliability Journal	}. Comparisons suggest that the subjects considered in our paper are among the largest considered so far, regarding the size of the 
SUT and of the test suite. Further, in most of related work, only a subset of the reported subject software had been analyzed. 
\JMR{1.3}{Finally, our benchmark is the only embedded software application among the large subjects considered in the literature.}

Based on our analysis, empirical studies on the applicability of mutation testing and analysis to safety-critical, industrial systems are very limited.
The most recent and relevant analyses on this topic are those of 
Ramler et al.~\cite{Ramler2017}, Delgado et. al~\cite{delgado2018evaluation}, and Baker et al.~\cite{Baker2013}.

Baker et al. applied mutation analysis to 22 blocks (20 LOC each) of C code and 25 blocks of Ada code belonging to control and diagnostic software in aircraft engines~\cite{Baker2013}.
Their main objectives were the study of compilation errors induced by mutation operators, the distribution of equivalent and live mutants across operators, and the identification of classes of test case deficiencies leading to live mutants. Delgado et al., instead, applied mutation analysis to 15 functions of a Commercial-Off-The-Shelf Component used in nuclear systems ~\cite{delgado2018evaluation}. According to their findings, 30\% of the live mutants are equivalent.
Both Baker et al. and Delgado et al. did not investigate mutation analysis scalability which is, in contrast, addressed by our work.

The largest cyber-physical system investigated in a mutation analysis study is the one considered by Ramler et al., who qualitatively assessed mutation analysis with 
the application components of a safety critical industrial system (around 60,000 LOC, in total)~\cite{Ramler2017}.
Their main outcomes included the identification of the
most frequent test suite weaknesses~\cite{Baker2013} identified by means of mutation analysis
(a) imprecise and over-optimistic oracles, (b) poor selection of test data conditions, and (c) child procedures not tested within the parent. 
Although Ramler et al. indicate that scalability issues prevent the practical applicability of mutation analysis, they do not address this challenge. In this paper, we complement the findings of Ramler et al. with a comprehensive solution for scaling mutation testing to industrial cyber-physical systems and a large-scale empirical evaluation.

Earlier investigations of mutation analysis for safety-critical systems focus on the representativeness of mutants, an objective that is orthogonal to ours. Daran et al. conducted a study to identify if mutations are correlated with real faults~\cite{daran1996software}; the experimentation was carried out on a critical software from the civil nuclear field. 
Andrews et al., instead, explored the relation between hand-seeded and real faults in the software \emph{Space}~\cite{andrews2005mutation}. \JMR{1.3}{Space is a space software development utility, verifying the consistency of a specification file with a reference grammar and constraints~\cite{SPACE}}, that was developed by ESA and has been used as case study subject in software engineering papers since 1998~\cite{frankl1998further}.

To summarize, our work includes the largest industrial embedded software benchmark ever considered 
in the scientific literature on mutation testing and analysis; in addition, it complements existing findings, focusing on the applicability of mutation analysis to embedded software in cyber-physical systems, with a comprehensive solution (complete mutation testing pipeline) for scalable mutation analysis and its extensive evaluation.

\begin{table}[h]
\tiny 
\caption{Subject systems in the mutation testing literature.}
\label{table:benchmarks} 
\UPDATE{
\begin{tabular}{|p{1.6cm}p{1.1cm}p{1cm}p{1cm}p{2.3cm}|}
\hline
\textbf{Case Study (SUT)}	& \textbf{SUT category}&	\textbf{Size of SUT}	&	\textbf{\# Test cases}&	\textbf{References}	\\
\hline
LLVM Framework*  			  & C & 18 kLOC & 625 & \cite{denisov2018mull}\\
OpenSSL*  		 			  & GL & 20 kLOC & 77 & \cite{denisov2018mull}\\
Codeflaws  	     			  & B & 68 LOC & 31 & \cite{papadakis2018mutant}\\
RODOS* 			 			  & OS & 3,510 & 48 & \cite{denisov2018mull}\\
Coreutils* 		 			  & CLT & 8-83 kLOC & 1,022-18,719 & \cite{hariri2019comparing,papadakis2018mutation,chekam2017empirical}\\
\textbf{\SAIL{}\emph{-CSW}}*  & FS & 74 kLOC & 384 & This paper.\\
Siemens* 					  & RPS & 1-69 kLOC & 1,531-22,138 & \cite{phan2018music,wang2017faster,papadakis2014mitigating,yao2014study,clark2013semantic}\\
Vim* 						  & SUI & 39-42 kLOC & 98 & \cite{wang2017faster,kintis2017detecting,papadakis2015trivial}\\
Make* 						  & CLT & 7-35 kLOC & 691 & \cite{papadakis2018mutation,chekam2017empirical,kintis2017detecting,papadakis2015trivial,yao2014study}\\
Memory Benchmark 			  & B & 32 kLOC & 503 & \cite{wu2017memory}\\
Findutils  					  & CLT & 18 kLOC & 4,931 &  \cite{papadakis2018mutation,chekam2017empirical}\\
KmyMoney 					  & SUI & 13 kLOC & 248 & \cite{delgado2017assessment}\\
Curl 						  & CLT & 12 kLOC & N/A & \cite{phan2018music}\\
\textbf{\UTIL{}} 			  & FS & 10 kLOC & 201& This paper.\\
\textbf{\GCSP{}} 			  & FS & 9 kLOC & 89 & This paper.\\
Grep 						  & CLT & 9 kLOC & 5,899 & \cite{papadakis2018mutation,chekam2017empirical}\\
Space* 						  & SDU & 9 kLOC & 13,585 & \cite{tokumoto2016muvm,papadakis2014mitigating,yao2014study,clark2013semantic}\\
Git* 						  & CLT & 8 kLOC & N/A & \cite{kintis2017detecting,papadakis2015trivial}\\
MSMTP* 						  & CLT & 6 kLOC & N/A & \cite{kintis2017detecting,papadakis2015trivial}\\
\textbf{\MLFS{}} 			  & FS & 5 kLOC & 4,042 & This paper.\\
Matrix TCL Pro 				  & GL &  3 kLOC & 24 & \cite{delgado2017assessment}\\
\textbf{\PARAM{}}  			  & FS & 3 kLOC & 170& This paper.\\
Dolphin 					  & SUI & 3 kLOC & 70 &  \cite{delgado2017assessment}\\
Deletion Bench.  		  & B   & 2,853 LOC & 814 & \cite{delamaro2014designing,delamaro2014experimental}\\
Gzip* 						  & CLT & 2,819 LOC & N/A & \cite{kintis2017detecting,papadakis2015trivial}\\
TinyXML2 					  & GL  & 2,620 LOC & 62 & \cite{delgado2017assessment,delgado2015class}\\
XmlRPC++ 					  & GL  & 2,194 LOC & 34 & \cite{delgado2017assessment,delgado2015class}\\
QtDom 						  & GL  & 2,117 LOC & 56 & \cite{delgado2017assessment}\\
Yao Benchmark  				  & B   & 1,208 LOC & N/A & \cite{yao2014study}\\
Flex* 						  & CLT &  10-14 kLOC & 567 & \cite{papadakis2014mitigating,yao2014study}\\
NuclearIndustrial  			  & ES & 484 LOC & 302 & \cite{delgado2018evaluation}\\
SafetyCritical				  & ES & 60 kLOC & N/A & \cite{Ramler2017}\\ 
Nequivack Bench.  		  & B & 403 LOC & N/A & \cite{holling2016nequivack}\\
MuVM Bench.  			  & B & 302 LOC & 2,256 & \cite{tokumoto2016muvm}\\
\hline                                                    
\end{tabular}
}

We use an asterisk (*) to indicate work that applied mutation analysis to a subset of the system.

\textbf{SUT category Legend:} C=Compiler, 
FS=Flight software (library or whole system running on components on-orbit), 
GL=Generic library, RPS=Research prototype software, SUI=Software with UI, CLT=Command line tool, OS=Operating system, B=Benchmark, ES=Embedded software, SDU=Space software development utility.

\end{table}

\section{Conclusion}
\label{sec:conclusion}

In this paper, we proposed and assessed a complete mutation analysis pipeline (\APPR) that enables the application of mutation analysis to space software. It is also expected to be widely applicable for many other types of embedded software, that is typically part of many cyber-physical systems, and that share similar characteristics. In particular, we address challenges related to scalability---probably the most acute problem regarding the application of mutation analysis---and the detection of equivalent and duplicate mutants. Our contributions include (1) the identification and integration in a complete tool pipeline of state-of-the-art mutation analysis optimizations tailored to the context of embedded software and cyber-physical systems, (2) a strategy for the prioritization and selection of test cases to speed up mutation analysis, (3) an approach for the sampling of mutants that provides accuracy guarantees, even when mutation analysis relies on a reduced test suite, (4) a strategy for the identification of nonequivalent and nonduplicate mutants based on code coverage information, (5) an extensive empirical evaluation 
(experiments took 1300 days of computation)
conducted on a representative space software benchmark provided by our industry partners. Our benchmark is the largest to date as far as embedded software is concerned.

Our empirical evaluation provides the following results: (1) different compiler optimization options are complementary for the detection of trivially equivalent and redundant mutants, (2) our mutants sampling strategy outperforms state-of-the-art strategies both in terms of mutation score accuracy and size of the selected mutant set, (3) the proposed test suite selection and prioritization approach enables an impressive reduction of mutation analysis time (above \JMRCHANGE{70\%}), thus helping make mutation analysis applicable to large systems, (4)
cosine distance applied to statement coverage, collected from mutated files, 
can be used to successfully identify nonequivalent mutants. 
The above results show that our mutation analysis pipeline can be integrated into the software quality assurance practices enforced by certification bodies and safety standards in space systems and many other cyber-physical systems with similar characteristics.

\appendices

\section{Distribution of Trivially Equivalent and Duplicate Mutants}
\label{appendix:equivalent}

To enable further comparison with related work~\cite{kintis2017detecting},
in Table~\ref{table:results:compilerOptimizationsProportionOperators}, we report the proportion of trivially equivalent and  duplicate mutants per mutation operator. The mutation operator causing the largest number of trivially equivalent mutants is UOI (e.g., 
it applies a post-increment operator to the last use of a variable),
followed by ROR, 
ROD, 
and ICR. 
Related studies are conducted with a smaller set of operators~\cite{kintis2017detecting}; however, the operators causing the largest numbers of trivially equivalent and duplicate mutants are the same. The main difference with related work~\cite{kintis2017detecting} concerns the ABS operator, which leads to a small set of equivalent mutants in our case, the main reason being that we rely on a definition of the ABS operator that minimizes the number of equivalent mutants by simply inverting the sign of the value instead of using the \emph{abs} function~\cite{Kintis2018}. Indeed, in functions with a positive integer domain, the replacement of a value with its absolute value trivially leads to equivalent mutants. Except for the ABS operator, \textbf{our study confirms the ranking observed in related work} despite different showing proportions. In addition, our results show that \textbf{the nature of the software affects the distribution of equivalent and duplicate mutants across operators}. Indeed, \MLFS{}{}, which focuses on mathematical functions, includes larger proportions of equivalent and duplicate mutants caused by ICR and AOR. In \PARAM{}, which does not deal with mathematical functions, the number of equivalent and duplicate mutants caused by these operators is much smaller. Finally, we notice that the \textbf{SDL and OODL operators lead to a minimal set of trivially equivalent and duplicate mutants, except for ROD}.

\begin{table*}[htb]
\caption{RQ1.  Proportion (\%) of Trivially Equivalent/Duplicate Mutants Detected per Mutation Operator.}
\label{table:results:compilerOptimizationsProportionOperators} 
\scriptsize
\centering
\begin{tabular}{|
@{\hspace{1pt}}p{11mm}|
@{\hspace{1pt}}>{\raggedleft\arraybackslash}p{5mm}@{\hspace{1pt}}|
@{\hspace{1pt}}>{\raggedleft\arraybackslash}p{6mm}@{\hspace{1pt}}|
@{\hspace{1pt}} >{\raggedleft\arraybackslash}p{6mm}@{\hspace{1pt}}|
@{\hspace{1pt}} >{\raggedleft\arraybackslash}p{5mm}@{\hspace{1pt}}|
@{\hspace{1pt}} >{\raggedleft\arraybackslash}p{6mm}@{\hspace{1pt}}|
@{\hspace{1pt}} >{\raggedleft\arraybackslash}p{5mm}@{\hspace{1pt}}|
@{\hspace{1pt}} >{\raggedleft\arraybackslash}p{6mm}@{\hspace{1pt}}|
@{\hspace{1pt}} >{\raggedleft\arraybackslash}p{6mm}@{\hspace{1pt}}|
@{\hspace{1pt}} >{\raggedleft\arraybackslash}p{5mm}@{\hspace{1pt}}|
@{\hspace{1pt}} >{\raggedleft\arraybackslash}p{6mm}@{\hspace{1pt}}|
@{\hspace{1pt}} >{\raggedleft\arraybackslash}p{6mm}@{\hspace{1pt}}|
@{\hspace{1pt}} >{\raggedleft\arraybackslash}p{6mm}@{\hspace{1pt}}|
@{\hspace{1pt}} >{\raggedleft\arraybackslash}p{5mm}@{\hspace{1pt}}|
 >{\raggedleft\arraybackslash}p{5mm}@{\hspace{1pt}}|
@{\hspace{1pt}}>{\raggedleft\arraybackslash}p{6mm}@{\hspace{1pt}}|
@{\hspace{1pt}} >{\raggedleft\arraybackslash}p{6mm}@{\hspace{1pt}}|
@{\hspace{1pt}} >{\raggedleft\arraybackslash}p{5mm}@{\hspace{1pt}}|
@{\hspace{1pt}} >{\raggedleft\arraybackslash}p{6mm}@{\hspace{1pt}}|
@{\hspace{1pt}} >{\raggedleft\arraybackslash}p{5mm}@{\hspace{1pt}}|
@{\hspace{1pt}} >{\raggedleft\arraybackslash}p{6mm}@{\hspace{1pt}}|
@{\hspace{1pt}} >{\raggedleft\arraybackslash}p{6mm}@{\hspace{1pt}}|
@{\hspace{1pt}} >{\raggedleft\arraybackslash}p{5mm}@{\hspace{1pt}}|
@{\hspace{1pt}} >{\raggedleft\arraybackslash}p{6mm}@{\hspace{1pt}}|
@{\hspace{1pt}} >{\raggedleft\arraybackslash}p{6mm}@{\hspace{1pt}}|
@{\hspace{1pt}} >{\raggedleft\arraybackslash}p{6mm}@{\hspace{1pt}}|
@{\hspace{1pt}} >{\raggedleft\arraybackslash}p{5mm}@{\hspace{1pt}}|
}

\hline
\textbf{}& \multicolumn{13}{c|}{\textbf{Trivially Equivalent}} &\multicolumn{13}{c|}{\textbf{Trivially Duplicate}}\\
\textbf{Subject}&
\textbf{ABS}&	\textbf{AOR}&	\textbf{ICR}&	\textbf{LCR}&	\textbf{ROR}&	\textbf{SDL}&	\textbf{UOI}&	\textbf{AOD}&	\textbf{LOD}&	\textbf{ROD}&	\textbf{BOD}&	\textbf{SOD}&	\textbf{LVR}&
\textbf{ABS}&	\textbf{AOR}&	\textbf{ICR}&	\textbf{LCR}&	\textbf{ROR}&	\textbf{SDL}&	\textbf{UOI}&	\textbf{AOD}&	\textbf{LOD}&	\textbf{ROD}&	\textbf{BOD}&	\textbf{SOD}&	\textbf{LVR}\\
\hline
\mbox{\SAIL{}\emph{-CSW}}
&3.65&1.85&6.74&2.09&18.00&1.60&52.34&1.53&0.05&7.64&3.42&0.51&0.59&1.52&6.20&18.79&1.05&23.89&8.04&20.23&4.83&1.06&8.86&2.20&1.96&1.39\\
\GCSP{}&8.84&0.00&2.28&1.14&17.40&3.14&56.35&0.00&0.14&9.84&0.86&0.00&0.00&2.11&2.37&10.77&1.24&31.45&9.04&23.92&1.69&1.36&13.48&1.58&0.79&0.19\\
\PARAM{}&8.89&0.00&0.00&0.22&25.56&6.22&55.33&0.00&0.00&3.56&0.22&0.00&0.00&0.96&3.47&7.95&1.01&36.71&13.58&12.52&1.49&2.46&18.45&0.92&0.00&0.48\\
\UTIL{}&6.66&0.07&3.95&1.24&15.96&7.91&58.86&0.00&0.37&3.81&0.95&0.00&0.22&0.34&7.38&15.64&0.87&30.35&12.61&13.52&1.66&2.66&12.77&0.89&0.52&0.77\\
\MLFS{}{}&6.37&0.55&11.63&0.28&27.98&2.77&39.61&0.00&0.00&9.42&1.39&0.00&0.00&5.03&10.12&25.36&0.87&23.77&4.14&6.73&8.51&0.88&9.44&2.33&1.90&0.91\\
\hline
\textbf{Total}&4.59&1.42&6.04&1.81&18.32&2.64&\textbf{53.06}&1.16&0.09&7.22&2.79&0.38&0.47&1.87&6.48&18.47&1.02&25.36&8.23&\textbf{17.83}&4.72&1.25&9.91&2.02&1.68&1.17\\
\hline
\end{tabular}

\end{table*}

\begin{figure}[tb]
\begin{center}
\includegraphics[width=9cm]{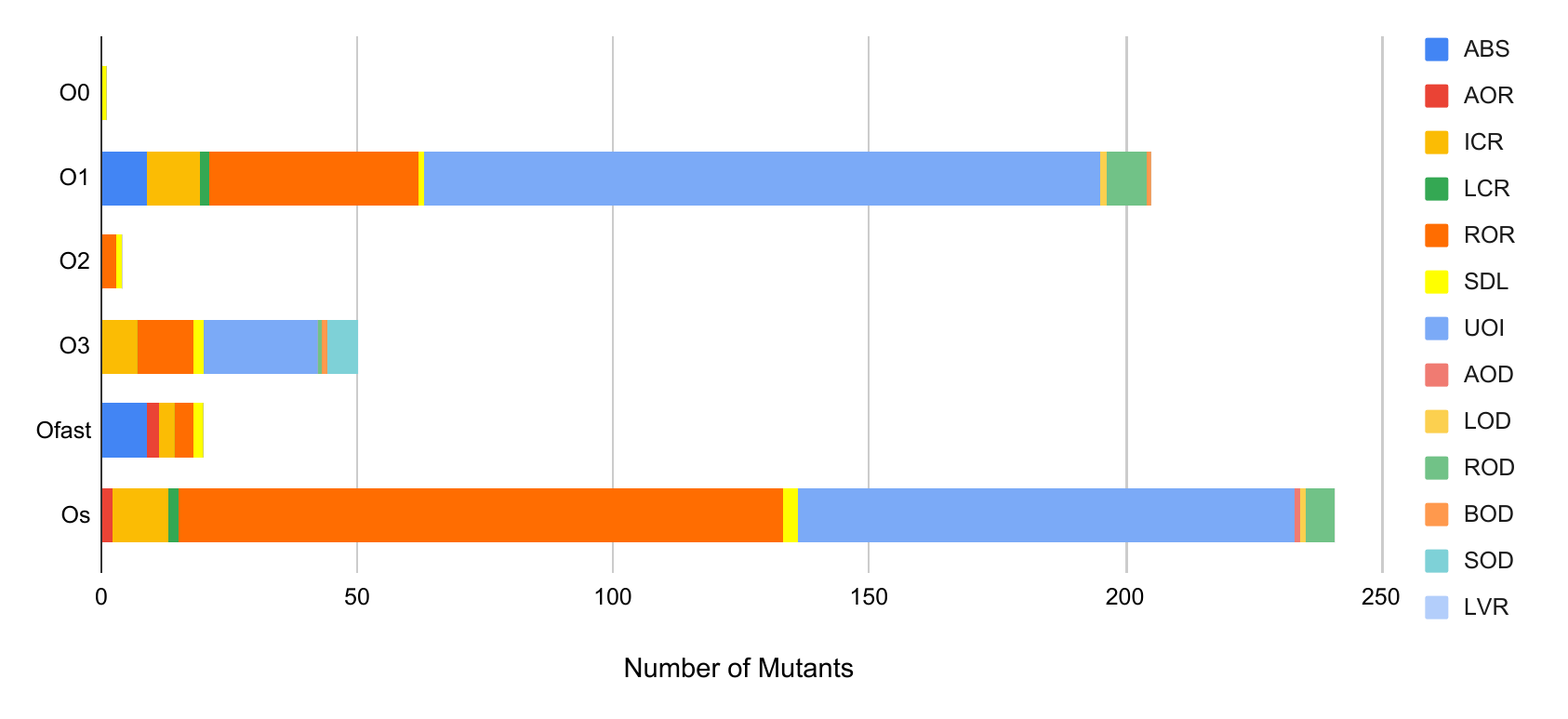}
\caption{Univocal, Trivially Equivalent Mutants detected by Compiler Optimizations.}
\label{fig:results:univeq}
\end{center}
\end{figure}

\begin{figure}[tb]
\begin{center}
\includegraphics[width=9cm]{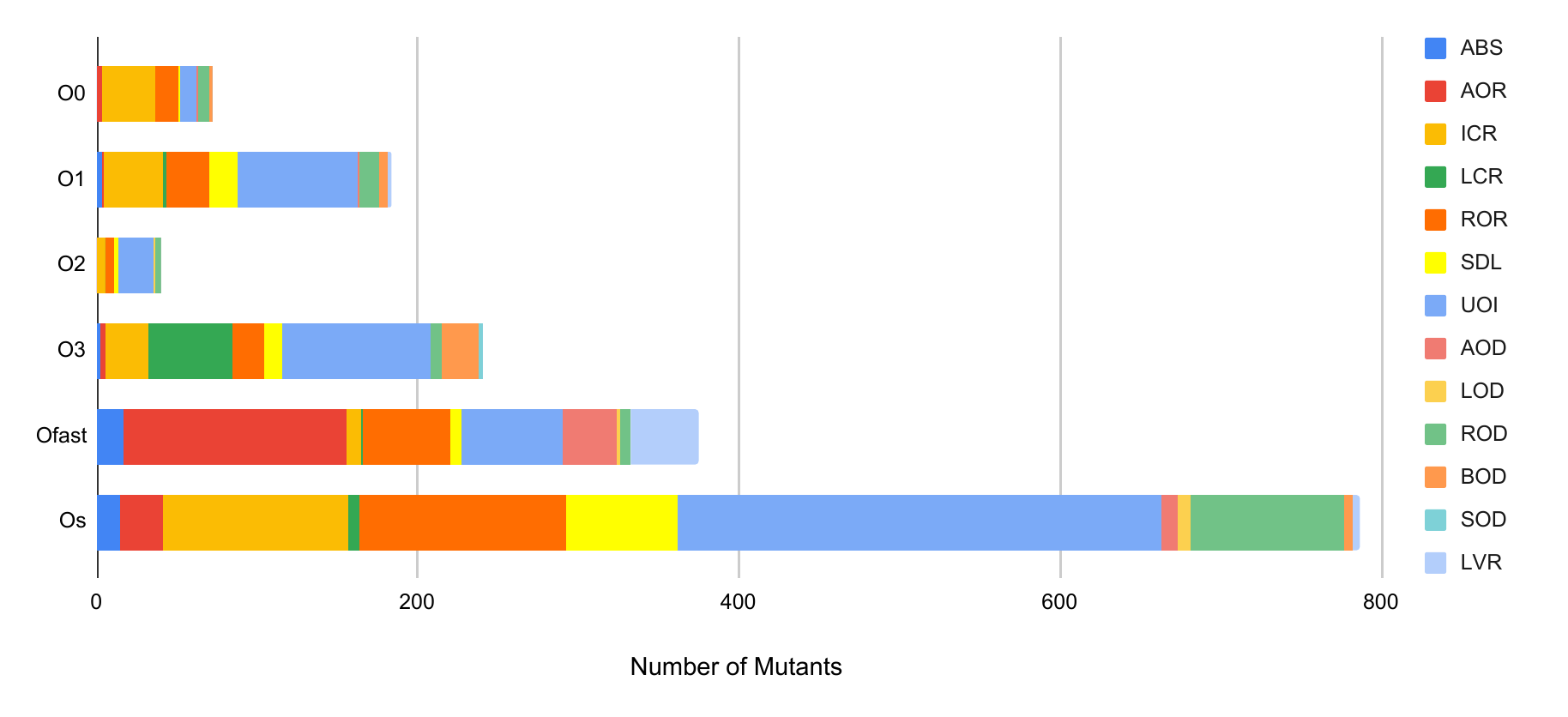}
\caption{Univocal, Trivially Duplicate Mutants detected by Compiler Optimizations.}
\label{fig:results:univred}
\end{center}
\end{figure}

Finally, to further characterize  the differences across different compiler optimization levels, we provide in Figures~\ref{fig:results:univeq} and~\ref{fig:results:univred}, for each compiler optimization level, the distribution of univocal, trivially equivalent and duplicate mutants across mutation operators. The optimization level \emph{-Os} is more effective in detecting trivially equivalent mutants caused by the ROR operator (a larger number of ROR mutants is associated to \emph{-Os} as captured by the length of the orange bar), while the option \emph{-O1} is more effective in detecting trivially equivalent mutants caused by the UOI operator. Concerning the detection of trivially duplicate mutants (Figure~\ref{fig:results:univred}), \emph{-Os} performs better in detecting the duplicate mutants caused by almost all the operators, except for the ones caused by operators affecting math expressions (i.e., AOR, AOD, and LVR), which are better detected by optimization level \emph{-Ofast}, probably because it includes additional math optimization options.

\section{Assessment of assumptions on limited correlation between mutants}
\label{appendix:correlation}
We aim to assess the correctness of the two assumptions underlying the adoption of the binomial distribution to estimate the mutation score:

\begin{itemize}
\item[A1] \emph{The correlation between mutants is limited}. We aim to measure and discuss the degree of association between the trials (i.e., the outcome of a sampled mutant) in our experiments. The degree of association should be low.

\item[A2] \emph{The binomial distribution enables the  estimation of the mutation score at the same level of accuracy as the correlated binomial distribution}. 
We compare the distribution of the mutation score observed in our experiments with the distribution of the mutation score  estimated by the probability mass function (PMF) for the binomial distribution (i.e., $\mathit{PMF}_B)$ and the PMF for the correlated binomial distribution (i.e., $\mathit{PMF}_C)$.
We aim to demonstrate that, in practice, in our context, there is no difference between the accuracy (i.e., the difference between the estimated mutation score and the mutation score of the sampled mutants) of $\mathit{PMF}_B$ and $\mathit{PMF}_C$.
\end{itemize}

\subsection{Assumption 1 - The correlation between mutants is limited.}
\subsubsection{Measurements}
The association between two trials (i.e., the $i^{th}$ and the $j^{th}$ mutant in our case) can be measured by computing a coefficient  of association from the contingency tables derived for every pair of trials~\cite{VanDerGeest:2005}. More precisely, we generate a $2X2$ contingency table for every pair of mutants $m_i$ and $m_j$. Each cell of the contingency table measures the frequency of occurrence, over $R$ mutation analysis runs ($R=100$ in our study), for the following situations: (a) mutants are both killed, (b) $m_i$ killed with $m_j$ live, (c) $m_i$ live with $m_j$ killed, (d) mutants are both live. %
Based on the contingency table, we derive two coefficients of association, Yule's Q~\cite{YuleQ} and the odds ratio.  
Yule's $Q_{ij}$ measures the association between the $i^{th}$ and the $j^{th}$ mutant:

\footnotesize
\begin{equation*} 
Q_{ij} = \frac{ad-bc}{ad+bc}
\end{equation*} 
\normalsize

$Q_{ij}$ ranges between -1 and +1, and for independent mutants it should be equal to zero.

The odds ratio is the ratio of (1) the odds of $m_i$ being killed when $m_j$ is killed and (2) the odds of $m_i$ being killed when $m_j$ is not killed:

\footnotesize
\begin{equation*} 
OR_{ij} = \frac{ad}{bc}
\end{equation*} 
\normalsize

Two mutants are independent when $OR_{ij}$ is equal to one.

We compute $Q_{ij}$ and $\mathit{OR}_{ij}$ for every pair of mutants considering the mutation analysis results collected when sampling 300 (i.e., less than the mutants sampled by FSCI), 400 (i.e., the upper bound observed for FSCI), and 1000 mutants (i.e., the number of mutants suggested by Gopinath et al.~\cite{gopinath2015hard}). To compute $Q_{ij}$ and $\mathit{OR}_{ij}$, we considered the data collected in the 100 experiments carried out for RQ2 (see Section 4.4 of the main manuscript). 

\subsubsection{Results}
\begin{figure*}
\begin{subfigure}{.3\textwidth}
  \includegraphics[width=6cm]{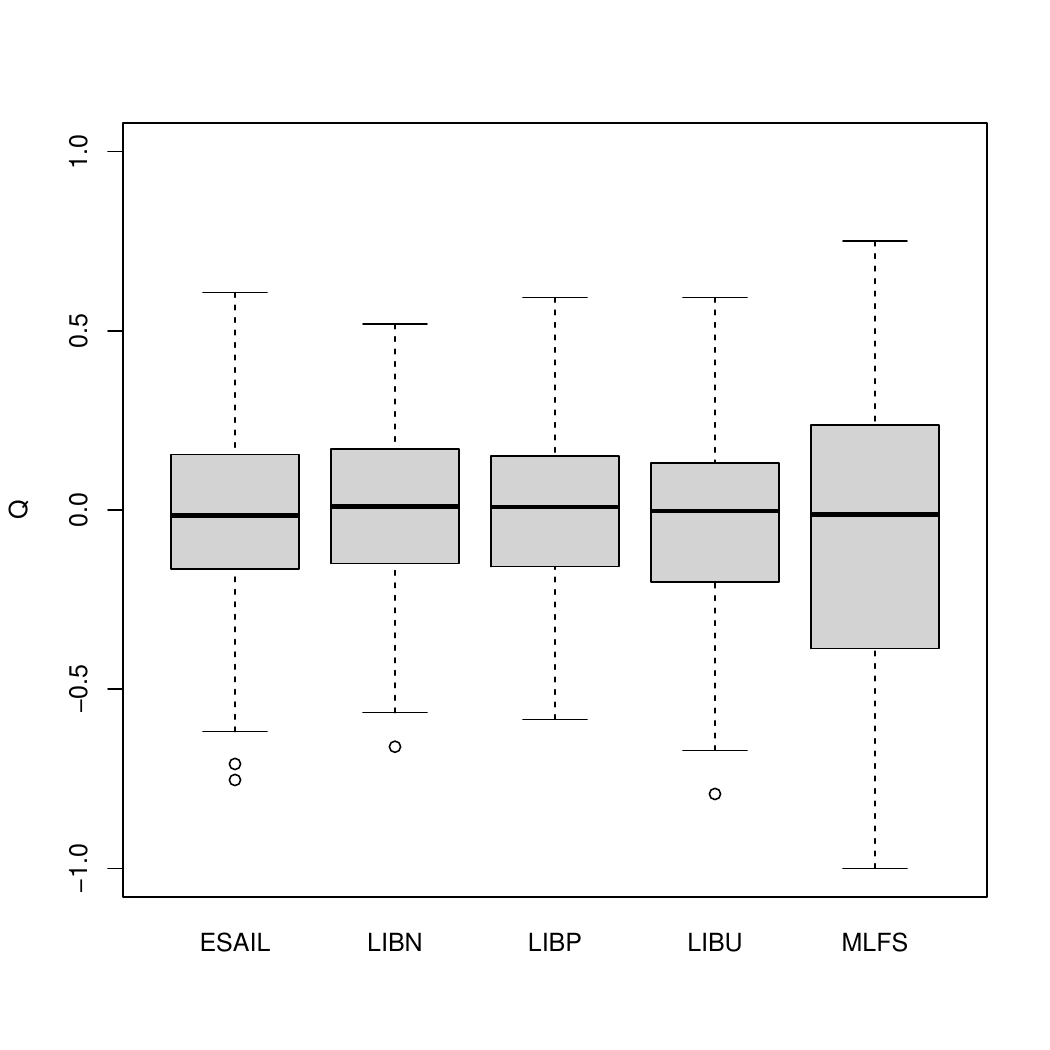}
  \caption{300 mutants}
  \label{fig:exp:q300}
\end{subfigure}%
\begin{subfigure}{.3\textwidth}
  \includegraphics[width=6cm]{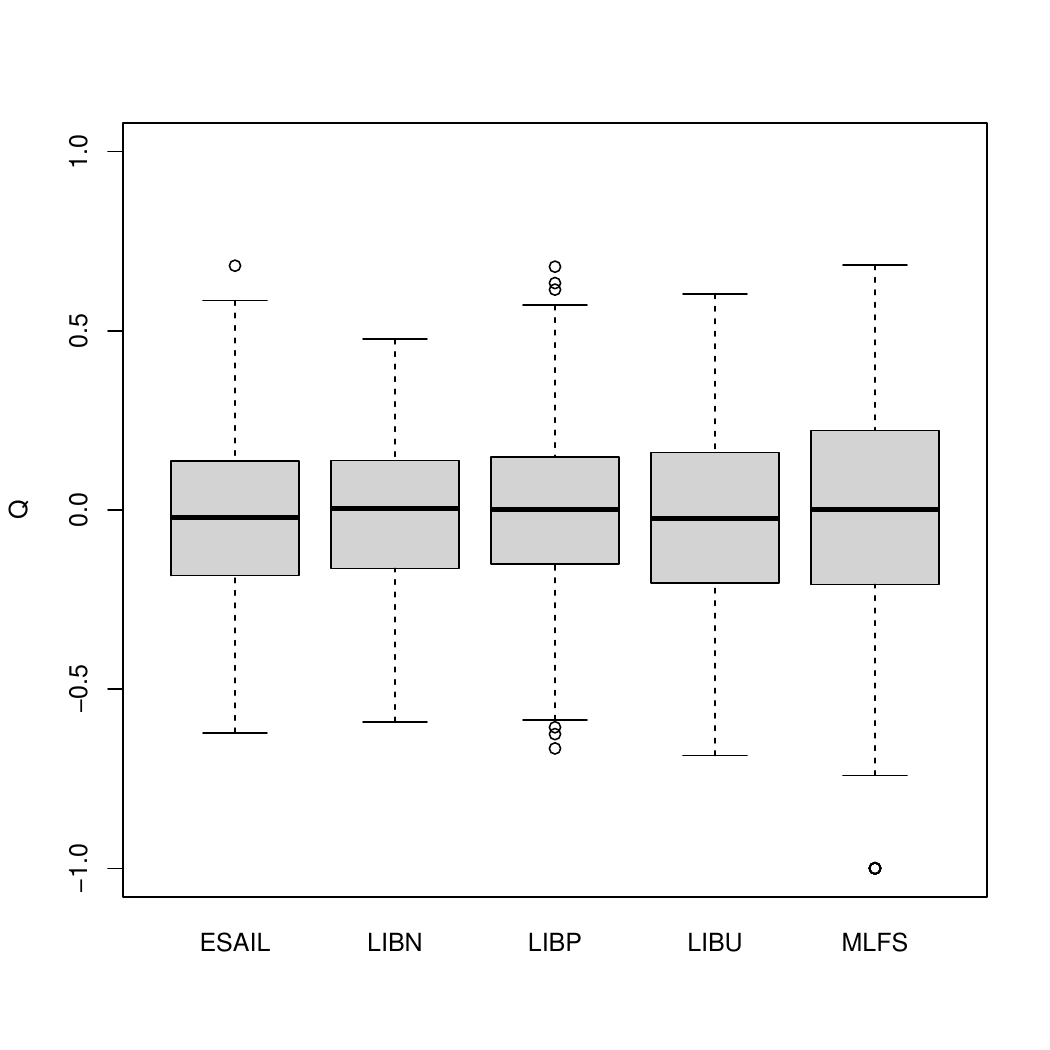}
  \caption{400 mutants}
  \label{fig:exp:q400}
\end{subfigure}
\begin{subfigure}{.3\textwidth}
  \includegraphics[width=6cm]{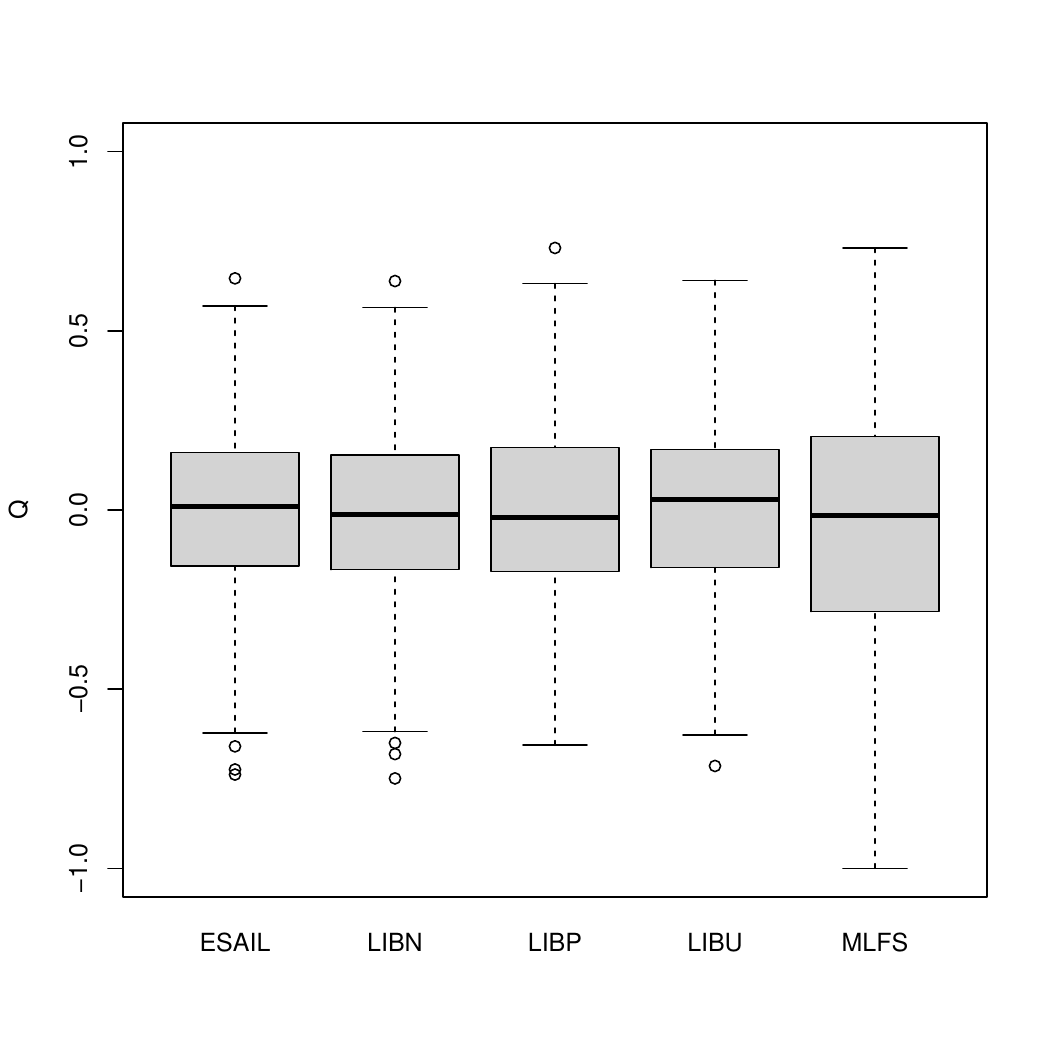}
  \caption{1000 mutants}
  \label{fig:exp:q1000}
\end{subfigure}
\caption{Yule's Q computed for experiments involving  different numbers of mutants.}
\label{fig:exp:Q}
\end{figure*}

Figures~\ref{fig:exp:q300} to~\ref{fig:exp:q1000} show boxplots of the distribution of $Q$ for every pair of mutants (i.e., every data point reports the value of $Q_{ij}$ computed for a specific pair of mutants). Across the three boxplots, the median is between 0.001 and 0.013, that is practically zero, thus showing that, overall, the mutants tend to be independent. 
Unsurprisingly, mutants are more correlated for the \MLFS{}{}, where the first and third quartiles
 (i.e., half of the pairs of mutants), 
are equal to $-0.386$ and $0.237$, respectively. This is due to \MLFS{}{} having the most exhaustive test suite (it achieves MC/DC coverage); indeed, in such a situation, it is very likely that pairs of mutants fail together because the test suite will likely kill them both. For the subjects, the first and third quartiles lie between $-0.203$ and $0.174$, thus showing limited association levels.

\begin{figure*}
\begin{subfigure}{.3\textwidth}
  \includegraphics[width=6cm]{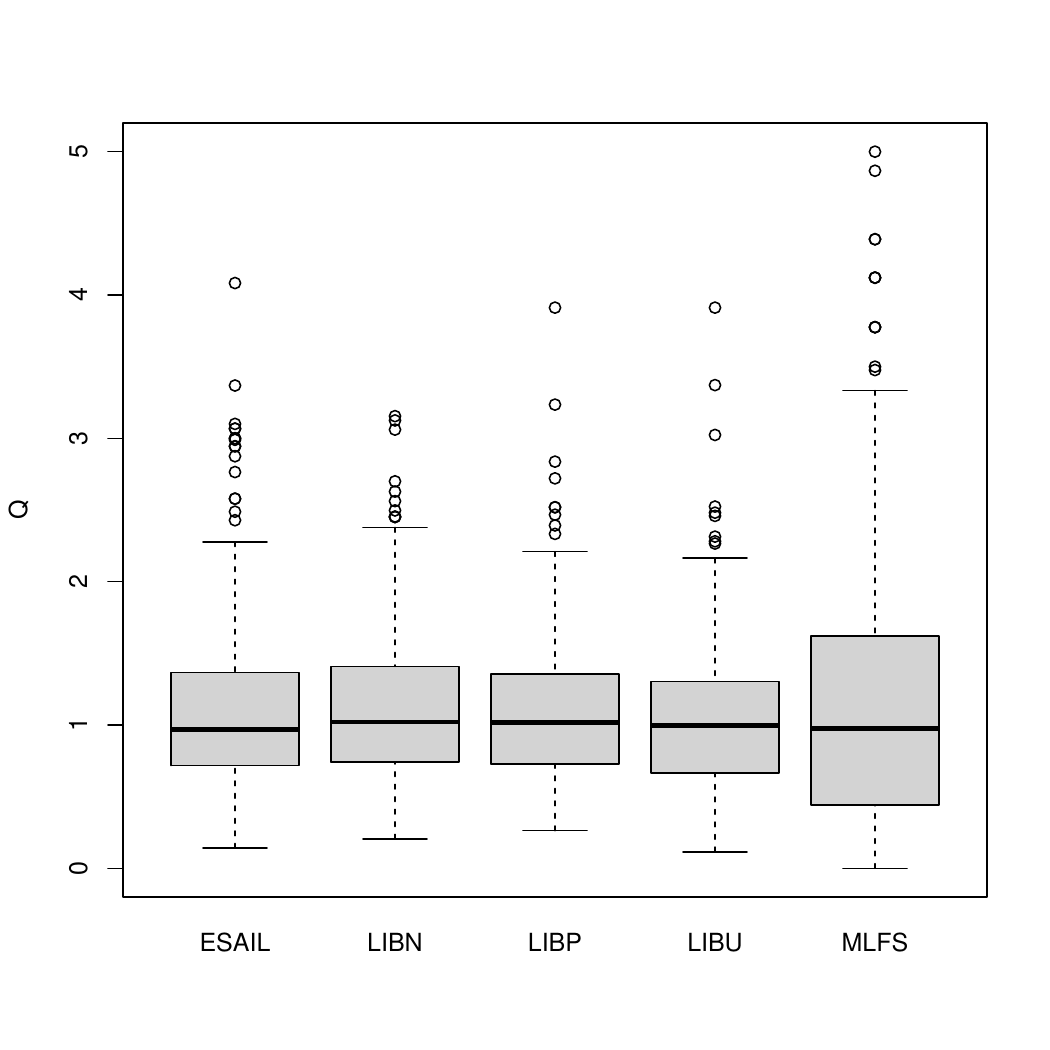}
  \caption{300 mutants}
  \label{fig:exp:O300}
\end{subfigure}%
\begin{subfigure}{.3\textwidth}
  \includegraphics[width=6cm]{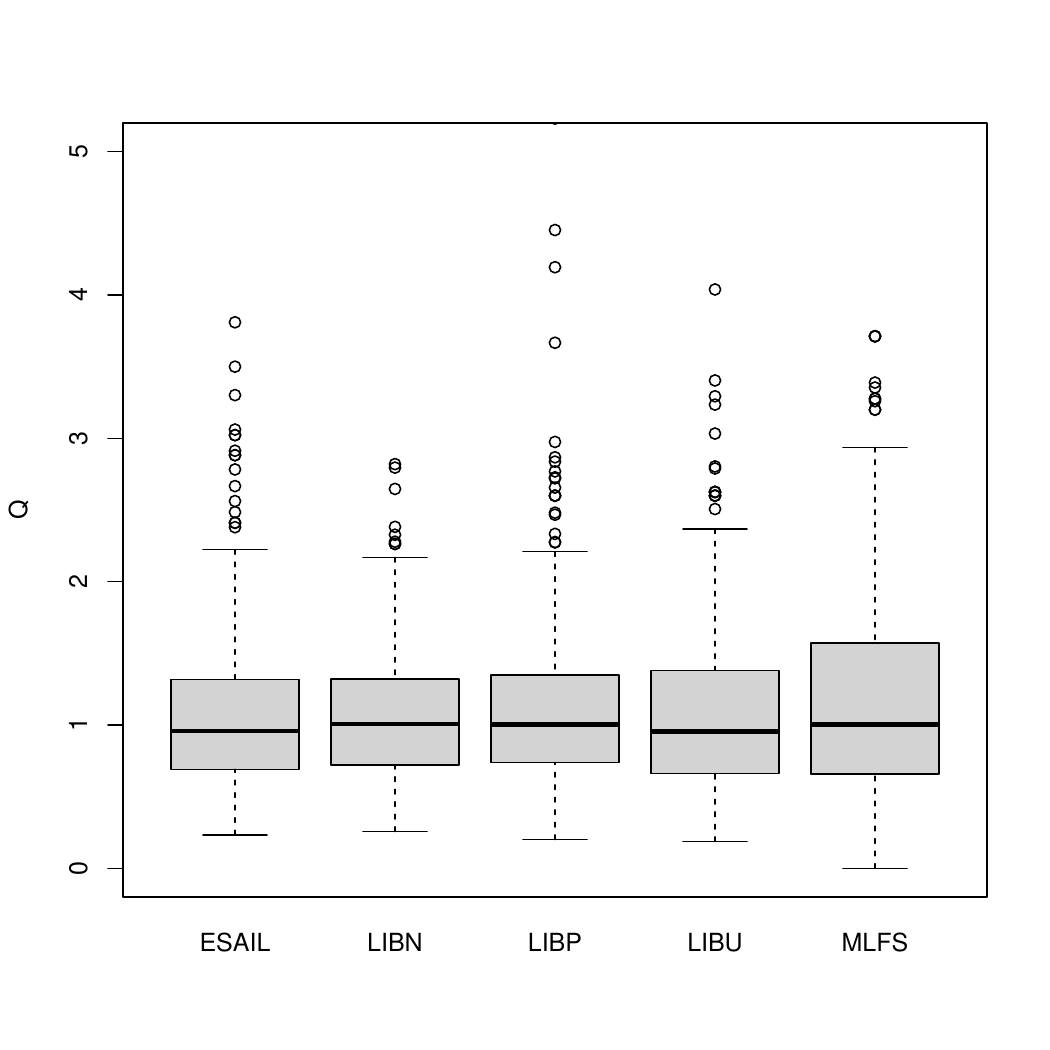}
  \caption{400 mutants}
  \label{fig:exp:O400}
\end{subfigure}
\begin{subfigure}{.3\textwidth}
  \includegraphics[width=6cm]{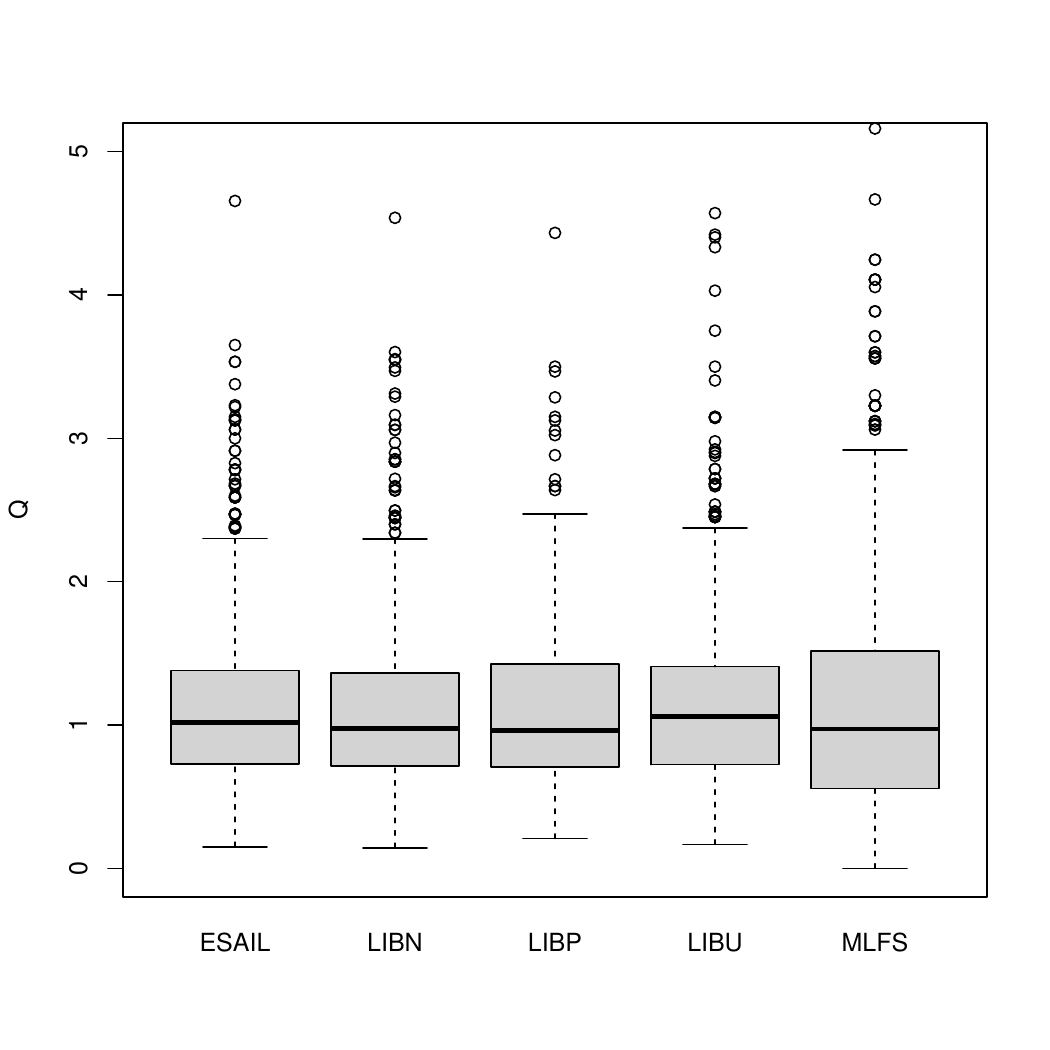}
  \caption{1000 mutants}
  \label{fig:exp:O1000}
\end{subfigure}
\caption{Odds ratio computed for experiments involving  different numbers of mutants.}
\label{fig:exp:O}
\end{figure*}

Figures~\ref{fig:exp:O300} to~\ref{fig:exp:O1000} show boxplots with the distribution of $OR$ for every pair of mutants. It once again shows that mutants tend to be independent; indeed, across the three boxplots, the median lies between 0.95 and 1.06, while the first and third quartiles lie between 0.44 and 1.62, respectively.

\subsection{Assumption 2 - The binomial distribution accurately estimates the mutation score.}

\subsubsection{Measurements}
 
 As mentioned in Section~2.4,
 when trials are not independent, the correlated binomial distribution
  (also known as Bahadur-Lazarsfeld distribution) can be used to
 estimate the outcome of a binomial experiment (i.e., the computation of the mutation score in our context).

In the presence of second order interactions\footnote{To simplify our discussion we ignore interactions above the second order, which is usual practice~\cite{Zhang:CrrelatedFirearm:NIST:2019}; the interested reader is referred to related work~\cite{VanDerGeest:2005}.}, the probability mass function 
of the correlated binomial distribution can be computed as

\begin{equation*} 
\footnotesize
P(Y,N)=P'(Y,N,p) * \Big{(}1+r_2*g_2(Y,N,p)\Big{)}
\end{equation*}

with $Y$ indicating the number of successful trials (i.e., the number of mutants killed, in our context) out of N trials, $P'(Y,N,p)$ being the probability mass function of the binomial distribution with probability of success $p$, $r_2$ being the correlation coefficient, and $g_2(Y,N,p)$ being a 
defined as

\footnotesize
\begin{equation*}
g_2(Y,N,p)=\frac{(Y-N*p)^2-(1-2*p)(Y-N*p)-N*p(1-p)}{2*p(1-p)}
\end{equation*}
\normalsize

The correlated binomial distribution reduces to the binomial distribution when $r_2$ is zero.

We aim to compare the distribution of the mutation score observed in our experiments (hereafter, $\mathit{PMF}_S$) with the distribution of the mutation score estimated by both the PMF of the binomial (hereafter, $\mathit{PMF}_B$) and the PMF of the correlated binomial distributions (hereafter, $\mathit{PMF}_C$).

To compute $\mathit{PMF}_C$, we should identify proper values for the parameters $p$ and $r_2$.
To this end, we follow the approach proposed by Zhang et al.~\cite{Zhang:CrrelatedFirearm:NIST:2019}, which consists in relying on a non-linear least squares algorithm\footnote{We rely on the Gauss-Newton algorithm provided by R~\cite{R:NLS}.} to identify the values that minimize the error sum of squares (ESS) computed as

\footnotesize
\begin{equation*}
\mathit{ESS}_C=\sum_{y=0}^{N} \bigg{(} h(y) - P'(y,N,\tilde{p}) * \Big{(}1 + \tilde{r_2}*g_2(y,N,\tilde{p})\Big{)} \bigg{)^2}
\end{equation*}
\normalsize

with $\tilde{p}$ and $\tilde{r_2}$ being the parameters estimated for $p$ and $r_2$. $h(y)$ is the proportion of runs in which, for our experiments, we observed $Y$ mutants being killed.

To determine if the correlated binomial distribution is more accurate than the binomial distribution, we can compare $\mathit{ESS}_C$ with the $\mathit{ESS}$ computed for $\mathit{PMF}_B$ 

\footnotesize
\begin{equation*}
ESS_B=\sum_{y=0}^{N} \Big{(} h(y) - P'(y,N,p) \Big{)^2}
\end{equation*}
\normalsize

The parameter $p$ is the actual mutation score.
If $ESS_B$ is close to $ESS_C$, the two PMF perform similarly in our context (i.e., the effect of covariance is negligible).
Also, the effect of covariance can be considered negligible if the estimated value for $r_2$ is close to zero. 

To further facilitate the understanding of the similarity between $\mathit{PMF}_B$ and $\mathit{PMF}_C$, we also discuss the shape of the curve of $\mathit{PMF}_B$ and $\mathit{PMF}_C$ with that of the curve that fits the distribution of the mutation score observed in our experiments (hereafter, $\mathit{PMF}_S$).
To derive $\mathit{PMF}_S$, we rely on a density estimation function based on a gaussian kernel~\cite{Venables:2010}.

Between $\mathit{PMF}_B$ and $\mathit{PMF}_C$, the curve that approximates better $\mathit{PMF}_S$ is the one with the largest intersection of their Area Under Curve (AUC). We thus compute, first, $\mathit{PAUC}_B$ as the proportion of the area under $\mathit{PMF}_S$ and $\mathit{PMF}_B$ that is under both curves (i.e.,  intersection over union). 
Then we do the same for $\mathit{PMF}_C$ by computing $\mathit{PAUC}_C$ as
the proportion of the area under $\mathit{PMF}_S$ and $\mathit{PMF}_C$ that is under both curves.
The curve that better approximates  $\mathit{PMF}_S$ is the one leading to the highest value for $\mathit{PAUC}$.
The binomial distribution is an appropriate approximation of the mutation score if $\mathit{PAUC}_B$ is close to $\mathit{PAUC}_C$.

\subsubsection{Results}

\begin{table}[tb]
\caption{Parameters estimated for $PMF_C$ considering different numbers (N) of sampled mutants.}
\label{table:results:PMFC} 
\scriptsize
\centering
\begin{tabular}{|
@{\hspace{1pt}}p{3mm}|
@{\hspace{1pt}}>{\raggedleft\arraybackslash}p{6mm}@{\hspace{1pt}}|
>{\raggedleft\arraybackslash}p{9mm}@{\hspace{1pt}}|
>{\raggedleft\arraybackslash}p{12mm}@{\hspace{1pt}}|
 >{\raggedleft\arraybackslash}p{12mm}@{\hspace{1pt}}|
  >{\raggedleft\arraybackslash}p{11mm}@{\hspace{1pt}}|
>{\raggedleft\arraybackslash}p{10mm}@{\hspace{1pt}}|
}
\hline
\textbf{N}   &  & \multicolumn{1}{c|}{\textbf{\SAIL{}}}  & \multicolumn{1}{c|}{\textbf{\GCSP{}}} & \multicolumn{1}{c|}{\textbf{\PARAM{}}} & \multicolumn{1}{c|}{\textbf{\UTIL{}}} & \multicolumn{1}{c|}{\textbf{\MLFS{}}} \\
\hline

\multirow{2}{*}{300}&
$\tilde{p}$&
65.26& 65.54& 68.92& 70.8  &81.80\\
&$\tilde{r_2}$&
-0.00029& 0.00000& -0.00013& -0.00036& 0.00005\\
&$ESS$&
0.01037& 0.01888& 0.00776& 0.00839& 0.00884\\
\hline

\multirow{2}{*}{400}&
$\tilde{p}$&
65.06& 65.73& 68.97& 71.02  &81.90\\
&$\tilde{r_2}$&
-0.00067& -0.00047& -0.00032& -0.00039& 0.00008\\
&$ESS$&
0.01002& 0.00955& 0.01194& 0.00905& 0.01017\\
\hline

\multirow{2}{*}{1000}&
$\tilde{p}$&
65.29& 65.43& 68.91& 71.01  &81.66\\
&$\tilde{r_2}$&
-0.00080& -0.00043& -0.00045& 0.00000& 0.00010\\
&$ESS$&
0.01337& 0.00548& 0.00739& 0.00754& 0.01088\\
\hline
\end{tabular}
\end{table}

\begin{table}[tb]
\caption{ESS obtained with $\mathit{PMF}_B$ and delta with $\mathit{PMF}_C$ (i.e., $\Delta=\mathit{PMF}_B-\mathit{PMF}_C$).}
\label{table:results:ESS:bin} 
\scriptsize
\begin{tabular}{|
@{\hspace{1pt}}p{4mm}|
@{\hspace{1pt}}>{\raggedleft\arraybackslash}p{6mm}@{\hspace{1pt}}|
>{\raggedleft\arraybackslash}p{8mm}@{\hspace{1pt}}|
>{\raggedleft\arraybackslash}p{12mm}@{\hspace{1pt}}|
 >{\raggedleft\arraybackslash}p{12mm}@{\hspace{1pt}}|
  >{\raggedleft\arraybackslash}p{11mm}@{\hspace{1pt}}|
>{\raggedleft\arraybackslash}p{10mm}@{\hspace{1pt}}|
}
\hline
\textbf{N}   &  & \multicolumn{1}{c|}{\textbf{\SAIL{}}}  & \multicolumn{1}{c|}{\textbf{\GCSP{}}} & \multicolumn{1}{c|}{\textbf{\PARAM{}}} & \multicolumn{1}{c|}{\textbf{\UTIL{}}} & \multicolumn{1}{c|}{\textbf{\MLFS{}}} \\
\hline
All&$p$&
65.36& 65.64& 69.12& 71.20& 81.80\\
\hline
300
&$ESS$&
0.01044& 0.01892& 0.00784& 0.008859& 0.008849\\
&$\Delta$&
0.00007& 0.00007& 0.00008& 0.00046& 0.0\\
\hline
400
&$ESS$&
0.01065& 0.00975& 0.01211& 0.00929& 0.0102\\
&$\Delta$&
0.00063& 0.00020& 0.00020& 0.00024& 0.00003\\
\hline
1000
&$ESS$&
0.01565& 0.00633& 0.00841& 0.00771& 0.01103\\
&$\Delta$&
0.00228& 0.00085& 0.00102& 0.00017& 0.00015\\
\hline
\end{tabular}
\end{table}

\begin{table}[tb]
\caption{Proportion (\%) of AUC for $PMF_S$ shared with $PMF_B$ and $PMF_C$. Higher values are highlighted.}
\label{table:results:AUC} 
\scriptsize
\centering
\begin{tabular}{|
@{\hspace{1pt}}p{4mm}|
@{\hspace{1pt}}>{\raggedleft\arraybackslash}p{10mm}@{\hspace{1pt}}|
>{\raggedleft\arraybackslash}p{3mm}@{\hspace{1pt}}|
>{\raggedleft\arraybackslash}p{8mm}@{\hspace{1pt}}|
 >{\raggedleft\arraybackslash}p{8mm}@{\hspace{1pt}}|
  >{\raggedleft\arraybackslash}p{8mm}@{\hspace{1pt}}|
>{\raggedleft\arraybackslash}p{7mm}@{\hspace{1pt}}|
}
\hline
\textbf{N}   &  \multicolumn{1}{c|}{\textbf{PAUC}}& \multicolumn{1}{c|}{\textbf{\SAIL{}}}  & \multicolumn{1}{c|}{\textbf{\GCSP{}}} & \multicolumn{1}{c|}{\textbf{\PARAM{}}} & \multicolumn{1}{c|}{\textbf{\UTIL{}}} & \multicolumn{1}{c|}{\textbf{\MLFS{}}} \\
\hline

300&
$\mathit{PAUC}_B$
&\textbf{92.49}
&94.60
&\textbf{93.49}
&\textbf{89.52}
&\textbf{92.89}\\

&
$\mathit{PAUC}_C$
&92.09
&\textbf{94.85}
&93.01
&87.44
&92.79\\
\hline

400&
$\mathit{PAUC}_B$
&92.88
&\textbf{95.14}
&94.22
&\textbf{94.82}
&\textbf{95.19}\\

&
$\mathit{PAUC}_C$
&\textbf{93.73}
&92.44
&\textbf{94.92}
&91.23
&94.19\\
\hline

1000&
$\mathit{PAUC}_B$
&\textbf{93.84}
&\textbf{89.79}
&\textbf{91.83}
&93.89
&\textbf{92.98}\\

&
$\mathit{PAUC}_C$
&84.13
&85.59
&85.73
&\textbf{94.58}
&92.08\\

\hline
\end{tabular}
\end{table}

\begin{figure}
  \includegraphics[width=8cm]{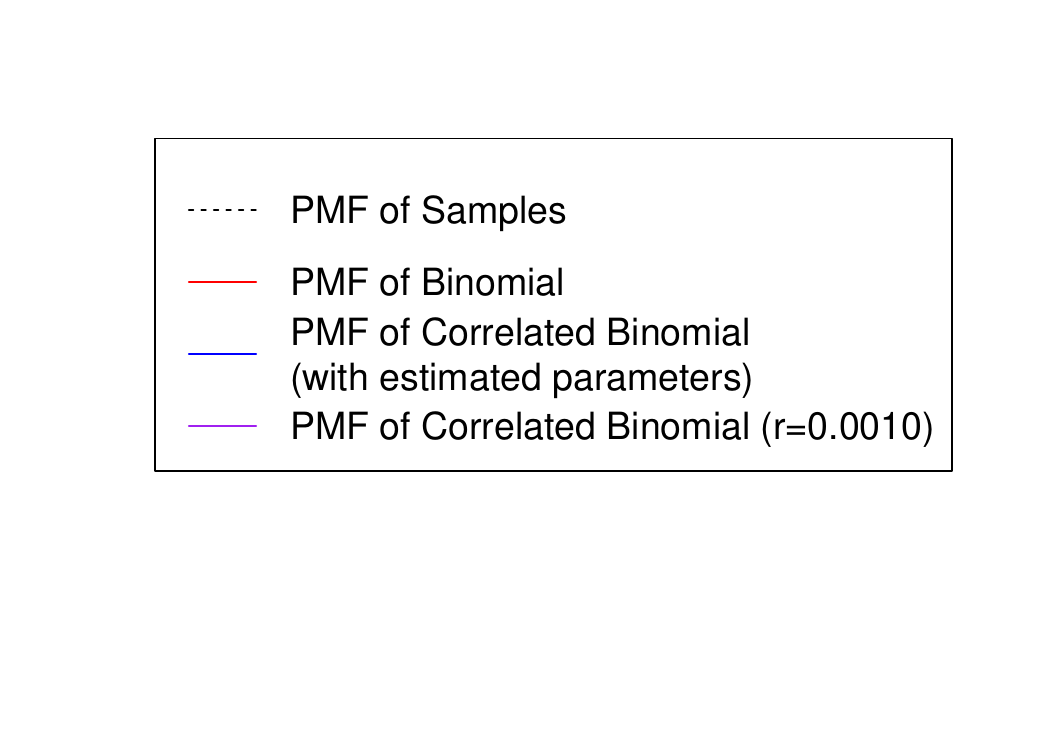}
 \caption{Legend for Figure~\ref{fig:exp:PMF}}
\label{fig:exp:pmf:legend}
\end{figure}

\newcommand{\PMFW}{5.5cm}
\begin{figure*}
\begin{subfigure}{.3\textwidth}
  \includegraphics[width=\PMFW]{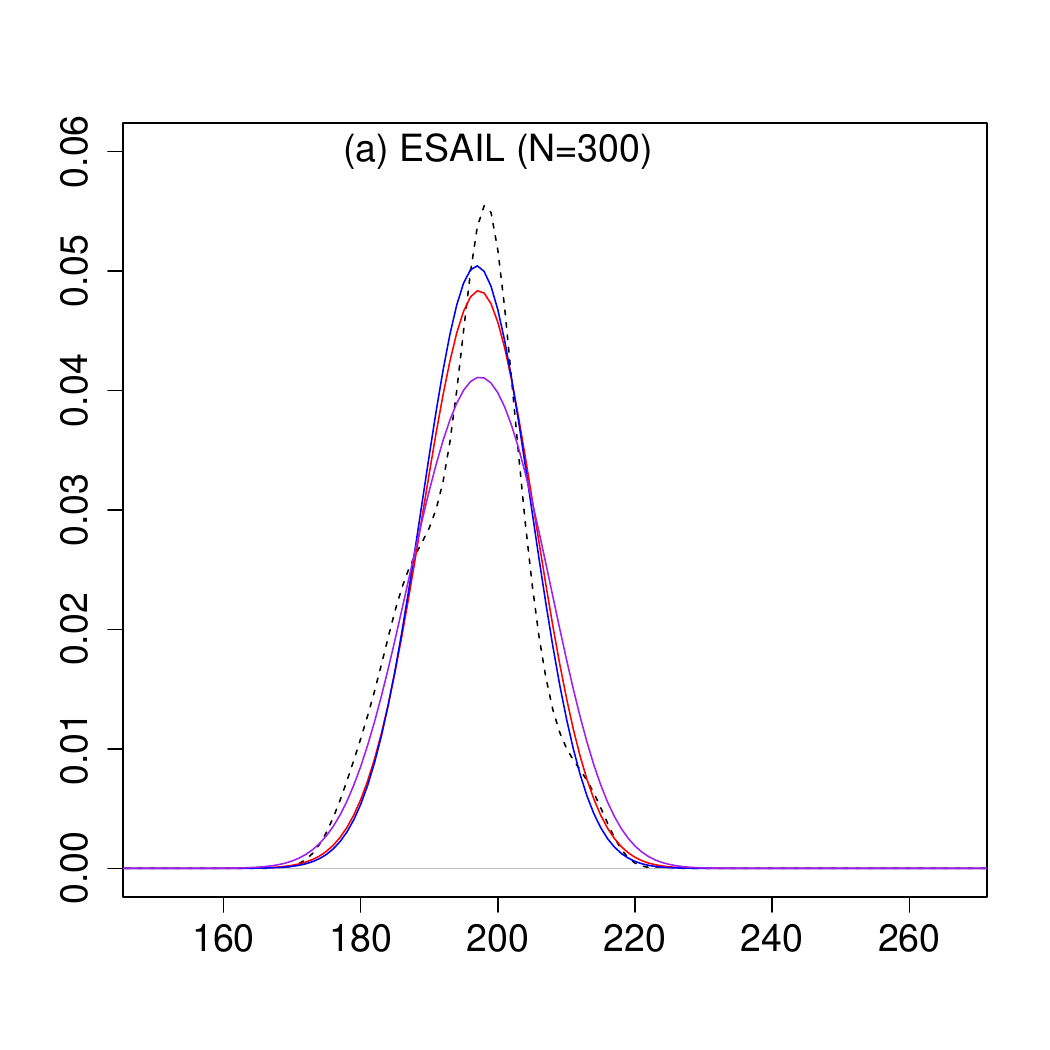}
\end{subfigure}%
\begin{subfigure}{.3\textwidth}
  \includegraphics[width=\PMFW]{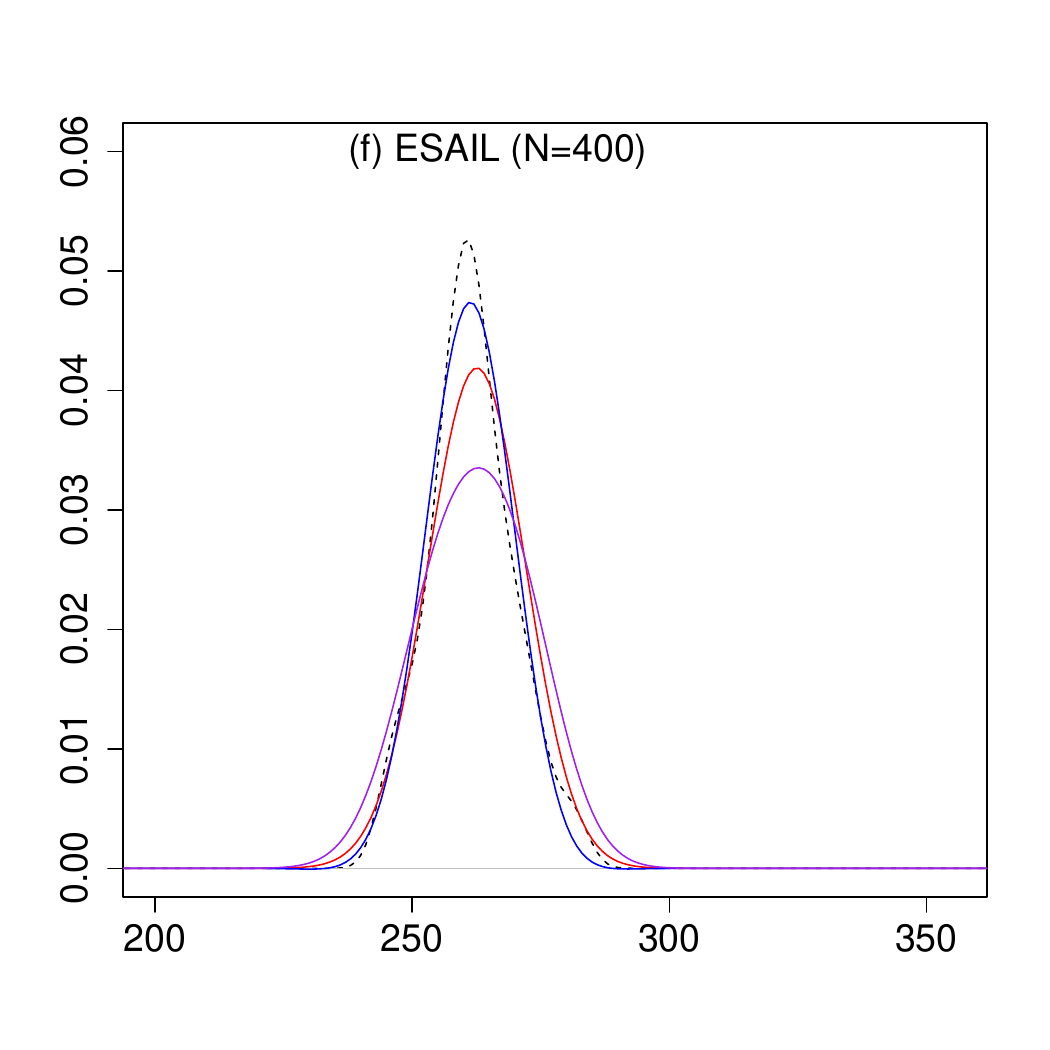}
\end{subfigure}%
\begin{subfigure}{.3\textwidth}
  \includegraphics[width=\PMFW]{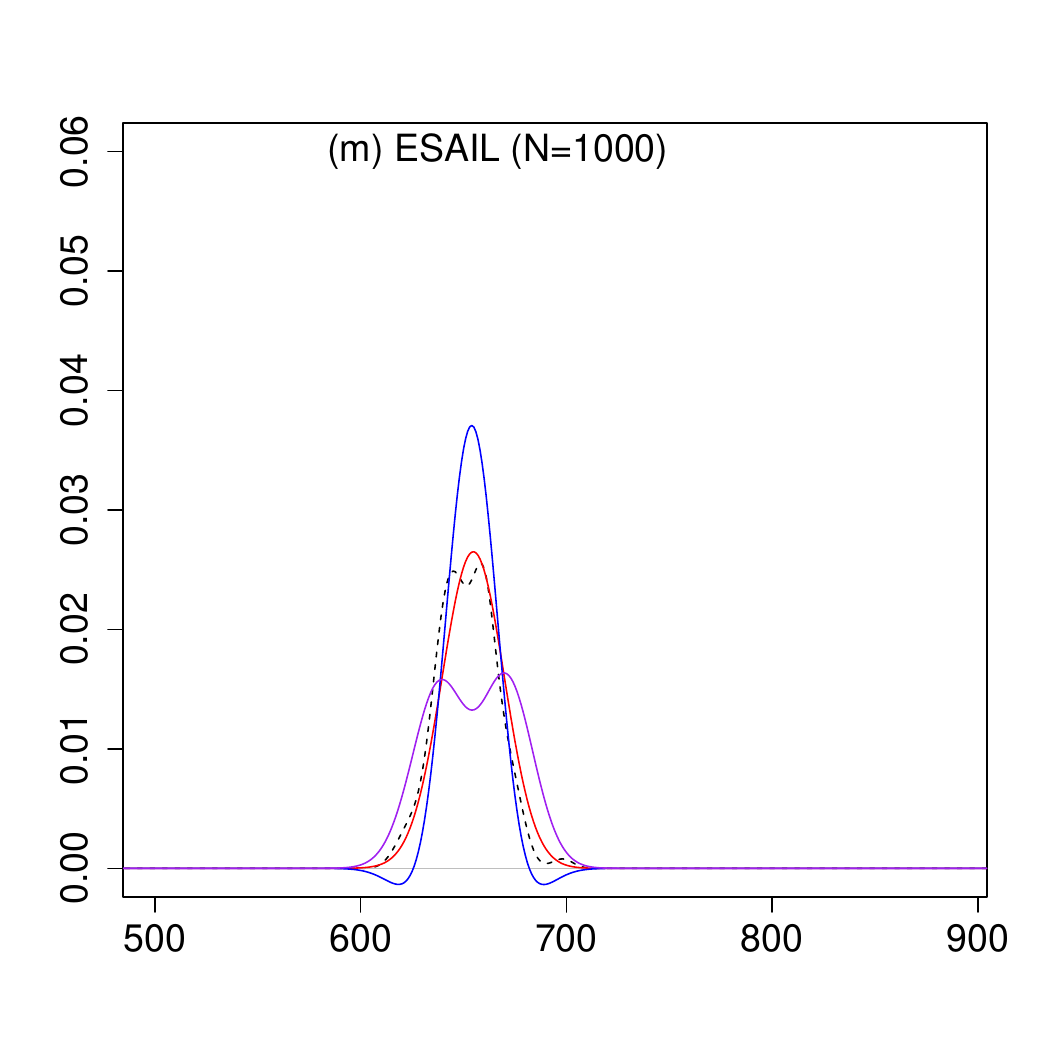}
\end{subfigure}%

\vspace{-1cm}
\begin{subfigure}{.3\textwidth}
  \includegraphics[width=\PMFW]{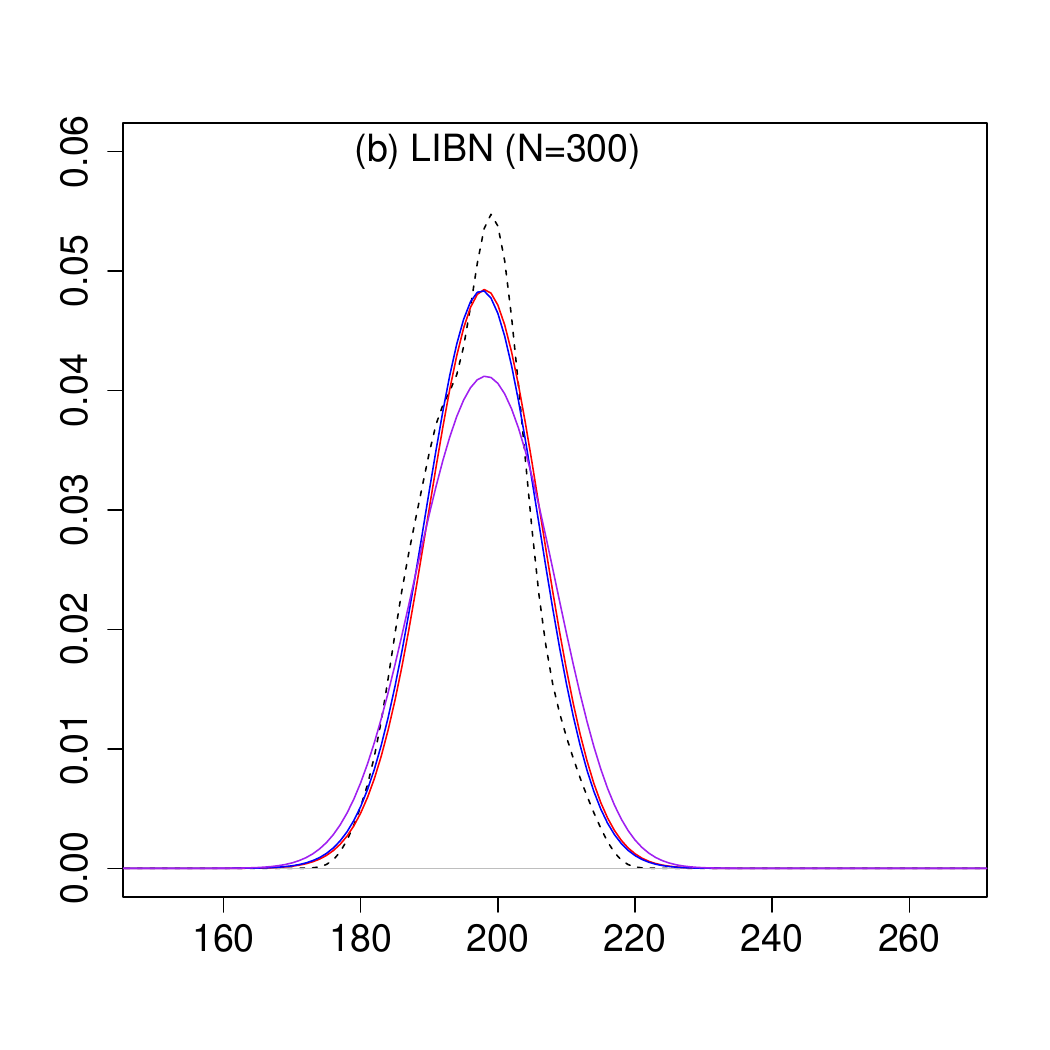}
\end{subfigure}
\begin{subfigure}{.3\textwidth}
  \includegraphics[width=\PMFW]{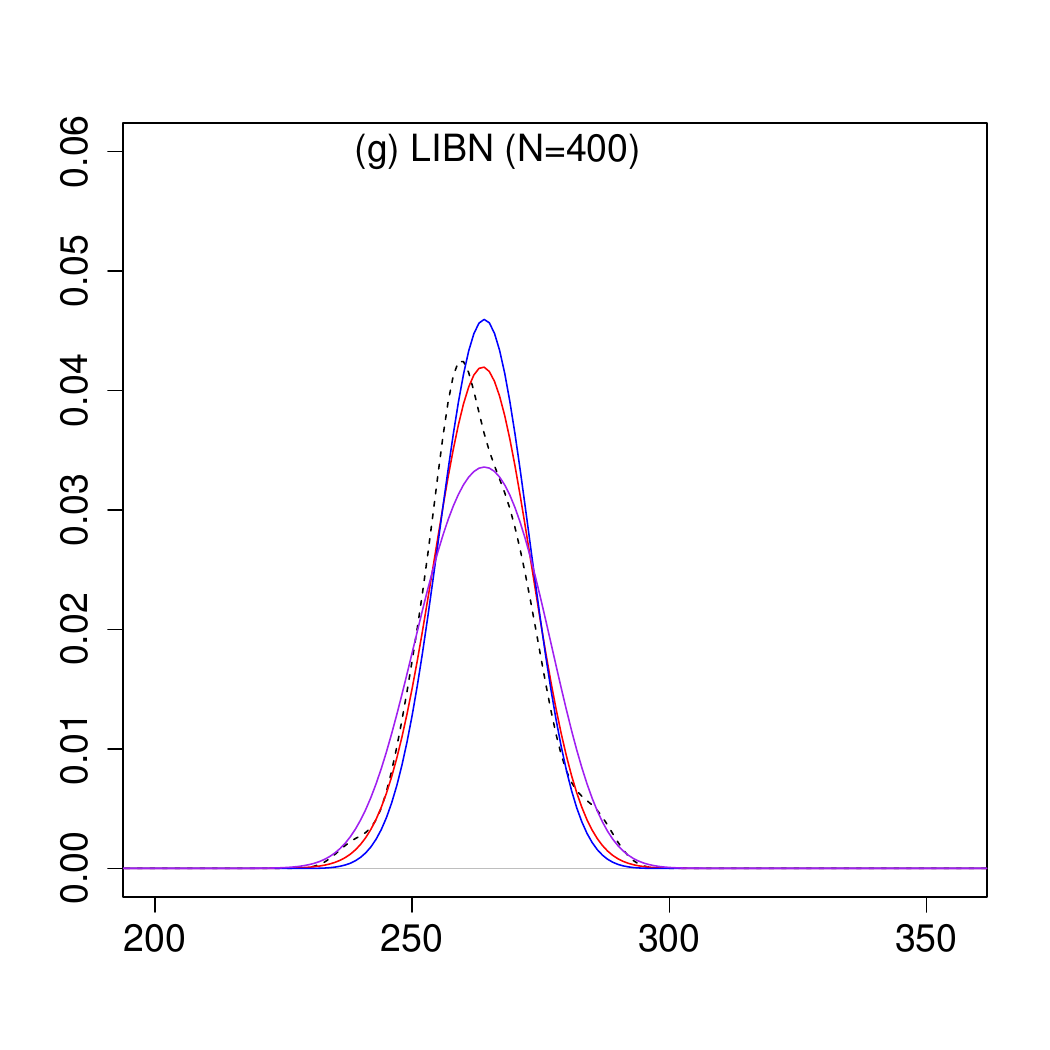}
\end{subfigure}
\begin{subfigure}{.3\textwidth}
  \includegraphics[width=\PMFW]{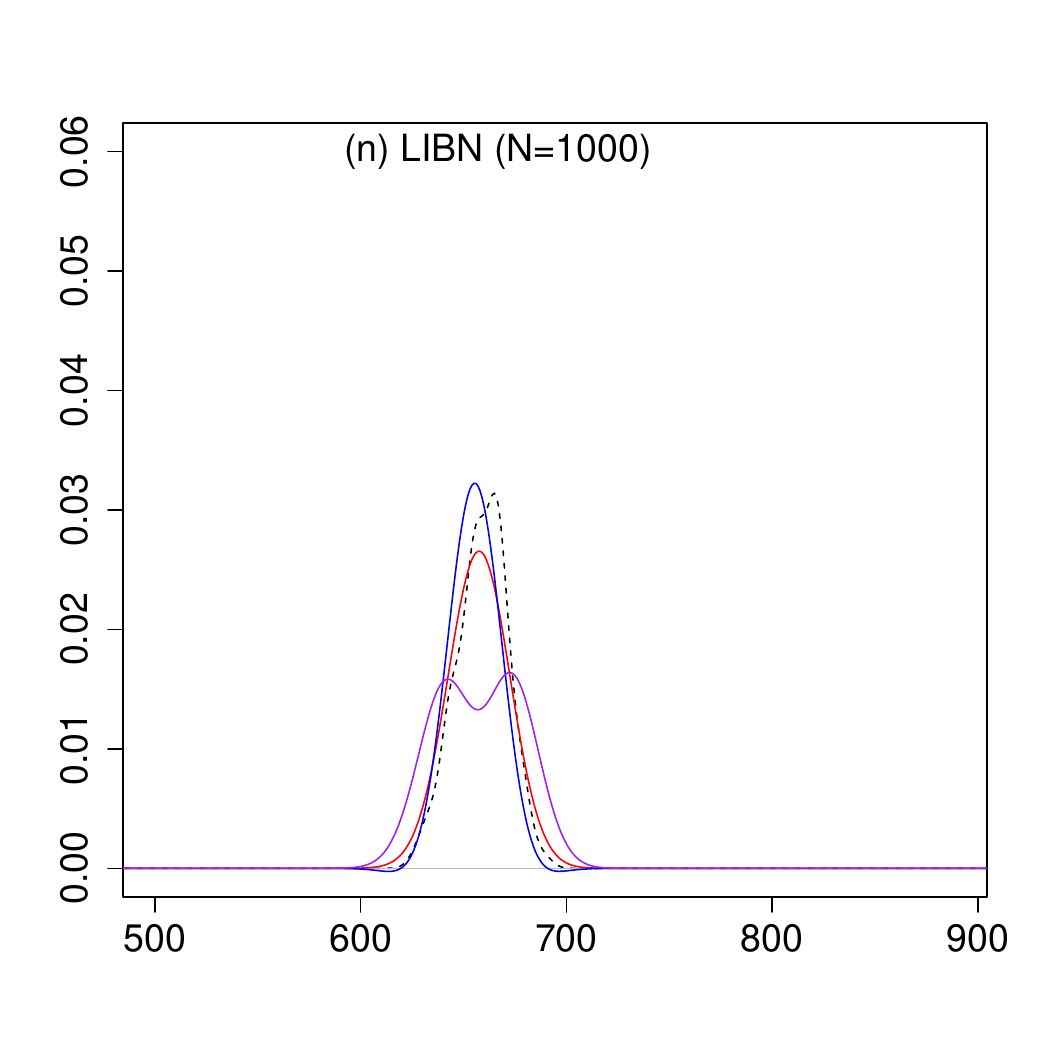}
\end{subfigure}

\vspace{-1cm}
\begin{subfigure}{.3\textwidth}
  \includegraphics[width=\PMFW]{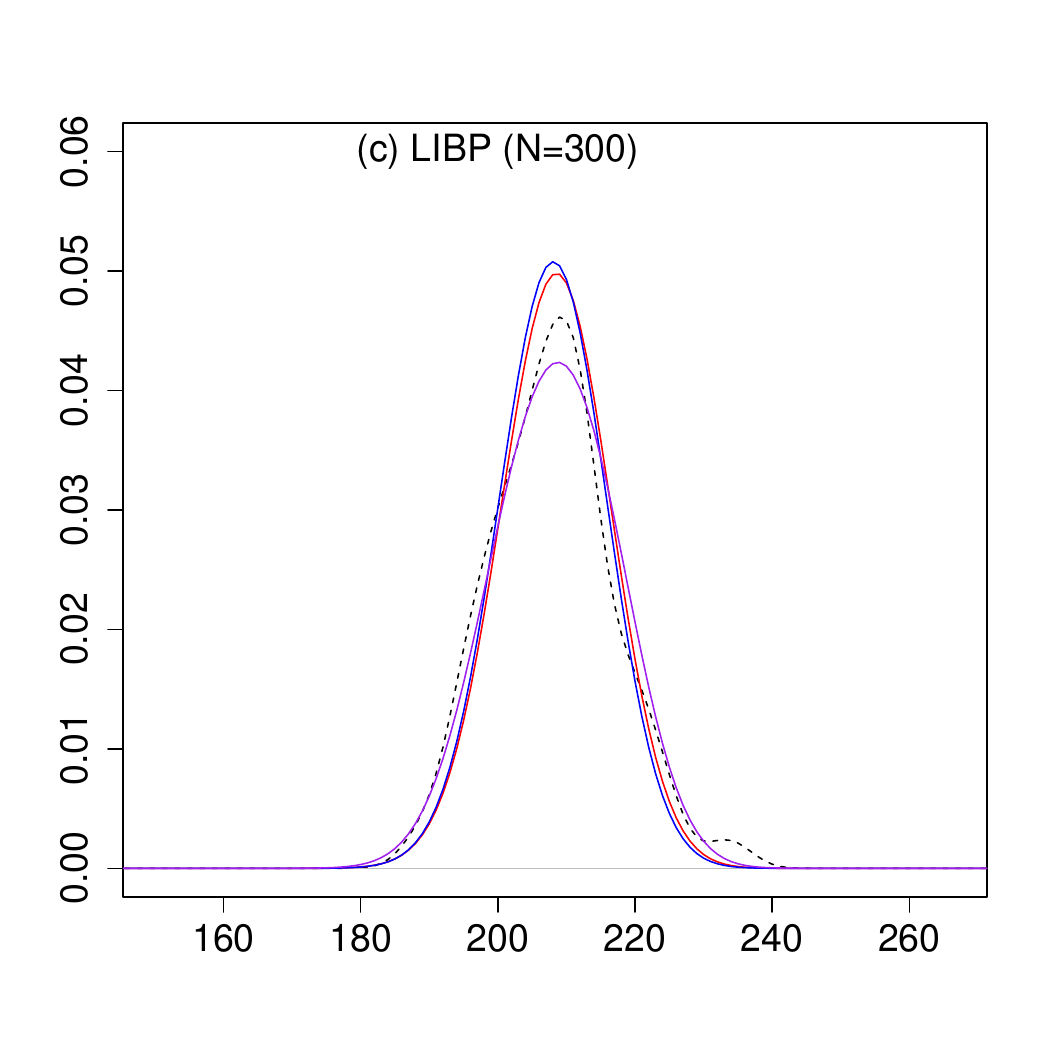}
\end{subfigure}
\begin{subfigure}{.3\textwidth}
  \includegraphics[width=\PMFW]{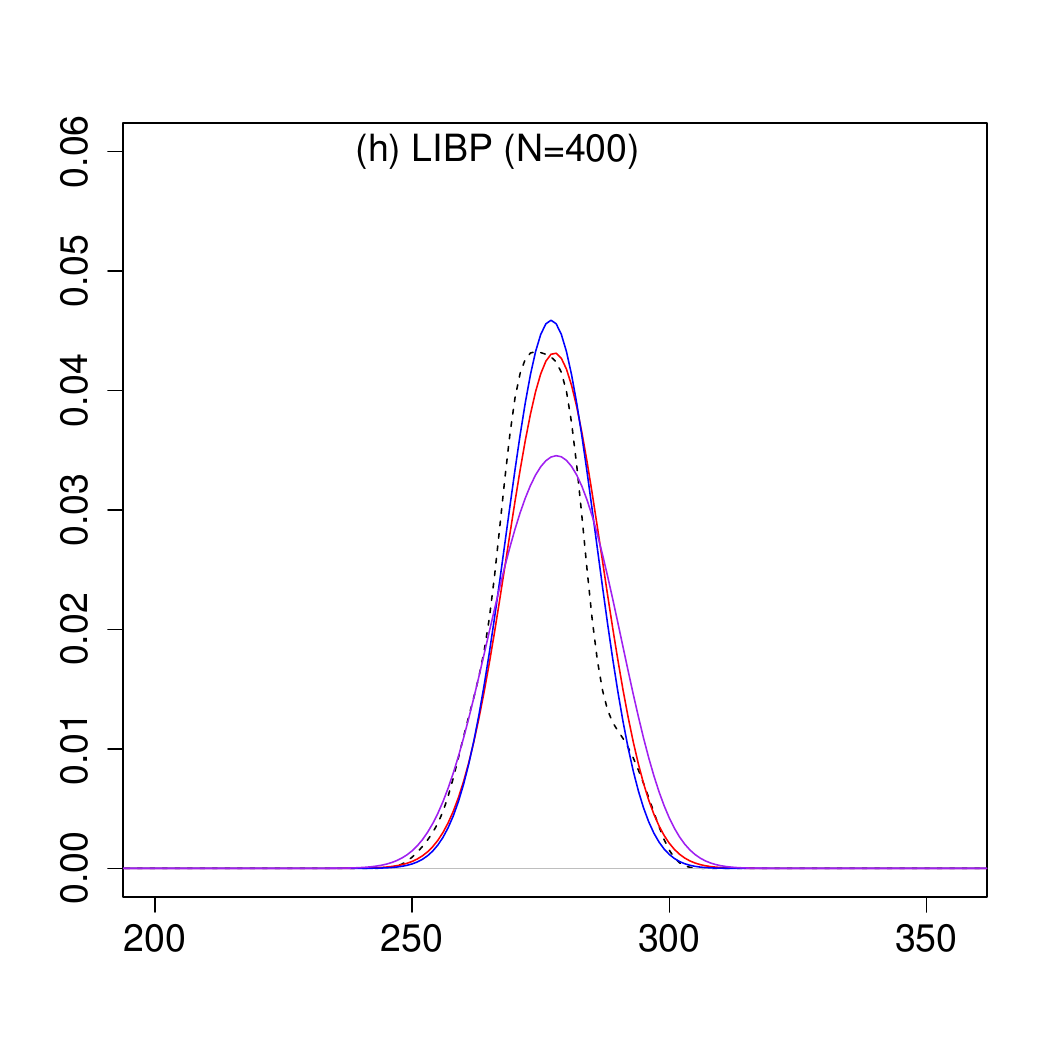}
\end{subfigure}
\begin{subfigure}{.3\textwidth}
  \includegraphics[width=\PMFW]{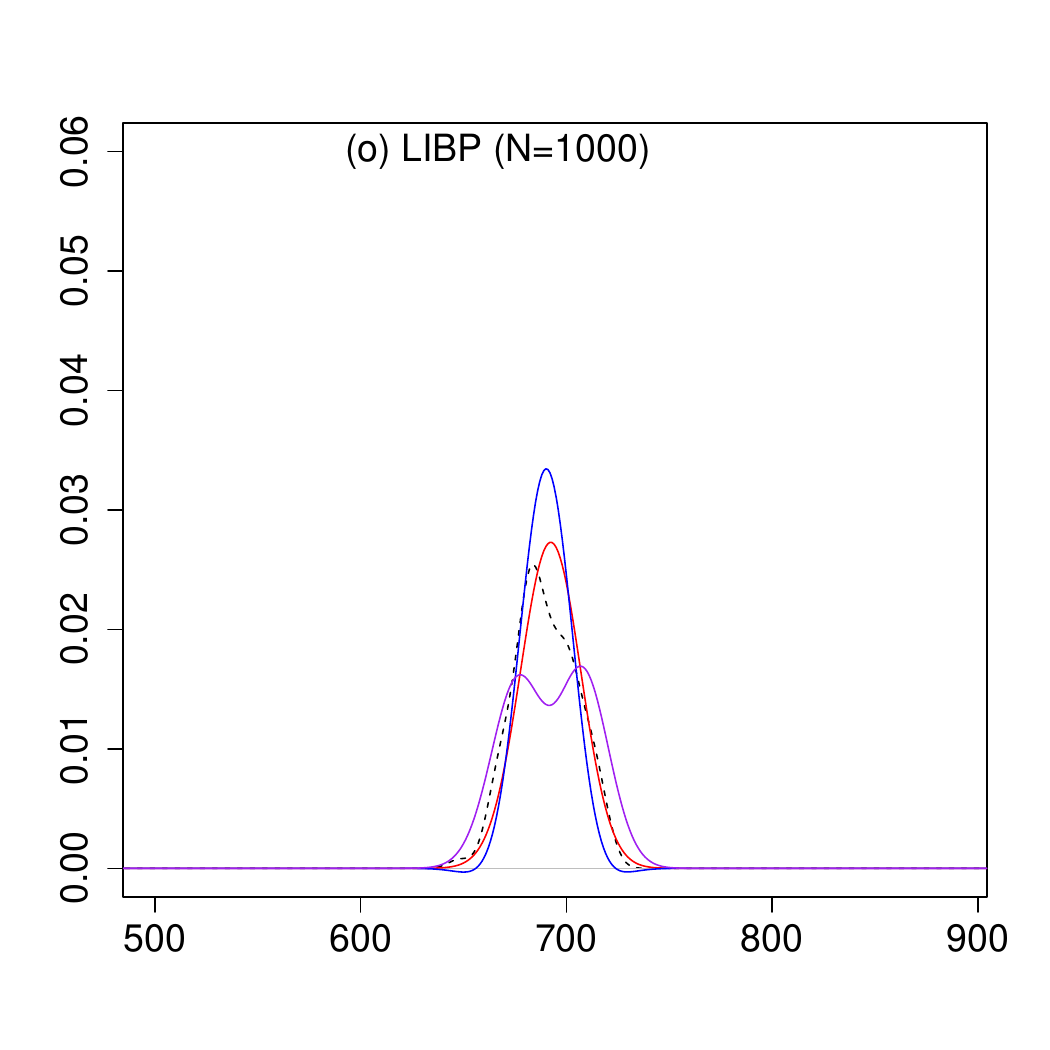}
\end{subfigure}

\vspace{-1cm}
\begin{subfigure}{.3\textwidth}
  \includegraphics[width=\PMFW]{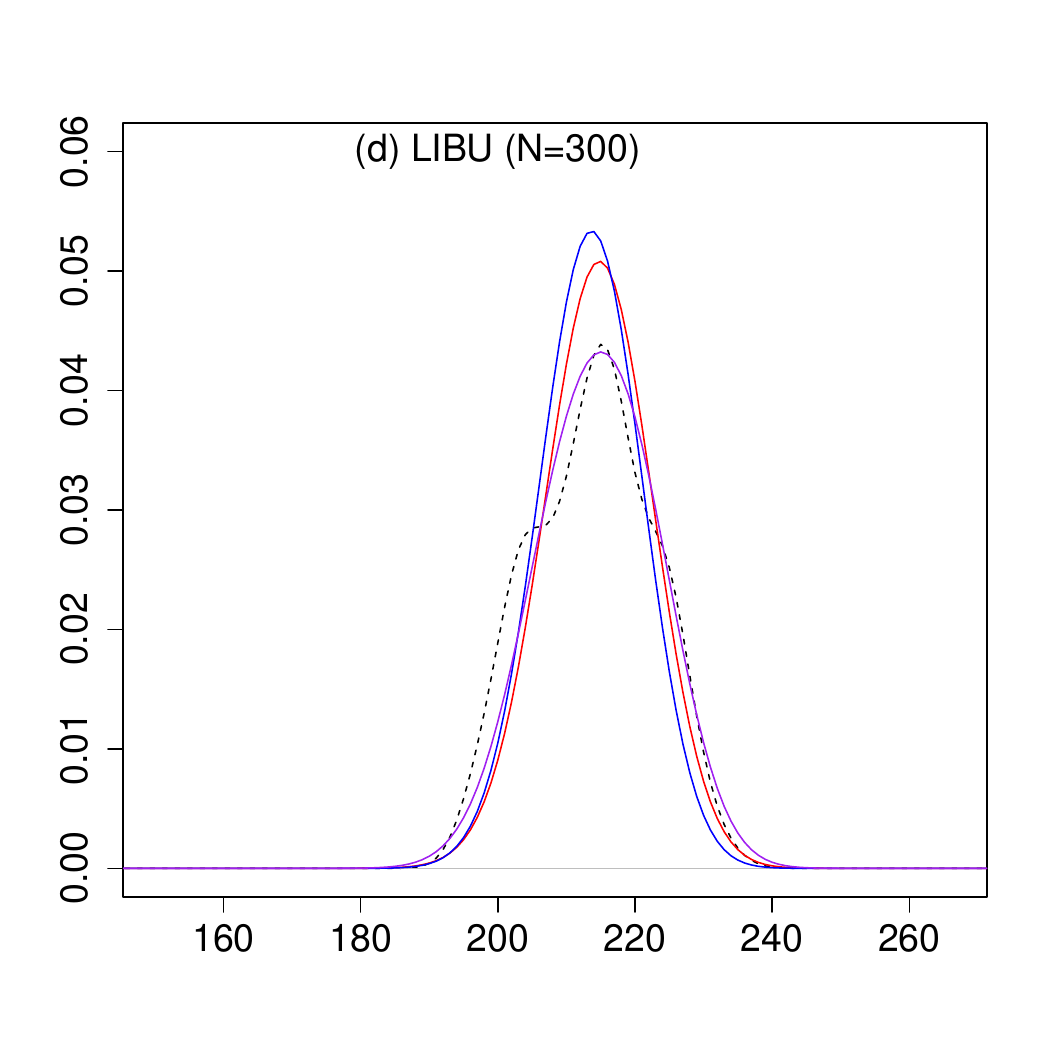}
\end{subfigure}
\begin{subfigure}{.3\textwidth}
  \includegraphics[width=\PMFW]{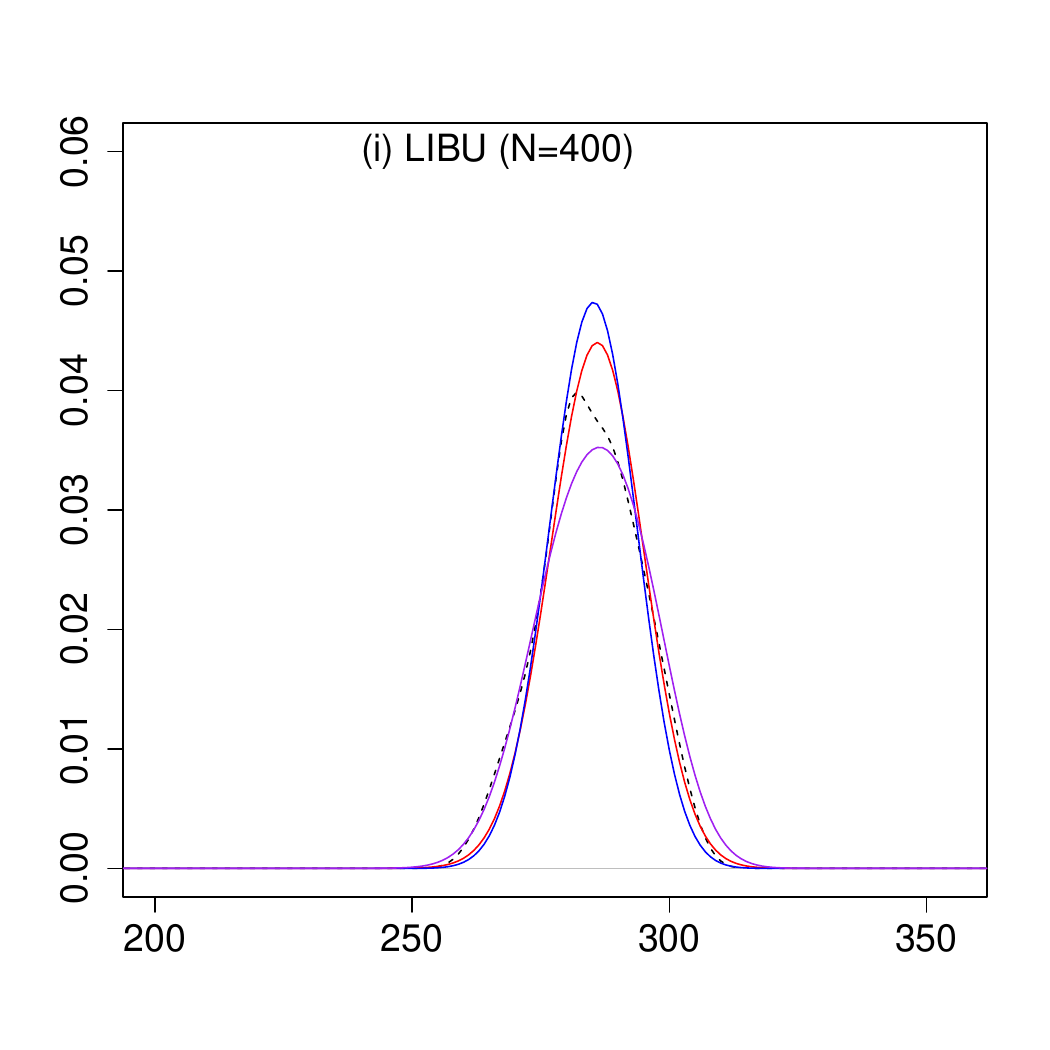}
\end{subfigure}
\begin{subfigure}{.3\textwidth}
  \includegraphics[width=\PMFW]{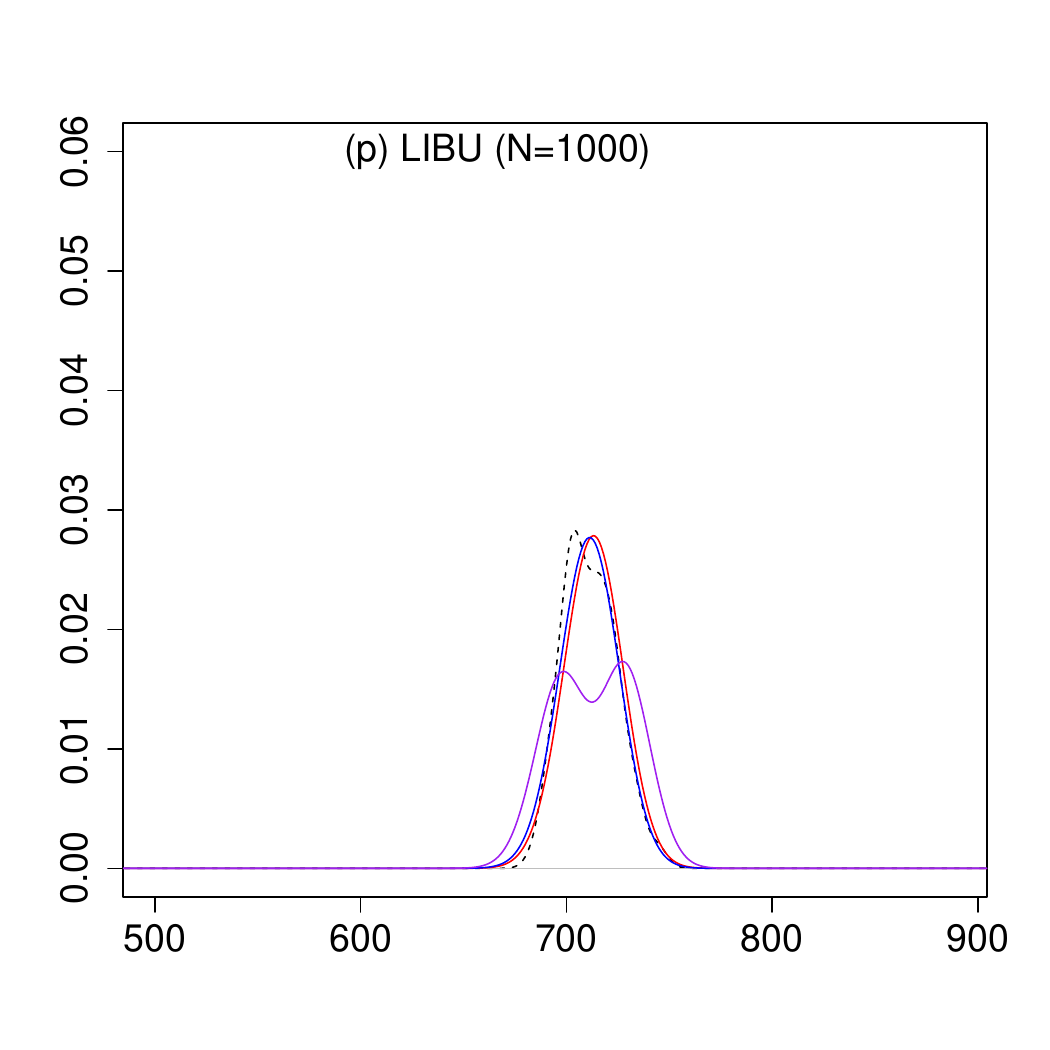}
\end{subfigure}

\vspace{-1cm}
\begin{subfigure}{.3\textwidth}
  \includegraphics[width=\PMFW]{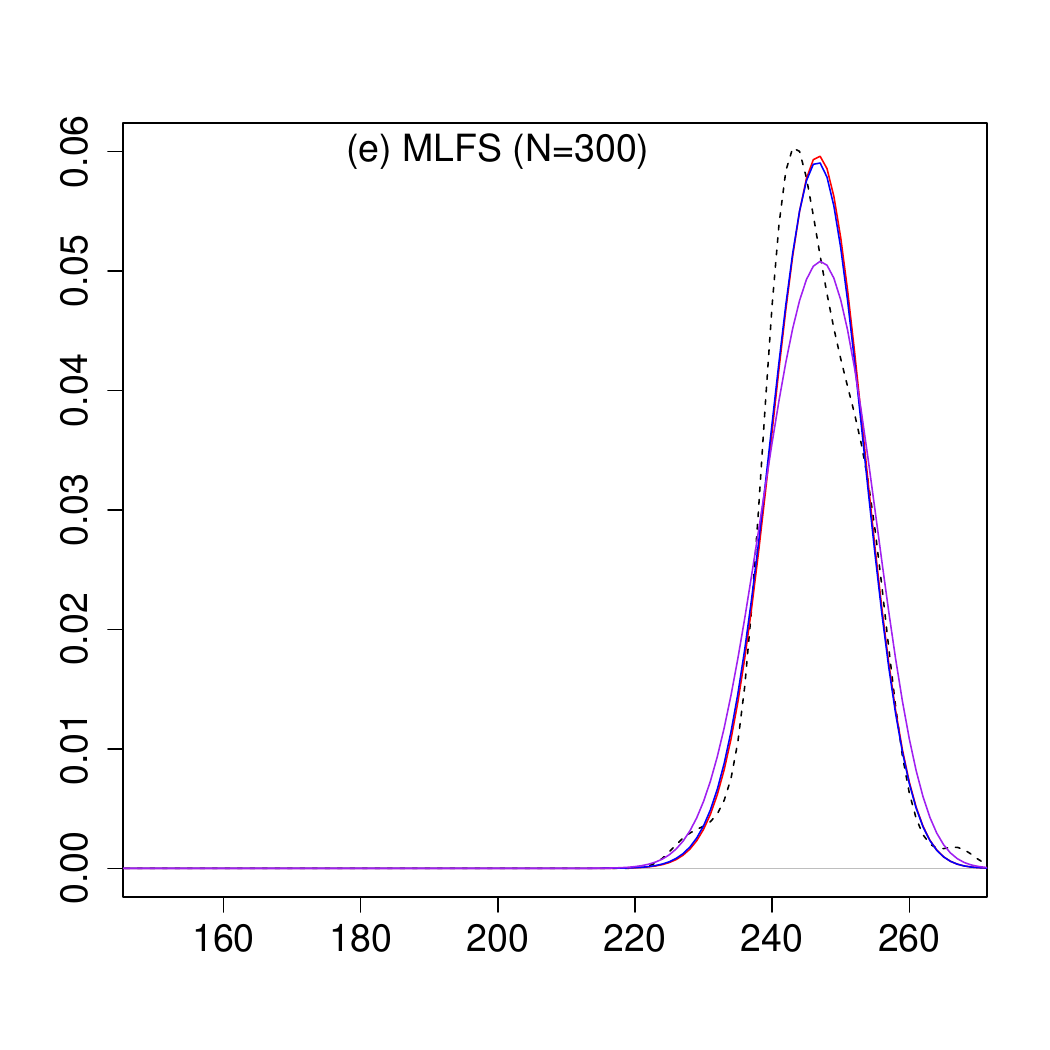}
\end{subfigure}
\begin{subfigure}{.3\textwidth}
  \includegraphics[width=\PMFW]{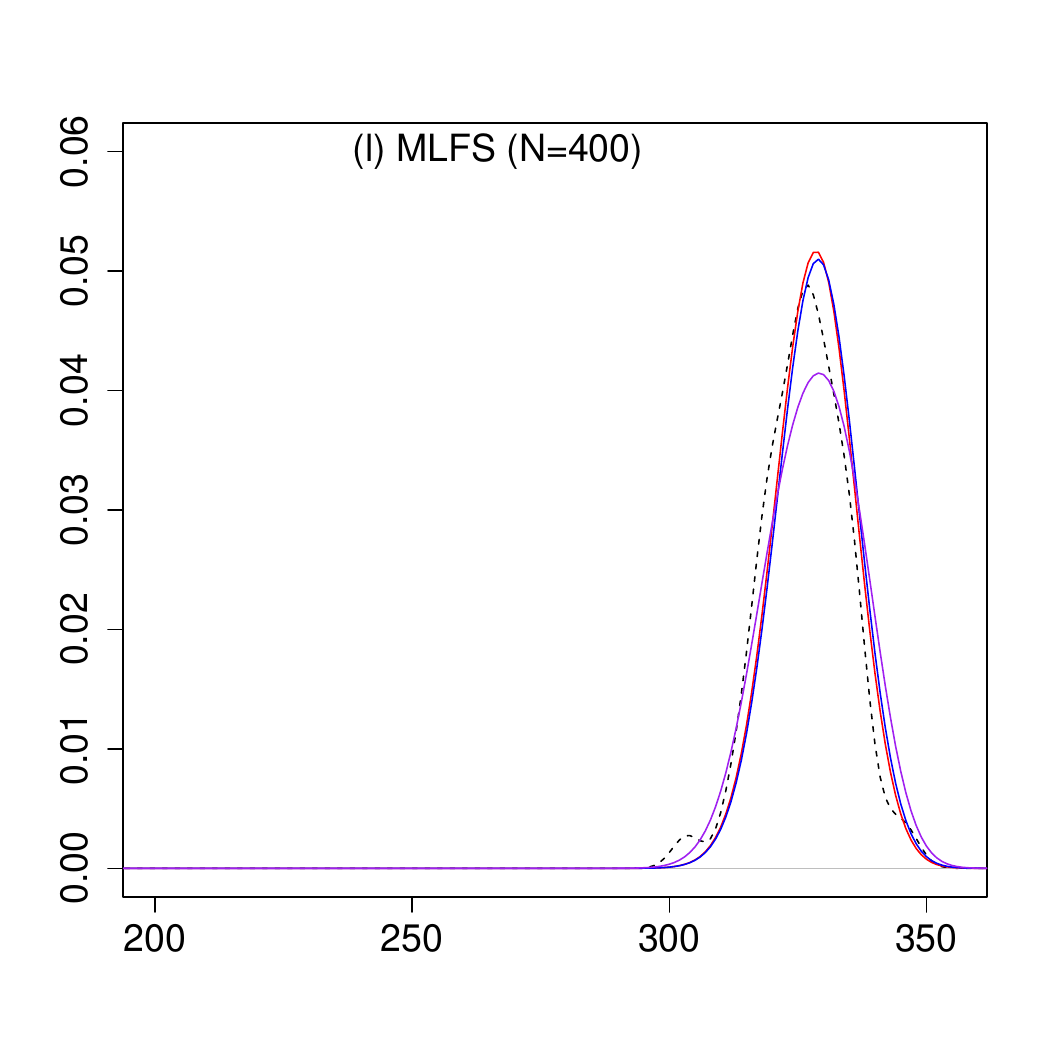}
\end{subfigure}
\begin{subfigure}{.3\textwidth}
  \includegraphics[width=\PMFW]{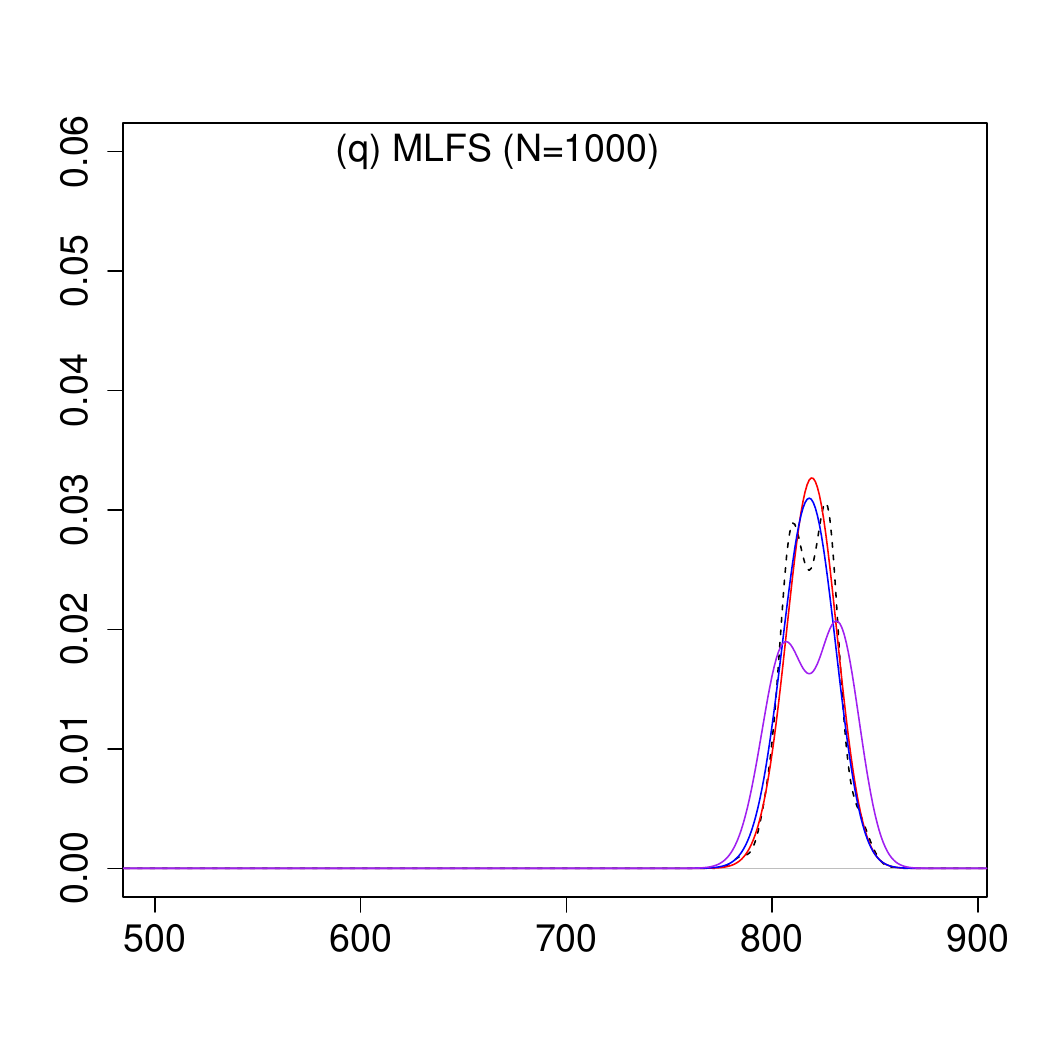}
\end{subfigure}

\caption{Probability distribution for different PMFs for a varying number (N) of mutants being sampled. The x-axis shows the number of mutants being killed, the y-axis the probability. Legend is provided in Figure~\ref{fig:exp:pmf:legend}. } 
\label{fig:exp:PMF}
\end{figure*}

Table~\ref{table:results:PMFC} provides the values estimated for $p$ and $r_2$ (i.e., $\tilde{p}$ and $\tilde{r_2}$) for $\mathit{PMF}_C$, considering the different subjects in our study, for different numbers of sampled mutants. We can notice that the value of  $\tilde{r_2}$ is practically zero in all the cases, which indicates that the effect of covariance is negligible. In most of the cases the covariance is slightly below zero (i.e., the detection of a mutant implies that another mutant is not detected), which we believe to be due to two reasons. 
First, the test suites of our subjects do not test all the implemented functions with the same degree of thoroughness, e.g., since our test suites focus on functional testing, they do not exercise well functions concerning real-time scheduling. Consequently, there is a certain likelihood that, after sampling a mutant belonging to a well tested function, we sample a mutant belonging to a function that is not thoroughly tested. To support this interpretation, we highlight the fact that  \MLFS{}{}, which is the benchmark component with the most thorough test suite, is not characterized by a negative $\tilde{r_2}$. The second reason for a negative $\tilde{r_2}$ is that the correlated binomial distribution may be suboptimal to model our data; we leave the analysis of other distributions to future work.

Table~\ref{table:results:ESS:bin} provides the ESS obtained with $\mathit{PMF}_B$ for different numbers of sampled mutants and the difference ($\Delta$) with respect to $\mathit{ESS}_C$. The difference is small (i.e., $\Delta$ is always below $0.0025$), which indicates that, in practice, \textbf{the binomial distribution approximates the mutation score as well as the correlated binomial distribution}.

Figures~\ref{fig:exp:PMF}-a to Figures~\ref{fig:exp:PMF}-q provide the plots of the curves obtained for $PMF_S$ (dotted black), $PMF_B$ (red), and $PMF_C$ (blue). 
The plots show that for all subjects, $PMF_B$ and $PMF_C$ largely overlap, with 
$PMF_C$ being slightly narrower and taller. The consequence is that $PMF_B$ tends to be more conservative than $PMF_C$.
Consequently, $PMF_C$ leads to a computation of a larger confidence interval, which, in turn, for FSCI, leads to the sampling of more mutants. Since this negatively only affects the scalability of mutation analysis (i.e., a slightly higher number of mutants will be tested) but does not negatively affect the quality of results, we can argue that $PMF_B$ estimates the mutation score as accurately as $PMF_C$.

Concerning the area under curve, we can observe that both $PMF_B$ and $PMF_C$ cover most of the AUC of $PMF_S$. 
Table~\ref{table:results:AUC} provides the values for $\mathit{PAUC}_B$ and $\mathit{PAUC}_C$; they are very close, with 
$PMF_B$ performing better than $PMF_C$ in 11 out of 15 cases, which shows that, in our context, the binomial distribution appropriately estimates the distribution of the mutation score.

To discuss what might happen when trials are affected by high covariance, in the charts of Figure~\ref{fig:exp:PMF} we also plotted the curve of $PMF_C$ (hereafter, $PMF_C^{10}$) obtained with $p$ equal to the population mean (i.e., the actual mutation score) and $r_2 = 0.0010$. 
We can notice that for $PMF_C^{10}$ the distribution is slightly more spread out, which is in line with the statistics literature indicating that (1) a distribution has more variance with a positive correlation among the Bernoulli trials~\cite{diniz2010}~\cite{Zhang:CrrelatedFirearm:NIST:2019} and (2) the sample size required in sequential tests should be increased proportionally to the value of the  correlation coefficient~\cite{Mingoti:2003}.
In practice, in the presence of higher correlations than the ones observed in experimental subjects, 
the confidence interval estimated by FSCI, which assumes a binomial distribution, could become unreliable (i.e., smaller than it should be); however, such situation does not occur in our context, where subject are characterized by limited correlations.

\section{Details about the savings obtained with the \APPR test suite}
\label{appendix:details:rq5}
\begin{table}[tb]
\caption{RQ5. Savings (Execution Time and Number of Test Cases) with the \APPR Test Suite.}
\label{table:results:reduction:prioritize} 
\scriptsize
\centering
\begin{tabular}{
|>{\raggedleft\arraybackslash}p{12mm}@{\hspace{1pt}}|
@{\hspace{1pt}}>{\raggedleft\arraybackslash}p{4mm}@{\hspace{1pt}}|
@{\hspace{1pt}}>{\raggedleft\arraybackslash}p{6mm}@{\hspace{1pt}}|
>{\raggedleft\arraybackslash}p{6mm}@{\hspace{1pt}}|
@{\hspace{1pt}}>{\raggedleft\arraybackslash}p{6mm}@{\hspace{1pt}}|
@{\hspace{1pt}}>{\raggedleft\arraybackslash}p{6mm}@{\hspace{1pt}}|
@{\hspace{1pt}}>{\raggedleft\arraybackslash}p{6mm}@{\hspace{1pt}}|
>{\raggedleft\arraybackslash}p{6mm}@{\hspace{1pt}}|
@{\hspace{1pt}}>{\raggedleft\arraybackslash}p{6mm}@{\hspace{1pt}}|
@{\hspace{1pt}}>{\raggedleft\arraybackslash}p{6mm}@{\hspace{1pt}}|
@{\hspace{1pt}}>{\raggedleft\arraybackslash}p{6mm}@{\hspace{1pt}}|
}
\hline
\textbf{}  &    &   &  \multicolumn{4}{c|}{\textbf{Time Savings [\%]}}    & \multicolumn{4}{c|}{\textbf{Test Savings [\%]}}  \\
\textbf{Subject} & \textbf{D} & \textbf{MS}
& \textbf{Min} & \textbf{Max} & \textbf{Med} & \textbf{Mn} 
&\textbf{Min} & \textbf{Max} & \textbf{Med} & \textbf{Mn} \\
\hline
\GCSP{} & $D_J$   & ALL      & 13.01 & 13.50 & 13.36 & 13.32 & 33.11 & 33.17 & 33.16 & 33.15 \\
\PARAM{} & $D_J$   & ALL      & 16.81 & 16.82 & 16.82 & 16.81 & 14.17 & 14.17 & 14.17 & 14.17 \\
\UTIL{}  & $D_J$   & ALL      & 4.05  & 4.29  & 4.18  & 4.17  & 9.41  & 9.44  & 9.42  & 9.42  \\
\MLFS{}{}     & $D_J$   & ALL      & 7.18  & 7.21  & 7.20  & 7.20  & 13.29 & 13.33 & 13.31 & 13.31 \\
\SAIL{}$_S$    & $D_J$   & ALL      & 52.78 & 52.85 & 52.82 & 52.82 & 91.79 & 91.81 & 91.79 & 91.80 \\
\GCSP{} & $D_O$    & ALL      & 13.07 & 13.49 & 13.40 & 13.37 & 33.11 & 33.17 & 33.15 & 33.14 \\
\PARAM{} & $D_O$    & ALL      & 16.81 & 16.82 & 16.81 & 16.81 & 14.17 & 14.17 & 14.17 & 14.17 \\
\UTIL{}  & $D_O$    & ALL      & 4.09  & 4.27  & 4.18  & 4.19  & 9.41  & 9.43  & 9.42  & 9.42  \\
\MLFS{}{}     & $D_O$    & ALL      & 7.18  & 7.20  & 7.19  & 7.19  & 13.29 & 13.33 & 13.30 & 13.31 \\
\SAIL{}$_S$    & $D_O$    & ALL      & 52.78 & 52.83 & 52.81 & 52.81 & 91.76 & 91.82 & 91.78 & 91.79 \\
\GCSP{} & $D_C$    & ALL      & -0.53 & -0.26 & -0.34 & -0.39 & 9.39  & 9.40  & 9.39  & 9.39  \\
\PARAM{} & $D_C$    & ALL      & 14.86 & 14.88 & 14.87 & 14.87 & 11.00 & 11.00 & 11.00 & 11.00 \\
\UTIL{}  & $D_C$    & ALL      & 2.03  & 2.25  & 2.20  & 2.17  & 4.80  & 4.84  & 4.81  & 4.82  \\
\MLFS{}{}     & $D_C$    & ALL      & 1.53  & 1.53  & 1.53  & 1.53  & 6.64  & 6.64  & 6.64  & 6.64  \\
\SAIL{}$_S$    & $D_C$    & ALL      & 5.97  & 6.44  & 6.24  & 6.27  & 43.05 & 43.27 & 43.20 & 43.20 \\
\GCSP{} & $D_E$ & ALL      & -0.21 & -0.03 & -0.08 & -0.10 & 10.37 & 10.37 & 10.37 & 10.37 \\
\PARAM{} & $D_E$ & ALL      & 14.66 & 14.69 & 14.68 & 14.68 & 10.44 & 10.44 & 10.44 & 10.44 \\
\UTIL{}  & $D_E$ & ALL      & 2.15  & 2.28  & 2.22  & 2.22  & 4.84  & 4.87  & 4.85  & 4.85  \\
\MLFS{}{}     & $D_E$ & ALL      & 0.71  & 0.73  & 0.72  & 0.72  & 5.41  & 5.46  & 5.45  & 5.45  \\
\SAIL{}$_S$    & $D_E$ & ALL      & -4.01 & -3.74 & -3.94 & -3.89 & 38.60 & 38.83 & 38.76 & 38.73 \\
\hline
\GCSP{} & $D_J$   & FSCI & 77.65 & 83.20 & 78.91 & 79.90 & 72.54 & 80.31 & 76.93 & 76.40 \\
\PARAM{} & $D_J$   & FSCI & 75.53 & 79.83 & 77.15 & 77.47 & 68.76 & 71.07 & 69.95 & 70.06 \\
\UTIL{}  & $D_J$   & FSCI & 88.54 & 90.33 & 89.50 & 89.40 & 77.38 & 79.53 & 77.87 & 78.13 \\
\MLFS{}{}     & $D_J$   & FSCI & 72.09 & 82.05 & 79.87 & 78.33 & 61.19 & 75.37 & 72.26 & 69.94\\
\SAIL{}$_S$    & $D_J$   & FSCI & 88.64 & 90.83 & 89.53 & 89.53 & 88.80 & 90.52 & 89.51 & 89.55 \\
\GCSP{} & $D_O$    & FSCI & 77.63 & 83.22 & 78.89 & 79.91 & 72.54 & 80.35 & 76.94 & 76.40 \\
\PARAM{} & $D_O$    & FSCI & 75.53 & 79.83 & 77.16 & 77.47 & 68.76 & 71.07 & 69.95 & 70.06 \\
\UTIL{}  & $D_O$    & FSCI & 88.54 & 90.33 & 89.50 & 89.40 & 77.38 & 79.53 & 77.87 & 78.13 \\
\MLFS{}{}     & $D_O$    & FSCI & 72.09 & 82.06 & 79.87 & 78.33 & 61.19 & 75.37 & 72.27 & 69.94 \\
\SAIL{}$_S$    & $D_O$    & FSCI & 88.64 & 90.83 & 89.53 & 89.57 & 88.80 & 90.52 & 89.51 & 89.55 \\
\GCSP{} & $D_C$    & FSCI & 76.24 & 82.85 & 77.57 & 78.77 & 70.60 & 78.03 & 73.72 & 74.29 \\
\PARAM{} & $D_C$    & FSCI & 75.25 & 79.63 & 76.94 & 77.22 & 68.28 & 70.69 & 69.53 & 69.65 \\
\UTIL{}  & $D_C$    & FSCI & 88.48 & 90.28 & 89.43 & 89.33 & 77.26 & 79.41 & 77.70 & 77.98 \\
\MLFS{}{}     & $D_C$    & FSCI & 80.00 & 82.28 & 81.45 & 81.16 & 72.36 & 75.40 & 74.14 & 73.86 \\
\SAIL{}$_S$    & $D_C$    & FSCI & 75.27 & 90.83 & 83.25 & 83.04 & 81.19 & 90.52 & 86.41 & 86.00 \\
\GCSP{} & $D_E$ & FSCI & 76.27 & 82.85 & 77.61 & 78.80 & 70.66 & 78.18 & 73.85 & 74.41 \\
\PARAM{} & $D_E$ & FSCI & 75.19 & 79.59 & 76.90 & 77.18 & 68.08 & 70.48 & 69.44 & 69.55 \\
\UTIL{}  & $D_E$ & FSCI & 88.48 & 90.28 & 89.42 & 89.33 & 77.24 & 79.42 & 77.69 & 77.98 \\
\MLFS{}{}     & $D_E$ & FSCI & 80.00 & 82.26 & 81.43 & 81.15 & 72.36 & 75.38 & 74.12 & 73.84 \\
\SAIL{}$_S$    & $D_E$ & FSCI & 73.53 & 90.83 & 82.40 & 82.14 & 80.50 & 90.52 & 86.03 & 85.69 \\

\GCSP{} & \textit{Full}    & FSCI & 80.35 & 82.35 & 81.55 & 81.53 & 67.25 & 71.68 & 69.44 & 69.50 \\
\PARAM{} & \textit{Full}    & FSCI & 65.76 & 72.03 & 67.86 & 68.40 & 49.98 & 58.91 & 53.39 & 54.31 \\
\UTIL{}  & \textit{Full}    & FSCI & 82.18 & 84.72 & 83.18 & 83.23 & 55.06 & 61.16 & 57.35 & 57.97 \\
\MLFS{}{}     & \textit{Full}    & FSCI & 66.59 & 76.03 & 70.29 & 70.89 & 52.56 & 65.73 & 57.69 & 58.54 \\
\SAIL{}$_S$    & \textit{Full}    & FSCI & 73.46 & 75.47 & 74.48 & 74.51 & 78.94 & 80.54 & 79.61 & 79.71 \\
\hline
\end{tabular}
\textbf{Legend:} D, distance; Full, no distance measure, the full test suite has been considered; MS, mutants set; Med, median; Mn, Mean.
\end{table}

Table~\ref{table:results:reduction:prioritize} provides additional details about the data plotted in
Figures 3 and 4 of the main manuscript.

\ifCLASSOPTIONcompsoc
  \section*{Acknowledgments}
\else
  \section*{Acknowledgment}
\fi

This work has been funded by the European Space Agency (ITT-1-9873/FAQAS),
the European Research Council (ERC) under the European Union’s Horizon 2020 research and innovation programme (grant agreement No 694277), and NSERC Discovery and Canada Research Chair programs. Authors would like to thank the ESA ESTEC officers, the \EduardoSpace team, 
\YAGO{}
 and \YagoSpace software engineers for their valuable support.

\ifCLASSOPTIONcaptionsoff
  \newpage
\fi

\bibliographystyle{./bibliography/IEEEtran}
\bibliography{./bibliography/mutationTesting}

\begin{IEEEbiography}[{\includegraphics[width=1in,height=1.25in,clip,keepaspectratio]{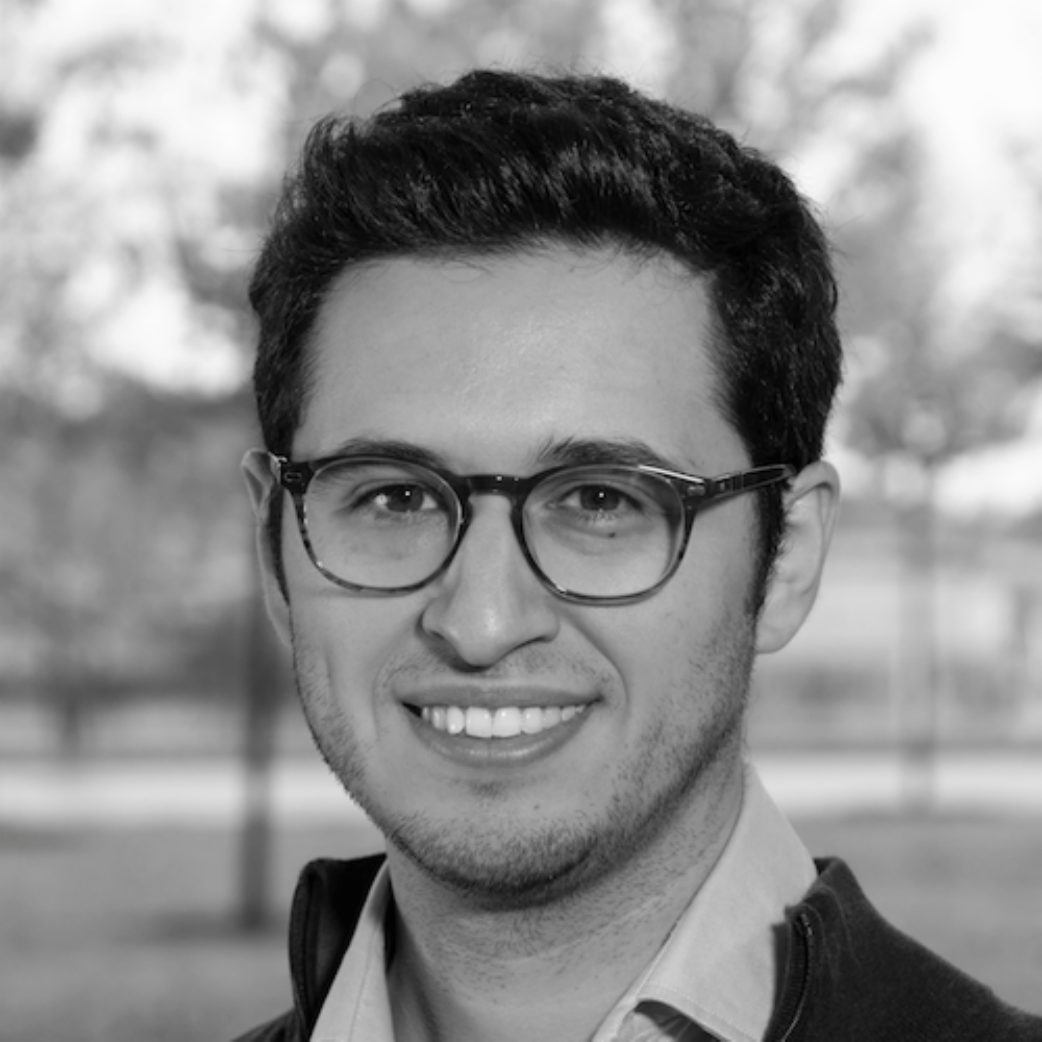}}]{Oscar Cornejo} 
is a Research Associate at the Interdisciplinary Centre for Security, Reliability and Trust (SnT), University of Luxembourg. He obtained his PhD degree in Computer Science from the University of Milano - Bicocca in 2019.

His research interests are in software engineering, focusing on automated software testing and program analysis. He is currently involved in research projects with industry partners from the space domain. 
\end{IEEEbiography}

\begin{IEEEbiography}[{\includegraphics[width=1in,height=1.25in,clip,keepaspectratio]{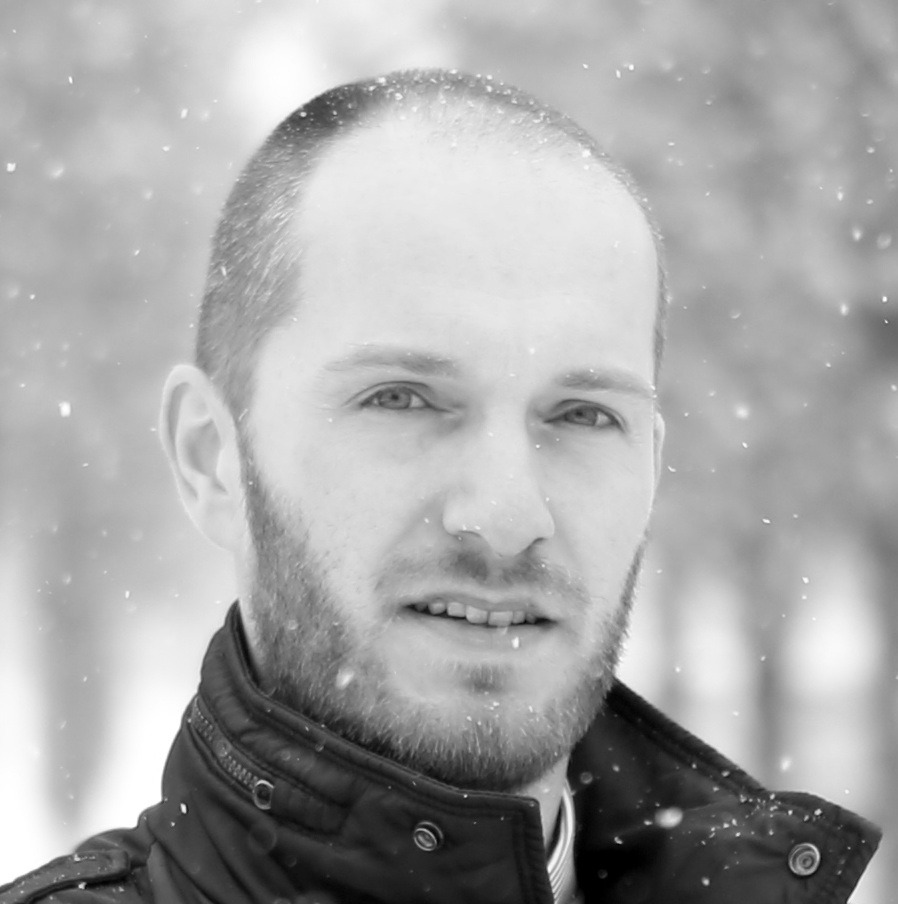}}]{Fabrizio Pastore}
is Chief Scientist II at the Interdisciplinary Centre for Security, Reliability and Trust (SnT), University of Luxembourg. He obtained his PhD in Computer Science in 2010 from the University of Milano - Bicocca.

His research interests concern automated software testing, including security testing and testing of AI-based systems; his work relies on the integrated analysis of different types of artefacts (e.g., requirements,  models, source code, and execution traces). He is active in several industry partnerships and national, ESA, and EU-funded research projects.
\end{IEEEbiography}

\begin{IEEEbiography}[{\includegraphics[width=1in,height=1.25in,clip,keepaspectratio]{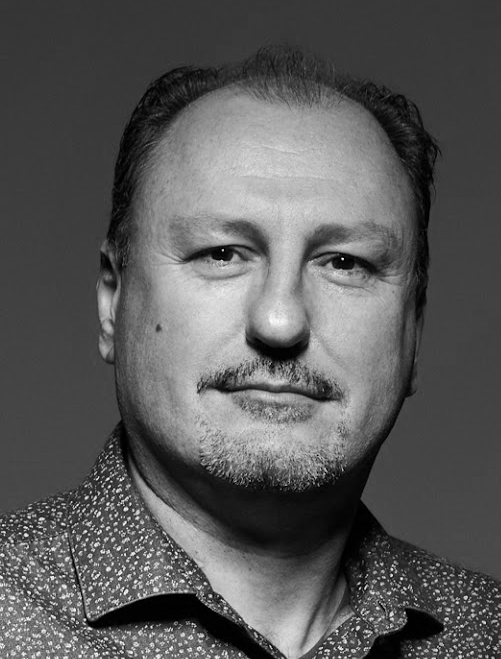}}]{Lionel C. Briand} is professor of software engineering and has shared appointments between (1) School of Electrical Engineering and Computer Science, University of Ottawa, Canada and (2) The SnT centre for Security, Reliability, and Trust, University of Luxembourg. He is the head of the SVV department at the SnT Centre and a Canada Research Chair in Intelligent Software Dependability and Compliance (Tier 1).

He holds an ERC Advanced Grant, the most prestigious European individual research award, and has conducted applied research in collaboration with industry for more than 25 years, including projects in the automotive, aerospace, manufacturing, financial, and energy domains. He was elevated to the grades of IEEE and ACM fellow, granted the IEEE Computer Society Harlan Mills award (2012) and the IEEE Reliability Society Engineer-of-the-year award (2013) for his work on software verification and testing. 
His research interests include: Testing and verification, search-based software engineering, model-driven development, requirements engineering, and empirical software engineering.
\end{IEEEbiography}

\end{document}